\documentclass[11pt]{article}%\pdfoutput=1
\usepackage{cite}
\usepackage{scalerel}
\usepackage{xcolor}
\usepackage{shuffle}
\usepackage{amsmath,amsfonts,amssymb}
\usepackage{latexsym,epsfig}
\usepackage{dsfont}
\usepackage{hyperref}
\usepackage{extarrows}
\usepackage{comment} % To comment several lines at once.
\usepackage{tikz-cd}
\usepackage{tikz}
\usepackage{calrsfs}
\DeclareMathAlphabet{\pazocal}{OMS}{zplm}{m}{n}
\usetikzlibrary{decorations.pathreplacing}
\def\hybrid{
        \topmargin -20pt
        \oddsidemargin 0pt
        \headheight 0pt \headsep 0pt
        \textwidth 6.25in % A4 paper
        \textheight 9.5in % A4 paper
        \marginparwidth .875in
        \parskip 5pt plus 1pt \jot = 1.5ex}

\hybrid

 \csname
@addtoreset\endcsname{equation}{section}
\def\cJ{\pazocal{J}}
\def\cD{\pazocal{D}}
\def\cO{\pazocal{O}}
\def\cA{\pazocal{A}}
\def\cE{\pazocal{E}}
\def\cM{\pazocal{M}}
\def\cN{\pazocal{N}}
\def\cK{\pazocal{K}}
\def\cX{\pazocal{X}}
\def\del{\partial}
\def\l{\langle}
\def\r{\rangle}

\def\Tr{{\rm Tr}}
\def\B{\square}

\def\bLambda{\boldsymbol{\Lambda}}

\allowdisplaybreaks

\def\bpm{\begin{pmatrix}}
\def\epm{\end{pmatrix}}

\begin{document}

\begin{titlepage}
\rightline{}
%\rightline\today
\rightline{December  2023}
\rightline{HU-EP-23/67-RTG}  
\begin{center}
\vskip 1.5cm
{\Large \bf{Tree-level  Scattering Amplitudes via Homotopy Transfer}}
\vskip 1.7cm

{\large\bf {Roberto Bonezzi, Christoph Chiaffrino, Felipe D\'iaz-Jaramillo\\[4mm]  and Olaf Hohm}}
\vskip 1.6cm

{\it  Institute for Physics, Humboldt University Berlin,\\
 Zum Gro\ss en Windkanal 6, D-12489 Berlin, Germany}\\[1.5ex] 
 ohohm@physik.hu-berlin.de, 
roberto.bonezzi@physik.hu-berlin.de, chiaffrc@physik.hu-berlin.de, felipe.diaz-jaramillo@hu-berlin.de
\vskip .1cm

\vskip .2cm

\end{center}

\bigskip\bigskip
\begin{center} 
\textbf{Abstract}

\end{center} 
\begin{quote}

We formalize the computation of tree-level scattering amplitudes   
in terms of the homotopy transfer of homotopy algebras, 
illustrating it with scalar $\phi^3$ and Yang-Mills theory. 
The data of a (gauge) field theory with an action is encoded in  a cyclic homotopy Lie or 
$L_{\infty}$ algebra defined on a chain complex including a space of  fields. This $L_{\infty}$ structure 
can be  transported, by means of homotopy transfer, to a smaller space that, in the massless case, 
consists  of harmonic  fields. The required homotopy maps are 
well-defined since we work  with the space of finite sums 
of plane-wave solutions. The resulting $L_{\infty}$ brackets  encode the tree-level scattering amplitudes 
and satisfy generalized Jacobi identities that  imply the Ward identities. 
We further present a method to compute color-ordered scattering amplitudes for Yang-Mills theory, 
using that its $L_{\infty}$ algebra is  the  tensor product 
of the color Lie algebra  with  a homotopy commutative associative or $C_{\infty}$ algebra. 
The color-ordered scattering amplitudes are then obtained by homotopy transfer of $C_{\infty}$ algebras.

\end{quote} 
\vfill
\setcounter{footnote}{0}
\end{titlepage}

\tableofcontents
%\newpage

\section{Introduction}

Scattering amplitudes are arguably the most important objects  of quantum field theory (QFT) 
due to their applications in particle physics. 
Therefore, any student taking a course on QFT is expected, first and foremost,  to learn 
how to compute scattering amplitudes starting  with a Lagrangian of a field theory. 
While the computation of loop level scattering amplitudes comes with additional conceptual 
complications, due to the ambiguities in renormalization schemes, one certainly expects the 
student to be able to derive tree-level scattering amplitudes. This tells us that, in contrast to the mathematical foundations of QFT in general which 
are still quite  mysterious, there must be  a mathematically precise and definite algorithm to pass from the 
classical Lagrangian of a field theory, which in the interesting cases are gauge theories, 
to the tree-level scattering amplitudes. One goal of this paper is to spell out this algorithm, 
which using the framework of homotopy algebras essentially exposes the well-known   Feynman diagram techniques  but in a manner that 
is  mathematically clear and self-contained, thereby removing  the bells and whistles 
of the standard QFT text book. (We refer also to the approach of Costello \cite{costellorenormalization}  and Costello-Gwilliam \cite{Costello:2021jvx}
for a related  axiomatization of perturbative QFT.)

In order to define  an algorithm to pass from a Lagrangian for a gauge theory 
to its tree-level scattering amplitudes one first has to decide  what kind of mathematical object one 
is given when presented with a Lagrangian. The viewpoint taken in this paper, which we believe is 
fully confirmed by the examples previously discussed in the 
literature \cite{Zwiebach:1992ie,Zeitlin:2007ttl,Zeitlin:2008cc,Zeitlin:2009zz,Hohm:2017pnh}, 
is that this object is a \textit{cyclic $L_{\infty}$ algebra} \cite{Lada:1994mn,Lada:1992wc,Alexandrov:1995kv}. 
These are generalizations of differential graded Lie algebras, defined on a graded vector space 
equipped with a nil-potent differential and a graded symmetric Lie bracket on which the differential 
acts via the Leibniz rule. Importantly, however, the Jacobi identity need not to hold 
exactly but only `up to homotopy', which in turn is controlled by higher maps beginning with 
a three-bracket. The classical action 
of a field theory is encoded in the cyclic $L_{\infty}$ algebra as 
 \begin{equation}\label{LinftActionIntro} 
  S = \frac{1}{2} \langle \cA , B_1(\cA)\rangle  + \frac{1}{3!} \langle \cA, B_2(\cA,\cA)\rangle 
  +\frac{1}{4!} \langle \cA, B_3(\cA,\cA,\cA)\rangle + \cdots\;, 
 \end{equation} 
where $\cA$ collectively denotes the fields, 
which we take to have degree zero, 
$B_1$, $B_2$ and $B_3$ denote the differential, two-bracket
and three-bracket of the $L_{\infty}$ algebra, and $\langle \, , \rangle$ denotes an inner product 
that renders the algebra cyclic. The ellipsis in (\ref{LinftActionIntro}) indicate that in general there 
are infinitely many terms, corresponding to higher $L_{\infty}$ brackets, but our core example will be 
Yang-Mills theory for which there are no higher brackets than a three-bracket. 
Other data of the gauge theory such as the non-abelian gauge transformations or the gauge algebra 
are also encoded in the $L_{\infty}$ brackets, just evaluated on inputs with different degrees, such 
as gauge parameters in degree $-1$, as will be reviewed below.

In the second step, the algorithm must then start from a cyclic $L_{\infty}$ algebra that 
encodes the given gauge field theory and provide the tree-level scattering amplitudes. 
This will be done via \textit{homotopy transfer}, 
an operation  originating in  algebraic topology \cite{Crainic:2004bxw,vallette2014algebra}. 
(See \cite{Erbin:2020eyc,Koyama:2020qfb,Arvanitakis:2020rrk,Chiaffrino:2020akd,Arvanitakis:2021ecw,Chiaffrino:2023wxk} for other applications 
of homotopy transfer.) 
Here an algebraic  structure on a  vector space  is transported to a generally much smaller space. 
Importantly, in the case of a proper subspace, for homotopy transfer the original two-bracket does \textit{not} need to close on 
this subspace, i.e., in general one does not have a subalgebra. Rather, the output of the restricted two-bracket 
is projected down to the subspace, which implies that the Jacobi identity 
in  general is violated. However, while a proper projector is of course not invertible it may be 
invertible `up to homotopy', in which case the algebraic structure on the original `large' space is 
transported to a homotopy algebra on the projected `small' space. 
An example is the minimal model, where the smaller space is the (co-)homology itself: 
the space of $B_1$-closed vectors modulo $B_1$-exact vectors. 
It was first noted by Kajiura in 2003 that homotopy transfer to the minimal model can be 
formulated diagrammatically in a way that precisely mimics  Feynman rules, so that the 
computation of tree-level scattering amplitudes can be viewed as homotopy transfer \cite{Kajiura:2003ax}. 
(See also \cite{Munster:2011ij,Doubek:2017naz,Nutzi:2018vkl,Arvanitakis:2019ald,Jurco:2019yfd,Lopez-Arcos:2019hvg,Saemann:2020oyz,Okawa:2022sjf,Konosu:2023pal} for papers following up on this observation.)

In this paper we aim to improve the existing treatments in the literature 
in the following two important respects: First, one needs to define the 
functional space in which the  fields live such  that a homotopy map 
is well-defined in terms of the Green's function of the kinetic operator (such as the d'Alembert  operator 
$\square$ for a massless scalar) and so that the cohomology is non-trivial. Second, the minimal model is 
insufficient in order to describe the gauge properties  of scattering amplitudes that 
are encoded in Ward identities. Pertaining to the first issue, we choose the functional space to be simply the 
vector space of \textit{finite linear combinations} of plane wave solutions $e^{ik\cdot x}$. 
As we will argue,  this  is sufficient for the  computation 
of tree-level scattering amplitudes. 
%oh Moreover, we believe that this is the first treatment that is rigorous in that 
%the data of the homotopy transfer (projector, inclusion and homotopy map) are well defined and not just formal. 
Furthermore, our treatment  is rigorous in that 
the data of the homotopy transfer (projector, inclusion and homotopy map) are well defined and not just formal. 
%Using this space of finite sums of plane-wave solutions one could define 
The space of finite sums of plane-wave solutions has been used before in \cite{Nutzi:2018vkl,Lopez-Arcos:2019hvg} 
for 
homotopy transfer to the minimal model, 
but in order to deal with  the second issue above we rather do
homotopy transfer to the  larger space ker$\,\B$, for which 
all data of the homotopy transfer are also well-defined. 
The homotopy transfer to ker$\,\B$ leaves a residual gauge 
redundancy, with parameters $\Lambda$ satisfying $\square\Lambda=0$.

Concretely, starting from $L_{\infty}$ maps  $B_1$, $B_2$, $B_3$, etc., homotopy transfer 
yields maps $\bar{B}_n$ on the smaller space which define an $L_{\infty}$ algebra. 
Importantly, even if the starting $L_{\infty}$ algebra only carries a finite number of higher brackets, 
such as in Yang-Mills theory where the highest bracket is a three-bracket, the 
transported  $L_{\infty}$ algebra generally carries brackets $\bar{B}_n$ for arbitrary $n$. 
Writing  the fields in ${\rm ker}\,\B$ as finite linear combinations  of 
 \begin{equation}\label{AbarINTRO} 
\bar \cA^\mu_i=\epsilon^\mu_i\,e^{ik_i\cdot x}\,T_i\,, 
 \end{equation} 
where $i$ labels the gluon state under consideration, $T_i$ denotes an element of the color Lie algebra 
(which generally is not an element of a preconceived basis), and $\epsilon^\mu_i$ denotes the polarization 
vector that depends on the on-shell momentum $k_i$ (satisfying $k_i^2=0$ in the massless case) 
and is transverse, $k_i\cdot\epsilon_i=0$. 
The $n-$gluon tree level amplitude is then determined by evaluating the transported $\bar{B}_{n-1}$ and inner product 
on the plane-wave solutions (\ref{AbarINTRO}) for $i=1,\ldots, n$: 
\begin{equation}\label{YM amplitude BbarINTRO}
(2\pi)^D\delta^D(k_1+\cdots+k_n)
\,\mathcal{A}_n^{\rm tree}=\Big\l\bar \cA_n,\bar B_{n-1}\big(\bar \cA_1,\ldots,\bar \cA_{n-1}\big)\Big\r
\;.  
\end{equation}
Thanks to the space to which we transport  being larger than the minimal model, 
the $\bar{B}_{n}$ obey non-trivial generalized Jacobi identities corresponding to a non-trivial residual gauge 
redundancy. These relations imply in particular 
the Ward identities.  

As an  aside let us note that there is an intriguing analogy  between this  non-linear gauge structure 
on the subspace ker$\,\B$, where the gauge parameters $\Lambda$ satisfy $\square\Lambda=0$, 
and the gauge structure of a so-called weakly constrained 
double field theory \cite{Hull:2009mi,Bonezzi:2023ced,Bonezzi:2023lkx}.  
The latter is a `stringy' version of gravity which in fact appears to be obtainable through a double copy 
construction of Yang-Mills theory \cite{Diaz-Jaramillo:2021wtl,Bonezzi:2022yuh,Bonezzi:2022bse,Bonezzi:2023ciu}, 
leading to a theory with fields depending on doubled coordinates $(x,\bar{x})$, 
subject to the level-matching constraint $\B-\bar{\B}=0$. In order  to define a gauge algebra, 
and a  consistent field theory in the form of an $L_{\infty}$ algebra more generally, one faces the problem 
that for two gauge parameters or fields satisfying such constraints the product generally does not satisfy 
the constraint, thereby posing an obstacle for defining an algebra structure. 
 It is thus intriguing that in Yang-Mills theory 
proper, thanks to the explicit homotopy transfer  to ker$\,\B$, one obtains such an  $L_{\infty}$ algebra
by construction. This observation was in fact the original motivation for the present study.

Apart from presenting an interpretation of Ward identities in terms of 
homotopy transfer to a larger space than the minimal model, 
another central goal  of this paper is to present 
homotopy algebras  as  a simple framework for 
\textit{color-ordered} scattering amplitudes. Here  the ingredients related 
to the gauge group (color-factors) are stripped off, 
which is  an important tool in the study of 
scattering amplitudes. One uses an insight  due to Zeitlin \cite{Zeitlin:2008cc}, 
that the $L_{\infty}$ algebra underlying 
a gauge theory such as Yang-Mills theory takes the form of a tensor product 
$\cX_{\rm YM} = \cK_{\rm YM} \otimes \mathfrak{g}$, where $\mathfrak{g}$ is the Lie algebra of the 
color gauge group and $\cK_{\rm YM}$ another kind of homotopy algebra encoding the 
kinematics. $\cK_{\rm YM}$ is a so-called 
$C_{\infty}$ algebra, the homotopy version of a graded \textit{commutative} associative algebra, 
which in turn is a special case of an $A_\infty$ algebra, the homotopy version of a 
general  \textit{associative}  algebra. The higher maps of a $C_{\infty}$ algebra exhibit precisely 
the symmetry properties of color-ordered  scattering amplitudes expressed in 
the so-called Kleiss-Kuijf relations. The relation between color-ordered scattering amplitudes 
and $C_{\infty}$ algebras has been pointed out before, see e.g.~\cite{Borsten:2021hua}, but 
in this paper  we go beyond that by giving  an explicit algorithm to compute 
color-ordered gluon  amplitudes, together with their Ward identities, 
in terms of the homotopy transfer of $C_{\infty}$ algebras.

Concretely, the $L_{\infty}$ algebra of Yang-Mills theory is related to its  kinematic $C_{\infty}$ algebra as follows: 
One writes  arbitrary elements of $\cX_{\rm YM}$ as $\Psi_i=\psi_i^a\,T_a$, where $\psi_i^a\in\cK_{\rm YM}$ and $T_a\in\mathfrak{g}$. 
The brackets $B_n$ of the $L_\infty$ algebra are then written  in terms of multilinear products $m_1$, $m_2$ and $m_3$ acting on $\cK_{\rm YM}$ and the color Lie bracket $[\cdot,\cdot]$ as
\begin{equation}\label{Bs from msINTRO}
\begin{split}
B_1(\Psi)&=m_1(\psi^a)\,T_a \;,\\
B_2(\Psi_1,\Psi_2)&=(-1)^{\Psi_1}g\,m_2(\psi_1^a,\psi_2^b)\,[T_a,T_b]\;,
\end{split}    
\end{equation}
where we refer to (\ref{Bs from ms}) below for the relation between $B_3$ and $m_3$, 
and for Yang-Mills theory there is no higher map than the three-map $m_3$. 
Next, one may perform homotopy transfer to the `on-shell subspace' directly at the level of $C_{\infty}$ algebras, 
which yields  higher maps $\bar{m}_n$ for arbitrary $n=1,2,3\ldots$. 
The color-ordered $n$-gluon amplitudes denoted by $A[1,2,\ldots,n-1,n]$ are then given by \cite{Borsten:2021hua} 
\begin{equation}\label{cycliccolorINTRO}
\begin{split}
A[1,2,\ldots,n-1,n]=  \epsilon_{n}\cdot \bar m_{n-1}(\epsilon_{1},\epsilon_{2},\ldots,\epsilon_{n-2},\epsilon_{n-1})\;,
\end{split}
\end{equation}
with $\bar A_{i}$ the color-stripped on-shell gauge fields $\bar A_{\mu\, i}=\epsilon_{\mu\, i}\, e^{ik_{i}\cdot x}$. 
We illustrate these methods by computing the color-ordered five-point amplitude by means of 
homotopy transfer and by deriving Ward identities for color-ordered amplitudes.

This paper is organized as follows. In section 2 we illustrate the 
homotopy transfer formulation for the toy model of scalar $\phi^3$ theory. Specifically, the computation 
of tree-level scattering amplitudes that lends itself most directly to 
homotopy transfer can be formulated in terms of the perturbative solutions 
of  the classical field equations  \cite{Boulware:1968zz}.  
While this latter approach  is reasonably well known among experts it has not quite made it to the text books, 
and so we use the opportunity to give a self-contained introduction. 
In sec.~3 we turn to our core example of Yang-Mills theory. We introduce 
its $L_{\infty}$ algebra, which we immediately write in terms of its underlying kinematic $C_{\infty}$ algebra, 
and we define the homotopy transfer to scattering amplitudes. Using this  we derive the Ward identities 
and discuss the non-linear gauge structure on ker$\,\B$. 
In sec.~4 we show how to compute color-ordered scattering amplitudes starting directly from 
the kinematic $C_{\infty}$ algebra, and we derive color-ordered Ward identities. 
In everything up to and including sec.~4 we develop, in a hopefully self-contained manner, 
the homotopy algebra  formulation 
along the lines of the concrete example of  Yang-Mills theory and its 4-point  and 5-point amplitudes. 
In sec.~5 we then change gears by giving  a mathematically extensive  discussion of homotopy algebras and its homotopy transfer,  
which applied to $C_{\infty}$ algebras yields a technique to compute arbitrary $n$-point 
color-ordered tree-level scattering amplitudes. 
While these subjects have been reviewed in quite a few places, see e.g.~\cite{Lada:1992wc,Hohm:2017pnh,Arvanitakis:2020rrk,Borsten:2021hua}, 
we aim to give a self-contained introduction emphasizing in particular the notion of equivalence for general homotopy algebras 
that seems particularly well suited for scattering amplitudes. 
We close in sec.~6 with an outlook. 
In an appendix we give a direct  proof of a recursive formula for the
homotopy transferred $L_{\infty}$ brackets that to the best of our knowledge has not appeared in the literature. 
(A more efficient proof requiring more sophisticated techniques is also given in sec.~5.)

\section{Amplitudes in scalar field theory}

In this section we will start by reviewing how the generating functional for tree-level scattering amplitudes is related to the on-shell classical action, for the case of a scalar field theory. We will then show how this is interpreted in terms of homotopy transfer of the associated $L_\infty$ algebra. To do so we will work with a plane wave basis, which makes direct contact with the standard Feynman rules for the $S-$matrix.

\subsection{Amplitudes from on-shell action}

For definiteness, in this section we work with a massless $\phi^3$ theory with action
\begin{equation}\label{phi3}
S[\phi]=\int dx\,\Big[-\tfrac12\,\del^\mu\phi\,\del_\mu\phi-\tfrac{1}{3!}\,g\,\phi^3\Big]\;,    
\end{equation}
with mostly plus convention for the Minkowski metric: $\eta_{\mu\nu}={\rm diag}(-1,+1,\cdots,+1)$, and we abbreviate $dx\equiv d^Dx$. Following the standard textbook approach, scattering amplitudes are computed by applying the LSZ reduction to connected correlation functions $\l\phi(x_1)\cdots\phi(x_n)\r_{\rm c}$. The generating functional for connected correlators is $W[J]$, defined by the path integral
\begin{equation}
e^{iW[J]}=\int \cD\phi\,e^{i\,S[\phi]+i\int  J\phi}\;,    
\end{equation}
where we normalize $\int \cD\phi\,e^{i\,S[\phi]}=1$ or, equivalently, $W[0]=0$, by omitting all vacuum bubbles from the diagrammatic expansion of $W[J]$. Given the generating functional $W[J]$, connected correlators are obtained by taking functional derivatives with respect to the sources:
\begin{equation}\label{conncorr}
\big\l\phi(x_1)\cdots\phi(x_n)\big\r_{\rm c}=(-i)^n\frac{\delta}{\delta J(x_1)}\cdots \frac{\delta}{\delta J(x_n)}\,iW[J]\Big\rvert_{J=0}\;.   
\end{equation}

To generate a loop expansion for $W$, we shall expand $S[\phi]+\int J\phi$ around the stationary point $\phi_*(J)$, defined by
\begin{equation}\label{defphistar}
\frac{\delta S}{\delta\phi}\Big\rvert_{\phi=\phi_*}+J=0\;.
\end{equation}
Writing $\phi=\phi_*(J)+\pi$ and integrating over the fluctuation $\pi$ one obtains
\begin{equation}
e^{iW[J]}=e^{i\,S[\phi_*(J)]+i\int  J\phi_*(J)}\int \cD\pi\exp\left\{\frac{i}{2}\int dxdy\,\pi(x)\,K_J(x,y)\,\pi(y)-\frac{ig}{3!}\int dx\,\pi^3(x)\right\}\;,    
\end{equation}
where we defined $K_J(x,y):=\frac{\delta^2S[\phi]}{\delta\phi(x)\delta\phi(y)}\Big\rvert_{\phi=\phi_*(J)}$. The prefactor outside the path integral gives the tree-level contribution to $W[J]$, which we are interested in, while the remaining integration over $\pi$ generates all loop diagrams \cite{Boulware:1968zz}.
The tree-level part $W_{\rm tree}[J]$ is thus given by the Legendre transform of the classical action:
\begin{equation}\label{Wtreedef}
W_{\rm tree}[J]=S[\phi_*(J)]+\int dx\,J\,\phi_*(J) \;, \quad\frac{\delta W_{\rm tree}}{\delta J}=\phi_*(J)\;,  
\end{equation}
where $\phi_*$ obeys \eqref{defphistar}, i.e. is a classical solution of the field equations with source $J$:
\begin{equation}\label{eomphistar}
\B\phi_*-\frac{g}{2}\,\phi_*^2+J=0 \;.  
\end{equation}

In order to solve the equations we introduce the Feynman propagator $G(x-y)$:
\begin{equation}
G(x-y)=\int dk\,\frac{1}{k^2-i\epsilon}\,e^{ik\cdot(x-y)}\;,\quad \B_xG(x-y)=-\delta^D(x-y)\;,   
\end{equation}
where $dk\equiv\frac{d^Dk}{(2\pi)^D}$.
One can solve \eqref{eomphistar} as a perturbation series in $g$ upon making the ansatz
\begin{equation}\label{gansatz}
\phi_*(J)=\sum_{n=0}^\infty g^n\,\phi_n(J)=\phi_0(J)+\phi_{\rm n.l.}(J)\;,    
\end{equation}
where $\phi_0$ solves the linearized equation $\B\phi_0=-J$ and $\phi_{\rm n.l.}=\sum_{n=1}^\infty g^n\phi_n$ is the sum of all contributions of order $g$ and higher, obeying $\B\phi_{\rm n.l.}=\frac{g}{2}(\phi_0+\phi_{\rm n.l.})^2$. This is solved order by order in $g$ through the recursion
\begin{equation}\label{recursive}
\phi_0=GJ\;,\quad \phi_n=-\frac{1}{2}\sum_{k+l=n-1}G\big(\phi_k\phi_l\big)\;,\quad n\geq1\;,    
\end{equation}
where we used a shorthand notation for the action of the propagator:
\begin{equation}
\begin{split}
GJ(x)&:=\int dy\,G(x-y)\,J(y)=\int dk\,\frac{e^{ik\cdot x}}{k^2-i\epsilon}\,\tilde J(k)\;,\\
G\big(fg\big)(x)&:=\int dy\,G(x-y)\,f(y)\,g(y)=\int dk_1dk_2\,\frac{e^{i(k_1+k_2)\cdot x}}{(k_1+k_2)^2-i\epsilon}\,\tilde f(k_1)\,\tilde g(k_2)\;.
\end{split}    
\end{equation}
By induction, the recursive relation \eqref{recursive} gives $\phi_{\rm n.l.}(J)$ as the sum of all cubic trees with one marked point, weighted by their symmetry factor. For instance, up to order $g^3$ one has
\begin{equation}\label{phistarexp}
\begin{split}
\phi_*(J)&=GJ-\frac{g}{2}\,G(GJ)^2+\frac{g^2}{2}\,G\big(GJ\,G(GJ)^2\big)\\
&-g^3\,\Big\{\frac12\,G\Big(GJ\,G\big(GJ\,G(GJ)^2\big)\Big)+\frac18\,G\Big(G(GJ)^2\,G(GJ)^2\Big)\Big\}+\cO(g^4)\\
&=
\begin{tikzpicture}[baseline={([yshift=-.5ex]current bounding box.center)},level distance=10mm,sibling distance=5mm]
\node{} child[grow=right]{node[scale=0.5,fill,circle,draw] {}} ;
\end{tikzpicture} \;-\frac{g}{2}\,\begin{tikzpicture}[baseline={([yshift=-.5ex]current bounding box.center)},level distance=10mm,sibling distance=10mm]
  \node {}
    child[grow=right] {
     child {node[scale=0.5,fill,circle,draw] {}}  child{node[scale=0.5,fill,circle,draw] {}}
    };
\end{tikzpicture}\;+\frac{g^2}{2}\,\begin{tikzpicture}[baseline={([yshift=-.5ex]current bounding box.center)},level distance=10mm,sibling distance=10mm]
  \node {}
    child[grow=right] {
     child{child {node[scale=0.5,fill,circle,draw] {}} child[missing]}    child {child {node[scale=0.5,fill,circle,draw] {}} child {node[scale=0.5,fill,circle,draw] {}}}
    };
\end{tikzpicture}\\
&-\frac{g^3}{2}\,\begin{tikzpicture}[baseline={([yshift=-.5ex]current bounding box.center)},level distance=10mm,sibling distance=10mm]
  \node{}[grow=right] child{
  child{[fill] circle (1mm)} child{child{[fill] circle (1mm)} child{child{[fill] circle (1mm)} child{[fill] circle (1mm)}}}
  } ;
\end{tikzpicture}
\;-\frac{g^3}{8}\,\begin{tikzpicture}[baseline={([yshift=-.5ex]current bounding box.center)},level distance=10mm]
  \tikzstyle{level 2}=[sibling distance=15mm]
  \tikzstyle{level 3}=[sibling distance=10mm]
\node{}[grow=right] child{child{child{[fill] circle (1mm)} child{[fill] circle (1mm)}} child{child{[fill] circle (1mm)} child{[fill] circle (1mm)}}} ;
\end{tikzpicture}\;+\cO(g^4)\;,
\end{split}    
\end{equation}
where we denoted the sources $J$ by filled dots and every line corresponds to a propagator. 

Inserting the expansion of $\phi_*(J)$ in the definition \eqref{Wtreedef} of $W_{\rm tree}[J]$ one obtains the implicit expression
\begin{equation}\label{W implicit}
W_{\rm tree}[J]=\int dx\,\Big\{\,\frac12\,J\,GJ+\frac12\,\phi_{\rm n.l.}(J)\,\B\phi_{\rm n.l.}(J)-\frac{g}{3!}\,\big(GJ+\phi_{\rm n.l.}(J)\big)^3\Big\} \;.   
\end{equation}
Using the perturbative expansion of $\phi_{\rm n.l.}(J)$ one recovers the standard diagrammatic expansion of the generating functional:
\begin{equation}
\begin{split}
W_{\rm tree}[J]&=\int dx\,\Big\{\,\frac12\,J\,GJ-\frac{g}{3!}\,\big(GJ\big)^3+\frac{g^2}{8}\,\big(GJ\big)^2G\big(GJ\big)^2-\frac{g^3}{8}\,\big(GJ\big)^2G\Big(GJ\,G\big(GJ\big)^2\Big)\Big\}\\
&+\cO(g^4)\\
&=\frac12\;\begin{tikzpicture}[baseline={([yshift=-.5ex]current bounding box.center)}]
\node[scale=0.5,fill,circle,draw] (1)  at  (0,0) {};
\node[scale=0.5,fill,circle,draw] (2) at  (1.5,0) {};
\draw (1) -- (2);
\end{tikzpicture}\;-\frac{g}{3!}\;
\begin{tikzpicture}[baseline={([yshift=-.5ex]current bounding box.center)}]
\node[scale=0.5,fill,circle,draw] (1)  at  (0,0) {};
\node[scale=0.5,fill,circle,draw] (2) at  (2,0) {};
\node[scale=0.5,fill,circle,draw] (3)  at  (1,1.732) {};
\draw (1) -- (1,0.577); \draw (2) -- (1,0.577); \draw (3) -- (1,0.577);
\end{tikzpicture}+\frac{g^2}{8}\,
\begin{tikzpicture}[baseline={([yshift=-.5ex]current bounding box.center)}]
\node[scale=0.5,fill,circle,draw] (1)  at  (0,0) {};
\node[scale=0.5,fill,circle,draw] (2) at  (0,2) {};
\node[scale=0.5,fill,circle,draw] (3)  at  (3,2) {};
\node[scale=0.5,fill,circle,draw] (4)  at  (3,0) {};
\draw (1) -- (0.6,1); \draw (2) -- (0.6,1); \draw (3) -- (2.4,1); \draw (4) -- (2.4,1); \draw (0.6,1) -- (2.4,1);
\end{tikzpicture}-\frac{g^3}{8}\,
\begin{tikzpicture}[baseline={([yshift=-.5ex]current bounding box.center)}]
\node[scale=0.5,fill,circle,draw] (1)  at  (0,0) {};
\node[scale=0.5,fill,circle,draw] (2) at  (0,2) {};
\node[scale=0.5,fill,circle,draw] (3)  at  (3,2) {};
\node[scale=0.5,fill,circle,draw] (4)  at  (3,0) {};
\node[scale=0.5,fill,circle,draw] (5)  at  (1.5,2) {};
\draw (1) -- (0.6,1); \draw (2) -- (0.6,1); \draw (3) -- (2.4,1); \draw (4) -- (2.4,1); \draw (0.6,1) -- (2.4,1); \draw (5) -- (1.5,1);
\end{tikzpicture}\\
&+\cO(g^4)\;,
\end{split}    
\end{equation}
given by the sum of all topologically distinct tree diagrams with no marked points and at least two sources:
\begin{equation}
W_{\rm tree}[J]=\sum_{n=2}^\infty(-1)^ng^{n-2}\sum_{\{\Gamma_n\}}\frac{1}{S_{\Gamma_n}}\,\Gamma_n \;,   
\end{equation}
where we denoted diagrams with $n$ indistinguishable external legs by $\Gamma_n$, and the corresponding symmetry factor by $S_{\Gamma_n}$. 

The LSZ reduction formula gives the $n-$point scattering amplitude $\mathcal{A}_n(k_1,\ldots,k_n)$ in terms of the connected correlator $\l\phi(x_1)\cdots\phi(x_n)\r_{\rm c}$ as\footnote{In this section we follow the conventions of \cite{Srednicki:2007qs}.}
\begin{equation}\label{LSZ}
\begin{split}
(2\pi)^D\delta^D\big(\scaleobj{0.7}{\sum_i}\,k_i\big)\,i\mathcal{A}_n(k_1,\ldots,k_n)=(-i)^n&\int dx_1\,e^{ik_1\cdot x_1}\B_{x_1}\cdots \int dx_n\,e^{ik_n\cdot x_n}\B_{x_n}\\
&\times\big\l\phi(x_1)\cdots\phi(x_n)\big\r_{\rm c}\;, 
\end{split}    
\end{equation}
where we treat all momenta $k^\mu_i$ as incoming and, since we are only interested in tree-level amplitudes, we do not discuss renormalization conditions.
Notice that all external momenta in \eqref{LSZ} are on-shell: $k_i^2=0$. This shows that the amplitude is a pure ``boundary'' object, since integrating naively the $\B_{x_i}$ by parts would make it vanish. Using \eqref{conncorr} for the correlators, the tree-level amplitude is given by
\begin{equation}
\delta(K)\,\mathcal{A}^{\rm tree}_n(k_1,\ldots,k_n)=(-1)^n\prod_{i=1}^n\int dx_i\,e^{ik_i\cdot x_i}\B_{x_i}\frac{\delta}{\delta J(x_i)}\,W_{\rm tree}[J]\Big\rvert_{J=0}\;,    
\end{equation}
where we denoted the momentum-conserving delta function by $\delta(K)\equiv(2\pi)^D\delta^D\big(\scaleobj{0.7}{\sum_i}\,k_i\big)$. One can further simplify the above expression by noting that the recursive solution $\phi_*(J)$ given by \eqref{gansatz} and \eqref{recursive}, as well as $W_{\rm tree}[J]$ 
in \eqref{W implicit}, depends on $J$ only through $\phi_0=GJ$. The operator removing a source and amputating the external propagator can thus be written as \cite{Monteiro:2011pc}
\begin{equation}
\B_{x_i}\frac{\delta}{\delta J(x_i)}=\B_{x_i}\int dy\,\frac{\delta\phi_0(y)}{\delta J(x_i)}\,\frac{\delta}{\delta\phi_0(y)}=-\frac{\delta}{\delta\phi_0(x_i)}\;,    
\end{equation}
acting on $W_{\rm tree}[J(\phi_0)]$. Using \eqref{W implicit} and $J=-\B\phi_0$ one has
\begin{equation}\label{Wphi0}
W_{\rm tree}[J(\phi_0)]=\int dx\,\Big\{-\frac12\,\phi_0\,\B\phi_0+\frac12\,\phi_{\rm n.l.}(\phi_0)\,\B\phi_{\rm n.l.}(\phi_0)-\frac{g}{3!}\,\big(\phi_0+\phi_{\rm n.l.}(\phi_0)\big)^3\Big\} \;.   
\end{equation}
Since the operators $\int dx_i\,e^{ik_i\cdot x_i}\frac{\delta}{\delta\phi_0(x_i)}$ substitute all occurrences of $\phi_0$ with on-shell plane waves, the first term $\phi_0\,\B\phi_0$ in \eqref{Wphi0} does not contribute (as there are no two-point amplitudes), and we can finally write the $n-$point amplitude as
\begin{equation}\label{ampGamma}
\delta(K)\,\mathcal{A}^{\rm tree}_n(k_1,\ldots,k_n)=\prod_{i=1}^n\int dx_i\,e^{ik_i\cdot x_i}\frac{\delta}{\delta \phi_0(x_i)}\,\Gamma_{\rm tree}[\phi_0]\Big\rvert_{\phi_0=0}\;,    
\end{equation}
where the generating functional $\Gamma_{\rm tree}[\phi_0]$ is given implicitly by
\begin{equation}\label{Gamma implicit}
\Gamma_{\rm tree}[\phi_0]=\int dx\,\Big\{\,\frac12\,\phi_{\rm n.l.}(\phi_0)\,\B\phi_{\rm n.l.}(\phi_0)-\frac{g}{3!}\,\big(\phi_0+\phi_{\rm n.l.}(\phi_0)\big)^3\Big\}  \;,  
\end{equation}
with $\phi_{\rm n.l.}=\sum_{n=1}^\infty g^n\,\phi_n$ still given recursively by \eqref{recursive}, except that $\phi_0$ is now the independent variable.
Comparing \eqref{W implicit} with \eqref{Gamma implicit} it is clear that the perturbative expansion of $\Gamma_{\rm tree}[\phi_0]$ has the same Feynman diagrams as the one of $W[J]$, except that the two-point function is absent and external propagators are amputated:
\begin{equation}\label{Gamma g4}
\begin{split}
\Gamma_{\rm tree}[\phi_0]&=\int dx\,\Big\{-\frac{g}{3!}\,\phi_0^3+\frac{g^2}{8}\,\phi_0^2\,G\big(\phi_0^2\big)-\frac{g^3}{8}\,\phi_0^2\,G\Big(\phi_0\,G\big(\phi_0^2\big)\Big)\Big\}+\cO(g^4)\\
&=-\frac{g}{3!}\;
\begin{tikzpicture}[baseline={([yshift=-.5ex]current bounding box.center)}]
\node (1)  at  (0,0) {};
\node (2) at  (2,0) {};
\node (3)  at  (1,1.732) {};
\draw (1) -- (1,0.577); \draw (2) -- (1,0.577); \draw (3) -- (1,0.577);
\end{tikzpicture}+\frac{g^2}{8}\,
\begin{tikzpicture}[baseline={([yshift=-.5ex]current bounding box.center)}]
\node (1)  at  (0,0) {};
\node (2) at  (0,2) {};
\node (3)  at  (3,2) {};
\node (4)  at  (3,0) {};
\draw (1) -- (0.6,1); \draw (2) -- (0.6,1); \draw (3) -- (2.4,1); \draw (4) -- (2.4,1); \draw (0.6,1) -- (2.4,1);
\end{tikzpicture}-\frac{g^3}{8}\,
\begin{tikzpicture}[baseline={([yshift=-.5ex]current bounding box.center)}]
\node (1)  at  (0,0) {};
\node (2) at  (0,2) {};
\node (3)  at  (3,2) {};
\node (4)  at  (3,0) {};
\node (5)  at  (1.5,2) {};
\draw (1) -- (0.6,1); \draw (2) -- (0.6,1); \draw (3) -- (2.4,1); \draw (4) -- (2.4,1); \draw (0.6,1) -- (2.4,1); \draw (5) -- (1.5,1);
\end{tikzpicture}+\cO(g^4)\;,
\end{split}    
\end{equation}
where in the above diagrams only internal lines are propagators, while external legs are factors of $\phi_0$.
As an example, applying \eqref{ampGamma} to the above expansion of $\Gamma_{\rm tree}$ gives the usual expression for the four-point amplitude:
\begin{equation}\label{4scalar amp}
\begin{split}
\mathcal{A}_4^{\rm tree}(k_i)&=g^2\,\Big\{\frac{1}{(k_1+k_2)^2}+\frac{1}{(k_2+k_3)^2}+\frac{1}{(k_3+k_1)^2}\Big\}\\
&=g^2\begin{tikzpicture}[baseline={([yshift=-.5ex]current bounding box.center)}]
\node[scale=0.7] (1)  at  (0,0) {2};
\node[scale=0.7] (2) at  (0,2) {1};
\node[scale=0.7] (3)  at  (3,2) {4};
\node[scale=0.7] (4)  at  (3,0) {3};
\draw (1) -- (0.6,1); \draw (2) -- (0.6,1); \draw (3) -- (2.4,1); \draw (4) -- (2.4,1); \draw (0.6,1) -- (2.4,1);
\end{tikzpicture}+g^2\begin{tikzpicture}[baseline={([yshift=-.5ex]current bounding box.center)}]
\node[scale=0.7] (1)  at  (0,0) {3};
\node[scale=0.7] (2) at  (0,2) {2};
\node[scale=0.7] (3)  at  (3,2) {4};
\node[scale=0.7] (4)  at  (3,0) {1};
\draw (1) -- (0.6,1); \draw (2) -- (0.6,1); \draw (3) -- (2.4,1); \draw (4) -- (2.4,1); \draw (0.6,1) -- (2.4,1);
\end{tikzpicture}+g^2\begin{tikzpicture}[baseline={([yshift=-.5ex]current bounding box.center)}]
\node[scale=0.7] (1)  at  (0,0) {1};
\node[scale=0.7] (2) at  (0,2) {3};
\node[scale=0.7] (3)  at  (3,2) {4};
\node[scale=0.7] (4)  at  (3,0) {2};
\draw (1) -- (0.6,1); \draw (2) -- (0.6,1); \draw (3) -- (2.4,1); \draw (4) -- (2.4,1); \draw (0.6,1) -- (2.4,1);
\end{tikzpicture}\;,
\end{split}
\end{equation}
in terms of the $s$, $t$ and $u-$channels.
Given that \eqref{ampGamma} substitutes all powers of $\phi_0$ with on-shell plane waves, it is natural to identify the functional $\Gamma_{\rm tree}[\phi_0]$ as the on-shell action $S[\phi_*(\phi_0)]$, for $\phi_*$ solving $\B\phi_*-\frac{g}{2}\,\phi_*^2=0$, written in terms of free fields obeying $\B\phi_0=0$. To do so, however, one has to supplement the action by total derivative terms which, as $\phi_0$ does not fall off at infinity, cannot be discarded \cite{Arefeva:1974jv}.

In the following we will present the $L_\infty$ interpretation of the functional $\Gamma_{\rm tree}[\phi_0]$ working directly with finite sets of plane waves. In doing so we will not encounter subtleties related to boundary terms, but the interpretation of the nonlinear field $\phi_*(\phi_0)$ will be slightly different. 

\subsection{$L_\infty$ algebra of scalar field theory and its on-shell action}

Homotopy Lie, or $L_\infty$, algebras are mathematical structures that can be naturally associated to any perturbative field theory \cite{Zwiebach:1992ie,Lada:1992wc,Hohm:2017pnh}. They consist of a $\mathbb{Z}-$graded vector space $\cX=\bigoplus_iX_i$ equipped with multilinear brackets $B_n:\cX^{\otimes n}\rightarrow\cX$ which, in our conventions, are graded symmetric and have intrinsic degree $|B_n|=1$. The brackets satisfy a set of quadratic relations which generalize the Jacobi identity of Lie algebras. 
In particular, the unary bracket $B_1$ of an $L_\infty$ algebra is required to be nilpotent: $B_1^2=0$, which makes the graded vector space into a chain complex $(\cX,B_1)$ with $B_1$ as differential:
\begin{equation}\label{chain complex}
\begin{tikzcd}
\cdots\arrow{r}&X_{-1}\arrow{r}{B_1}&X_0\arrow{r}{B_1}&X_1\arrow{r}{B_1}&X_2\arrow{r}&\cdots    
\end{tikzcd} \;.   
\end{equation}

In applications to field theory the graded vector space is associated to gauge parameters, fields, equations of motion and so on, assigned to subspaces $X_i$ of different degree. Nilpotency of the differential $B_1$ states gauge invariance of the free theory, while higher quadratic relations encode, for instance, the consistency of perturbative interactions with gauge symmetries, their deformations and algebra.

For the case at hand of a scalar field theory there is no gauge symmetry to constrain the interactions which, in turn, implies that all quadratic relations are trivial. Nonetheless, as we shall see, the procedure of going perturbatively on-shell and generating scattering amplitudes has a natural algebraic interpretation. In a scalar theory, the graded vector space $\cX$ consists of two spaces:
\begin{equation}
\cX=X_0\oplus X_1 \;,   
\end{equation}
where elements $\psi\in X_k$ are assigned degree $|\psi|=k$.
In here $X_0$ is the space of fields $\phi(x)$, while $X_1$ is the space of field equations, meaning that Euler-Lagrange variations like $\B\phi-\frac{g}{2}\,\phi^2$, as well as external sources $J$ take values in $X_1$. With a slight abuse of terminology we still name elements of $X_1$ equations or sources. 

 Since there are no elements in degrees other than zero and one, non-vanishing brackets are only allowed between fields\footnote{In a gauge theory, where $\cX$ consists of more than two spaces, this is not true and one in general has brackets between gauge parameters, parameters and fields etc.}: $B_n(\phi_1,\ldots,\phi_n)\in X_1$, which are thus totally symmetric. Similarly, all quadratic relations are trivial, since any expression of the form $B_k\big(B_l(\psi_1,\ldots,\psi_l),\ldots,\psi_{k+l-1}\big)$ would have at least degree 2.
The $L_\infty$ brackets between fields encode the interactions of the theory and can be read off from expanding the field equations in powers of $\phi$. In doing so, the latter take the form of a generalized Maurer-Cartan equation:
\begin{equation}\label{MC general}
\frac{\delta S}{\delta\phi}=0\quad\longrightarrow\quad B_1(\phi)+\tfrac12\,B_2(\phi,\phi)+\cdots+\tfrac{1}{n!}\,B_n(\phi,\cdots,\phi)=0 \;.   
\end{equation}
For our example of massless $\phi^3$ theory, the field equation $\B\phi-\frac{g}{2}\,\phi^2=0$ fixes the only non-vanishing brackets to be
\begin{equation}\label{Bs phi3}
B_1(\phi)=\B\phi\;,\quad B_2(\phi_1,\phi_2)=-g\,\phi_1\phi_2\;,    
\end{equation}
where we wrote the generic action of $B_2$ on two distinct inputs. The chain complex $(\cX,B_1)$ associated to the scalar is thus given by
\begin{equation}\label{chain scalar}
\begin{tikzcd}[row sep=5mm]
0\arrow{r}{0}&X_0\arrow{r}{B_1}&X_1\arrow{r}{0}&0\\
&\phi\arrow[r,mapsto,"\B"]&E
\end{tikzcd}    \;,
\end{equation}
and nilpotency of $B_1$ is indeed trivial, since it is extended by zero beyond $X_0$ and $X_1$.

The last ingredient for the $L_\infty$ description of a Lagrangian field theory is a degree $-1$ inner product $\l\,,\,\r:\cX\otimes\cX\rightarrow\mathbb{C}$, which we take to be graded antisymmetric:
\begin{equation}
\big\l\psi_1,\psi_2\big\r=(-1)^{1+\psi_1\psi_2}\big\l\psi_2,\psi_1\big\r \;,   
\end{equation}
where $\psi_i$ in phase factors always stand for their degree $|\psi_i|$.
A \emph{cyclic} $L_\infty$ algebra has such inner product and brackets $B_n$ obeying
\begin{equation}\label{cyclicity general}
\big\l B_n(\psi_1,\cdots,\psi_n),\psi_{n+1}\big\r=(-1)^{\psi_{n+1}\sum_{i=1}^n\psi_i}\big\l B_n(\psi_{n+1},\psi_1,\cdots,\psi_{n-1}),\psi_{n}\big\r\;,    
\end{equation}
so that the expression in \eqref{cyclicity general} is graded symmetric in all inputs $\psi_1,\cdots,\psi_{n+1}$. Given an inner product obeying \eqref{cyclicity general}, one can indeed define an action as
\begin{equation}\label{MC action scalar}
S[\phi]=\tfrac{1}{2}\,\big\l\phi,B_1(\phi)\big\r+\tfrac{1}{3!}\,\big\l\phi,B_2(\phi,\phi)\big\r+\cdots+\tfrac{1}{(n+1)!}\,\big\l\phi,B_n(\phi,\cdots,\phi)\big\r \;,  
\end{equation}
which yields \eqref{MC general} under variation, upon recalling that \eqref{cyclicity general} amounts to total symmetry for inputs in degree zero.

In a scalar field theory, the only non-vanishing component of the inner product pairs fields in $X_0$ with equations in $X_1$ via
\begin{equation}\label{inner scalar}
\big\l\phi,E\big\r=-\big\l E,\phi\big\r:=\int dx\,\phi(x)\,E(x)\;. 
\end{equation}
In the case of polynomial interactions like $\phi^3$, cyclicity \eqref{cyclicity general} of $B_2$, $B_3$ etc. is trivially guaranteed by the fact that $\int dx\,\phi_1\,\phi_2\cdots\phi_n$
is totally symmetric in $1,2,\cdots,n$. The situation is more subtle for the kinetic operator $B_1$, as demanding \eqref{cyclicity general} amounts to integrating $\B$ by parts \emph{without} boundary contributions\footnote{See for instance \cite{Chiaffrino:2023wxk} for a different treatment, where the inner product has to be supplemented by boundary terms.}, given that $\l\phi_1,\B\phi_2\r=\int dx\,\phi_1\,\B\phi_2$. 

Having in mind applications to scattering amplitudes,
to address this issue we will henceforth restrict our fields and equations (or sources) to be given by finite sums of plane waves:
\begin{equation}\label{planephi}
\phi(x)=\sum_i\phi_i(k_i)\,e^{ik_i\cdot x}\;,\quad E(x)=\sum_i \cE_i(k_i)\,e^{ik_i\cdot x}\;,\quad \phi_i,\cE_i\in\mathbb{C}\;, 
\end{equation}
where we write the Fourier coefficients as e.g.~$\phi(k)$ to keep track of the momentum of the corresponding plane wave.
Strictly speaking, plane waves (which are fields with delta function Fourier transforms) are not admissible field configurations in an unbounded space. Nonetheless, one can avoid using wave packets at the price of dealing with delta function normalizations for the inner product. Namely, the inner product \eqref{inner scalar} between fields and equations of the form \eqref{planephi}
reads
\begin{equation}\label{inner plane waves}
\big\l\phi,E\big\r=(2\pi)^D\sum_{ij}\delta^D(k_i+k_j)\,\phi_i(k_i)\,\cE_j(k_j)\;.  
\end{equation}
Momentum conservation is expressed by a delta function, as in the $S-$matrix, which is the standard normalization for exact momentum eigenstates in an infinite volume. Notice that the corresponding coefficient $\phi_i(k_i)\,\cE_j(-k_i)$ is well-defined and carries all the relevant information.
Upon restricting the functional space, cyclicity of $B_1$ is ensured, since
\begin{equation}
\begin{split}
\big\l\phi_1,B_1(\phi_2)\big\r&=\sum_{ij}\phi_{1i}\phi_{2j}\int dx\,e^{ik_{1i}\cdot x}\,\B e^{ik_{2j}\cdot x}\\
&=(2\pi)^D\sum_{ij}-k_{2j}^2\delta^D(k_{1i}+k_{2j})\,\phi_{1i}\phi_{2j}=(2\pi)^D\sum_{ij}-k_{1i}^2\delta^D(k_{1i}+k_{2j})\,\phi_{1i}\phi_{2j}\\
&=\big\l\phi_2,B_1(\phi_1)\big\r\;.
\end{split}    
\end{equation}

Given the $L_\infty$ description of $\phi^3$ theory, we now revisit the procedure for solving the field equations perturbatively:
\begin{equation}\label{Eom}
B_1(\phi)+\frac12\,B_2(\phi,\phi)=0\quad\longleftrightarrow\quad\B\phi-\frac{g}{2}\,\phi^2=0\;.    
\end{equation}
An important difference in working with plane waves, rather than functions with smooth Fourier transforms, is that the propagator $G$ is not defined on arbitrary sources  of the form \eqref{planephi}.
The reason is that, when acting with $G$, one usually integrates the distribution $\frac{1}{p^2-i\epsilon
}$ over a continuous function of momentum. With a plane wave $e^{ik\cdot x}$, one would rather evaluate it as $\frac{1}{k^2-i\epsilon}$,
which is nonsensical if $k^\mu$ is on-shell. To remedy this, we introduce a harmonic projector $P_{\rm h}$ and its orthogonal $(1-P_{\rm h})$, which are well-defined for sums of plane waves:
\begin{equation}\label{Ph}
\begin{split}
P_{\rm h}\psi(x)&=\sum_i\delta_{k_i^2,0}\, \psi_i\,e^{ik_i\cdot x}=\sum_{i,\,k_i^2=0} \psi_i\,e^{ik_i\cdot x}    \;, \\
(1-P_{\rm h})\psi(x)&=\sum_i(1-\delta_{k_i^2,0})\, \psi_i\,e^{ik_i\cdot x}=\sum_{i,\,k_i^2\neq0} \psi_i\,e^{ik_i\cdot x}    \;, 
\end{split}
\end{equation}
where $\psi$ is either a field $\phi$ or an equation $E$, and $\delta_{k^2,0}$ is a Kronecker delta.
Such a projector obeys 
\begin{equation}\label{Ph property}
P_{\rm h}^2=P_{\rm h}\;,\quad P_{\rm h}\B=\B P_{\rm h}=0\;,    
\end{equation}
and can be used to separate on-shell from off-shell plane waves. The propagator $G$ can act on sources $E$ \emph{if} projected with $1-P_{\rm h}$:
\begin{equation}
G\big((1-P_{\rm h})E\big)=\sum_{i,\,k_i^2\neq0}\frac{1}{k_i^2}\, \cE_i\,e^{ik_i\cdot x}  \;. 
\end{equation}
Since the propagator is only defined on $(1-P_{\rm h})$, we shall use the combination
\begin{equation}\label{h}
h:=G\,(1-P_{\rm h})\;,  
\end{equation}
which can act on any source $E(x)$. $h$ gives the usual propagator contribution for off-shell momenta, and zero for on-shell ones.
It is thus (minus) the inverse of $\B$ away from its kernel, and obeys
\begin{equation}\label{h property heuristic}
\B h=h\B=P_{\rm h}-1 \;,\quad P_{\rm h}h=hP_{\rm h}=0\;.   
\end{equation}
In $L_\infty$ terms, the kinetic operator $\B=B_1$ is a degree $+1$ map from the space of fields $X_0$ to the space of equations $X_1$. Since $h$ is a partial inverse of $B_1$, it is to be viewed as a degree $-1$ operator that maps equations back to the space of fields: $h:X_1\rightarrow X_0$.

With these definitions at hand, we try to solve the field equation \eqref{Eom} by the power series ansatz
\begin{equation}
\phi_*(\phi_0)=\sum_{n=0}^\infty g^n\phi_n(\phi_0)=\phi_0+\phi_{\rm n.l.}(\phi_0)\;,  \end{equation}
where $\phi_0=P_{\rm h}\phi_0$ is an on-shell free field obeying $\B\phi_0=0$, while the non-linear contributions $\phi_n$ for $n\geq1$ are given recursively 
by the analog of \eqref{recursive}:
\begin{equation}\label{recursion with h}
\phi_*=\phi_0-\frac{g}{2}\,h\big(\phi_*^2\big)\quad\longrightarrow\quad\phi_n=-\frac{1}{2}\sum_{k+l=n-1}h\big(\phi_k\phi_l\big)\;,    
\end{equation}
which would give an exact solution \emph{if} one could invert $\B$.
The first few orders of $\phi_{\rm n.l.}$ are given by
\begin{equation}
\phi_1=-\frac{1}{2}\,h(\phi_0^2)\;,\quad\phi_2=\frac{1}{2}\,h\big(\phi_0\,h(\phi_0^2)\big)\;,\quad\phi_3=-\frac12\,h\Big(\phi_0\,h\big(\phi_0\,h(\phi_0^2)\big)\Big)-\frac18\,h\Big(h(\phi_0^2)\,h(\phi_0^2)\Big)\;.
\end{equation}
The only difference compared to \eqref{phistarexp}, apart from $GJ\to\phi_0$, is that here the ``propagator'' $h$ gives zero whenever a total momentum of the product $\phi_k\phi_l$ is on-shell.
This difference has an important consequence: the field $\phi_*(\phi_0)$ does \emph{not} solve the equations exactly, since $\B$ is not invertible on arbitrary plane waves. 
Applying $\B$ to the recursion \eqref{recursion with h} we obtain
\begin{equation}\label{obstruction}
\B\phi_*-\frac{g}{2}\big(\phi_*^2\big)=-\frac{g}{2}\,P_{\rm h}\big(\phi_*^2\big)\;,
\end{equation}
while applying the projector we get $P_{\rm h}\phi_*=\phi_0$. We thus see that the field equation is not obeyed exactly by $\phi_*(\phi_0)$, with an obstruction that takes into account the possibility for certain sums of momenta to be on-shell. 

We shall now show that the action $S[\phi_*(\phi_0)]$ evaluated on this obstructed solution\footnote{Let us mention that if one were to impose $P_{\rm h}(\phi_*^2)=0$ to force $\phi_*$ to be a solution, $S[\phi_*]$ would vanish, as fits for an on-shell action with trivial boundary conditions.} is the generating functional of tree-level amplitudes in momentum space. To do so, let us examine how the projector and $h$ behave in the inner product. On sums of plane waves \eqref{planephi}, the inner product \eqref{inner plane waves} shows that off-shell momenta are paired with off-shell momenta and vice versa. This leads to
\begin{equation}\label{P symmetry scalar}
\big\l P_{\rm h}\phi,E\big\r=\big\l P_{\rm h}\phi,P_{\rm h}E\big\r=\big\l \phi,P_{\rm h}E\big\r \;,   
\end{equation}
with analogous expression for $(1-P_{\rm h})$. Similarly, using the definition of $h$ and \eqref{inner plane waves} one finds
\begin{equation}\label{h symmetry scalar}
\big\l hE_1,E_2\big\r=\big\l hE_2,E_1\big\r\;. 
\end{equation}
We can now insert $\phi_*(\phi_0)=\phi_0+\phi_{\rm n.l.}(\phi_0)$ into the action, written as $ S[\phi]=\frac12\big\l\phi,\B\phi\big\r-\frac{g}{3!}\big\l\phi,\phi^2\big\r$.
Using symmetry of the projector, together with $P_{\rm h}\phi_0=\phi_0$ and $P_{\rm h}\phi_{\rm n.l.}=0$, we obtain
the implicit expression
\begin{equation}
\bar S[\phi_0]=\frac12\Big\l\phi_{\rm n.l.}(\phi_0),\B\phi_{\rm n.l.}(\phi_0)\Big\r-\frac{g}{3!}\Big\l\phi_0+\phi_{\rm n.l.}(\phi_0),\big(\phi_0+\phi_{\rm n.l.}(\phi_0)\big)^2\Big\r\;,
\end{equation}
where we named $\bar S[\phi_0]:=S[\phi_*(\phi_0)]$ and $\phi_{\rm n.l.}$ is given iteratively by \eqref{recursion with h}.
Comparing this with \eqref{Gamma implicit} we see that $\bar S[\phi_0]$ is the same functional as $\Gamma_{\rm tree}[\phi_0]$, modulo the fact that now $\phi_0$ is given by sums of plane waves, and consequently the propagators $G$ are replaced by the well-defined operator $h$. 

In order to compute scattering amplitudes, one considers the expansion of $\bar S[\phi_0]$ in powers of $\phi_0$, which formally coincides with \eqref{Gamma g4}:
\begin{equation}\label{Sbar g4}
\bar S[\phi_0]=\int dx\,\Big\{-\frac{g}{3!}\,\phi_0^3+\frac{g^2}{8}\,\phi_0^2\,h\big(\phi_0^2\big)-\frac{g^3}{8}\,\phi_0^2\,h\Big(\phi_0\,h\big(\phi_0^2\big)\Big)\Big\}+\cO(g^4)\;,
\end{equation}
where we used the symmetry properties \eqref{P symmetry scalar} and \eqref{h symmetry scalar}, together with $\phi_0=P_{\rm h}\phi_0$ and $h=(1-P_{\rm h})h$. The $n-$point amplitude is then computed by taking the $\cO(\phi_0^n)$ (or $\cO(g^{n-2})$) contribution to $\bar S[\phi_0]$, setting $\phi_0=\sum_{i=1}^ne^{ik_i\cdot x}$ and restricting the result to the multilinear contribution in $k_1,k_2,\cdots ,k_n$, which coincides with the operation \eqref{ampGamma} on $\Gamma_{\rm tree}[\phi_0]$. 

Let us show with a standard example that the presence of the projector $(1-P_{\rm h})$ in $h$ does not change the usual form of the amplitudes. 
To compute the four-point amplitude, one takes the $\cO(g^2)$ contribution above and inserts
\begin{equation}
\phi_0(x)=e^{ik_1\cdot x}+e^{ik_2\cdot x}+e^{ik_3\cdot x}+e^{ik_4\cdot x}  \;.  
\end{equation}
Keeping the multilinear contribution in the four momenta we have
\begin{equation}
\begin{split}
\delta(K)\,\mathcal{A}_4^{\rm tree}(k_i)&=g^2\int dx\,\Big\{e^{i(k_1+k_2)\cdot x}\,h\Big(e^{i(k_3+k_4)\cdot x}\Big)+e^{i(k_2+k_3)\cdot x}\,h\Big(e^{i(k_1+k_4)\cdot x}\Big)\\
&\hspace{20mm}+e^{i(k_3+k_1)\cdot x}\,h\Big(e^{i(k_2+k_4)\cdot x}\Big)\Big\}\;,       
\end{split}
\end{equation}
yielding the standard $s$, $t$ and $u$ channels. At this point, one has to consider that the external momenta $k_i^\mu$ are fixed, so that their sums $k^\mu_i+k^\mu_j$ are either on-shell or not. In a four-point amplitude, any sum of two momenta is necessarily off-shell\footnote{Here we are not discussing what happens in the limit $(k_i+k_j)^2\to0$, the projector $P_{\rm h}$ only prevents from taking $(k_i+k_j)^2\equiv0$ in a four-point amplitude.}. This translates into
\begin{equation}
P_{\rm h}\Big(e^{i(k_i+k_j)\cdot x}\Big)=0\quad\longrightarrow\quad h\Big(e^{i(k_i+k_j)\cdot x}\Big)=\frac{1}{(k_i+k_j)^2}\Big(e^{i(k_i+k_j)\cdot x}\Big) \;, \end{equation}
which yields the familiar expression \eqref{4scalar amp} for the amplitude.

We conclude this subsection by observing that the functional $\bar S[\phi_0]$ ought to define the brackets $\bar B_n$ of a new $L_\infty$ algebra. Just as the original action $S[\phi]$ fixes the brackets $B_1$ and $B_2$ of the $\phi^3$ theory via \eqref{MC general}, we can extract infinitely many brackets $\bar B_n$ from $\bar S[\phi_0]$ by its variation:
\begin{equation}
\frac{\delta\bar S}{\delta\phi_0}=:\tfrac12\,\bar B_2(\phi_0,\phi_0)+\tfrac{1}{3!}\,\bar B_3(\phi_0,\phi_0,\phi_0)+\cdots+\tfrac{1}{n!}\,\bar B_n(\phi_0,\ldots,\phi_0)+\cdots\;.  \end{equation}
The first $\bar B_n$ can be obtained by directly varying \eqref{Sbar g4} and read
\begin{equation}\label{on-shell Bs}
\begin{split}
\bar B_2(\phi_0,\phi_0)&=-g\,P_{\rm h}(\phi_0^2)\;,\quad\bar B_3(\phi_0,\phi_0,\phi_0)=3\,g^2\,P_{\rm h}\big(\phi_0\,h(\phi_0^2)\big)\\
\bar B_4(\phi_0,\phi_0,\phi_0,\phi_0)
&=-g^3\,P_{\rm h}\Big\{12\,\phi_0\,h\big(\phi_0\,h(\phi_0^2)\big)+3\,h(\phi_0^2)\,h(\phi_0^2)\Big\}\;.
\end{split}    
\end{equation}
Notice the on-shell projector $P_{\rm h}$ in front of every $\bar B_n$. This is due to $\phi_0=P_{\rm h}\phi_0$, which forces the projection on the variation $\frac{\delta \bar S}{\delta\phi_0}$ by means of \eqref{P symmetry scalar}. At a diagrammatic level, this simply states that amputating an on-shell external leg produces an on-shell current.

While the original brackets $B_1$ and $B_2$ encoded the kinetic term and cubic interaction of the theory, the infinitely many brackets $\bar B_n$ derived from the on-shell action are associated to all tree-level scattering amplitudes.
As we shall see in the following, this new $L_\infty$ algebra
is naturally related to the original one by a procedure known as homotopy transfer.

\subsection{Homotopy transfer and amplitudes}

Homotopy transfer is a procedure that allows one to transport an $L_\infty$ algebra structure from one space to another under certain conditions. In this section we start by a brief review of the setup (for a field theory oriented presentation see e.g. \cite{Arvanitakis:2020rrk,Arvanitakis:2021ecw}), and then show that the $L_\infty$ algebra generated by the on-shell functional $\bar S[\phi_0]$ is the one obtained by homotopy transfer to the space of on-shell free fields.

Let us begin by considering two generic chain complexes\footnote{At this stage $B_1$ and $\bar B_1$ are just nilpotent differentials, not implying the existence of any nonlinear structure. We keep the notation $B_1$ just to avoid proliferation of symbols.} $(\cX, B_1)$ and $(\bar\cX,\bar B_1)$, with differentials $B_1$ and $\bar B_1$ of degree $+1$. In this context $\bar\cX$ can be usually viewed as a vector subspace of $\cX$, but it is not necessary.
Suppose that one has a projection $p:\cX\rightarrow\bar\cX$ and inclusion $\iota:\bar\cX\rightarrow\cX$ of intrinsic degree zero:
\begin{equation}\label{hom transf diagram}
\begin{tikzcd}[row sep=10mm]
\cdots\arrow{r}{B_1}&X_0\arrow[shift left=1mm]{d}{p}\arrow{r}{B_1}&X_1\arrow[shift left=1mm]{d}{p}\arrow{r}{B_1}&\cdots\\
\cdots\arrow{r}{\bar B_1}&\bar X_0\arrow[bend left=20, shift left=1mm]{u}{\iota}\arrow{r}{\bar B_1}&\bar X_1\arrow[bend left=20, shift left=1mm]{u}{\iota}\arrow{r}{\bar B_1}&\cdots\;,
\end{tikzcd}    
\end{equation}
and that they are chain maps, meaning they preserve the differential: 
\begin{equation}\label{chain map}
p\,B_1=\bar B_1\, p \;,\quad \iota\,\bar B_1=B_1\,\iota\;.   
\end{equation}
Let us further assume that their compositions obey
\begin{equation}\label{homotopy relation}
p\,\iota=1_{\bar\cX}\;,\quad\iota\, p=1_\cX+B_1\,h+h\,B_1\;,    
\end{equation}
for a degree $-1$ operator $h$ called a homotopy. The first relation above is the statement that $p$ is indeed a projection, while the second expresses that $\iota$ inverts $p$ only up to homotopy. Such triplet $(p,\iota,h)$ is referred to as the ``homotopy data''.
Under these assumptions, the homotopy transfer theorem ensures that if $\cX$ carries an $L_\infty$ structure with brackets $B_1,B_2, \cdots$, one can define an $L_\infty$ algebra on $\bar \cX$ with brackets $\bar B_1,\bar B_2, \cdots$, where all the higher brackets $\bar B_n$, $n\geq2$ are constructed explicitly from the $B_n$ and the triplet $(p,\iota,h)$. We will now specialize to our example of scalar $\phi^3$ theory and see how the projection to on-shell free fields fulfills the requirements for homotopy transfer.

As before, we restrict all functions in both spaces $\cX$ and $\bar\cX$ to finite sums of plane waves as in \eqref{planephi}. The goal is to find a projector $p$ to the space of free fields $\bar\phi$ obeying $\B\bar\phi=0$, i.e. to on-shell plane waves: $\bar\phi(x)=\sum_{i}\bar\phi_{i}\,e^{ik_i\cdot x}$, with $k_i^2=0$. This is given by the harmonic projector \eqref{Ph}, so that $p(\phi)=P_{\rm h}(\phi)$. 
Given that the differential in $\cX$ is $B_1=\B$, we have that the projected space has zero differential: $\bar B_1=0$.
Requiring the chain map condition \eqref{chain map} for $p$ then implies that $P_{\rm h}$ is also the projector for equations: $p(E)=P_{\rm h}(E)$. It is thus clear that $\bar\cX\simeq{\rm ker}\B\subset\cX$ is a linear subspace of $\cX$. The diagram \eqref{hom transf diagram} looks like  
\begin{equation}
\begin{tikzcd}[row sep=10mm]
0\arrow{r}{0}&\phi\arrow[shift left=1mm]{d}{P_{\rm h}}\arrow{r}{\B}&E\arrow[shift left=1mm]{d}{P_{\rm h}}\arrow{r}{0}&0\\
0\arrow{r}{0}&\bar\phi\arrow[bend left=20, shift left=1mm]{u}{\iota}\arrow{r}{0}&\bar E\arrow[bend left=20, shift left=1mm]{u}{\iota}\arrow{r}{0}&0\;,
\end{tikzcd}    
\end{equation}
where for $\iota$ we take the trivial inclusion, which views on-shell plane waves as a subset of generic ones: $\iota(\bar\cX)={\rm ker}\B\subset\cX$.

Since the inclusion map is trivial, from now on we will work directly on the space $\cX$ upon identifying $\bar\cX\simeq\iota(\bar\cX)\subset\cX$. In particular, we identify $\bar\phi\simeq\iota(\bar\phi)=\phi_0$, where fields $\phi_0\in X_0$ obey $\B\phi_0=0$ and coincide with the free fields used in the previous section.
Similarly, we identify the projector as an endomorphism in $\cX$:
\begin{equation}
P_{\rm h}\simeq\iota\, p:\cX\longrightarrow\cX\;.    
\end{equation}
The chain map conditions \eqref{chain map} and homotopy relations \eqref{homotopy relation} then reduce to
\begin{equation}
\begin{split}
P_{\rm h}\B=\B P_{\rm h}=0\;,\quad P_{\rm h}^2=P_{\rm h}\;,\quad \B h=\big(P_{\rm h}-1\big)_{X_1}\;,\quad h\B=\big(P_{\rm h}-1\big)_{X_0}\;,   
\end{split}    
\end{equation}
which have already been established in the previous section. Notice that the generic homotopy relation reduces to the above in a scalar theory, since $h$ acts non trivially only on $X_1$, and $B_1$ only on $X_0$. Let us mention that the homotopy \eqref{h} also obeys the so-called side conditions, which in this case read
\begin{equation}\label{side conditions scalar}
P_{\rm h}h=hP_{\rm h}=0\;,\quad h^2=0\;,    
\end{equation}
where nilpotency of $h$ is trivially ensured by degree.

Given that the conditions for homotopy transfer are satisfied, the transported brackets $\bar B_n$ are given (see e.g.\cite{Berglund2009HomologicalPT,Chuang2017OnTP,Arvanitakis:2020rrk} and Appendix \ref{Sec:Appendix Loo}) by the sum of all inequivalent tree diagrams (cubic in this case) with $n$ distinct inputs and one output, each counted once. As a concrete example, the first few brackets $\bar B_n$ read\footnote{To be precise these are $\iota\,\bar B_n$, embedded in $\cX$ via the trivial inclusion.}
\begin{equation}\label{bar bs homtransf}
\begin{split}
\bar B_2(\bar\phi_1,\bar\phi_2)&=P_{\rm h}B_2(\bar\phi_1,\bar\phi_2)=\begin{tikzpicture}[baseline={([yshift=-.5ex]current bounding box.center)},level distance=7mm,sibling distance=10mm]
  \node[scale=0.5,circle, draw] {}
    child[grow=up] {
     child {node[scale=0.7] {2}}  child{node[scale=0.7] {1}}
    };
\end{tikzpicture}\;,\\
\bar B_3(\bar\phi_1,\bar\phi_2,\bar\phi_3)&=P_{\rm h}\Big\{B_2\big(hB_2(\bar\phi_1,\bar\phi_2),\bar\phi_3\big)+B_2\big(hB_2(\bar\phi_1,\bar\phi_3),\bar\phi_2\big)+B_2\big(hB_2(\bar\phi_2,\bar\phi_3),\bar\phi_1\big)\Big\}\\
&=\begin{tikzpicture}[baseline={([yshift=-.5ex]current bounding box.center)},level distance=7mm,sibling distance=10mm]
  \node[scale=0.5,circle, draw] {}
    child[grow=up] {
     child{child {node[scale=0.7] {3}} child[missing]}    child {child {node[scale=0.7] {2}} child {node[scale=0.7] {1}}}
    };
\end{tikzpicture}+\begin{tikzpicture}[baseline={([yshift=-.5ex]current bounding box.center)},level distance=7mm,sibling distance=10mm]
  \node[scale=0.5,circle, draw] {}
    child[grow=up] {
     child{child {node[scale=0.7] {2}} child[missing]}    child {child {node[scale=0.7] {3}} child {node[scale=0.7] {1}}}
    };
\end{tikzpicture}+\begin{tikzpicture}[baseline={([yshift=-.5ex]current bounding box.center)},level distance=7mm,sibling distance=10mm]
  \node[scale=0.5,circle, draw] {}
    child[grow=up] {
     child{child {node[scale=0.7] {1}} child[missing]}    child {child {node[scale=0.7] {3}} child {node[scale=0.7] {2}}}
    };
\end{tikzpicture}\;,\\
\bar B_4(\bar\phi_1,\bar\phi_2,\bar\phi_3,\bar\phi_4)&=P_{\rm h}\Big\{B_2\Big(hB_2\big(hB_2(\bar\phi_1,\bar\phi_2),\bar\phi_3\big),\bar\phi_4\Big)+\text{11 permutations}\Big\}\\
&+P_{\rm h}\Big\{B_2\big(hB_2(\bar\phi_1,\bar\phi_2),hB_2(\bar\phi_3,\bar\phi_4)\big)+\text{2 permutations}\Big\}\\
&=\begin{tikzpicture}[baseline={([yshift=-.5ex]current bounding box.center)},level distance=7mm,sibling distance=10mm]
  \node[scale=0.5,circle, draw] {}
    child[grow=up]{child{child{child{node[scale=0.7] {4}} child[missing]} child[missing]} child{
     child{child {node[scale=0.7] {3}} child[missing]}    child {child {node[scale=0.7] {2}} child {node[scale=0.7] {1}}}
    }};
\end{tikzpicture}+\text{11 more}
+\begin{tikzpicture}[baseline={([yshift=-.5ex]current bounding box.center)},level distance=10mm]
\tikzstyle{level 2}=[sibling distance=15mm]
  \tikzstyle{level 3}=[sibling distance=10mm]
  \node[scale=0.5,circle, draw]{}[grow=up] child{child{child{node[scale=0.7] {4}} child{node[scale=0.7] {3}}} child{child{node[scale=0.7] {2}} child{node[scale=0.7] {1}}}} ;
\end{tikzpicture}+\text{2 more}\;,    
\end{split}    
\end{equation}
which, using $B_2(\phi_1,\phi_2)=-g\,\phi_1\phi_2$, agree with the explicit brackets \eqref{on-shell Bs} derived from  $\bar S[\phi_0]$.
In the above diagrams the external $i-$th leg stands for an on-shell field $\bar\phi_i$, vertices are factors of $B_2$ (each carrying a power of $g$) and internal lines are homotopies $h$, as one can see by comparison with the analytic expressions. The empty circle at the root of the trees signifies that the output is an element of the space of on-shell sources, i.e.~$\iota(\bar X_1)\simeq\bar X_1$. Notice that the counting of inequivalent diagrams takes into account the symmetry of the cubic vertex as $B_2(\Phi_1,\Phi_2)=B_2(\Phi_2,\Phi_1)$, where inputs $\Phi_{1,2}$ can be either external legs $\bar\phi_i$ or generic subtrees in $X_0$ as, for instance $hB_2(\bar\phi_1,\bar\phi_2)$.

Given the original inner product \eqref{inner scalar}, we can define an inner product on $\bar\cX$ between on-shell fields and sources via the trivial inclusion:
\begin{equation}\label{inner transf}
\big\l\bar\phi,\bar E\big\r_{\bar\cX}:=\big\l\iota(\bar\phi), \iota(\bar E)\big\r_{\cX}=\int dx\,\bar\phi(x)\,\bar E(x)\;.  
\end{equation}
The transferred brackets $\bar B_n$ are cyclic with respect to the above inner product, thanks to the cyclicity of $B_2$, the symmetry of the homotopy \eqref{h symmetry scalar} and the side conditions \eqref{side conditions scalar}. In practice, this means that the pairing $\l\bar\phi_1,\bar B_n(\bar\phi_2,\ldots,\bar\phi_{n+1})\r$ is totally symmetric in the $n+1$ inputs. 

From the diagrammatic construction of $\bar B_n(\bar\phi_1,\ldots,\bar\phi_n)$ it follows that the pairing with $\bar\phi_{n+1}$ is given by the sum of all trees with $n+1$ labeled external legs, each counted once.
These are precisely the diagrams contributing to the $(n+1)-$point amplitude, recalling that the symmetry factor of a labeled tree diagram is 1. We can thus identify
\begin{equation}
\delta(K)\,\mathcal{A}_n^{\rm tree}(k_1,\ldots,k_n)=\Big\l \bar\phi_1,\bar B_{n-1}\big(\bar\phi_2,\ldots,\bar\phi_n\big)\Big\r\;,\quad\bar\phi_i(x)=e^{ik_i\cdot x}  \;,\quad k_i^2=0\;,  
\end{equation}
where total Bose symmetry in the external legs is ensured by cyclicity of $\bar B_{n-1}$. The homotopies $h$ in the internal lines reduce to standard propagators if one assumes, as always in writing down an amplitude, that  exchanged momenta are off-shell. 
As an example, pairing $\bar\phi_4$ with $\bar B_3(\bar\phi_1,\bar\phi_2,\bar\phi_3)$ one obtains
\begin{equation}
\begin{split}
\big\l\bar\phi_4,\bar B_3(\bar\phi_1,\bar\phi_2,\bar\phi_3)\big\r&=\left\l\!\!
\begin{tikzpicture}[baseline={([yshift=-.5ex]current bounding box.center)},level distance=10mm,sibling distance=5mm]
\node[scale=0.7]{4} child[grow=right] ;
\end{tikzpicture}\;
,\begin{tikzpicture}[baseline={([yshift=-.5ex]current bounding box.center)},level distance=7mm,sibling distance=5mm]
  \node[scale=0.5,circle, draw] {}
    child[grow=right] {
     child{child {node[scale=0.7] {3}} child[missing]}    child {child {node[scale=0.7] {2}} child {node[scale=0.7] {1}}}
    };
\end{tikzpicture}+\begin{tikzpicture}[baseline={([yshift=-.5ex]current bounding box.center)},level distance=7mm,sibling distance=5mm]
  \node[scale=0.5,circle, draw] {}
    child[grow=right] {
     child{child {node[scale=0.7] {2}} child[missing]}    child {child {node[scale=0.7] {3}} child {node[scale=0.7] {1}}}
    };
\end{tikzpicture}+\begin{tikzpicture}[baseline={([yshift=-.5ex]current bounding box.center)},level distance=7mm,sibling distance=5mm]
  \node[scale=0.5,circle, draw] {}
    child[grow=right] {
     child{child {node[scale=0.7] {1}} child[missing]}    child {child {node[scale=0.7] {3}} child {node[scale=0.7] {2}}}
    };
\end{tikzpicture}\right\r \\
&=\begin{tikzpicture}[baseline={([yshift=-.5ex]current bounding box.center)}]
\node[scale=0.7] (1)  at  (0,0) {3};
\node[scale=0.7] (2) at  (0,2) {4};
\node[scale=0.7] (3)  at  (3,2) {1};
\node[scale=0.7] (4)  at  (3,0) {2};
\draw (1) -- (0.6,1); \draw (2) -- (0.6,1); \draw (3) -- (2.4,1); \draw (4) -- (2.4,1); \draw (0.6,1) -- (2.4,1);
\end{tikzpicture}+\begin{tikzpicture}[baseline={([yshift=-.5ex]current bounding box.center)}]
\node[scale=0.7] (1)  at  (0,0) {2};
\node[scale=0.7] (2) at  (0,2) {4};
\node[scale=0.7] (3)  at  (3,2) {1};
\node[scale=0.7] (4)  at  (3,0) {3};
\draw (1) -- (0.6,1); \draw (2) -- (0.6,1); \draw (3) -- (2.4,1); \draw (4) -- (2.4,1); \draw (0.6,1) -- (2.4,1);
\end{tikzpicture}+\begin{tikzpicture}[baseline={([yshift=-.5ex]current bounding box.center)}]
\node[scale=0.7] (1)  at  (0,0) {1};
\node[scale=0.7] (2) at  (0,2) {4};
\node[scale=0.7] (3)  at  (3,2) {2};
\node[scale=0.7] (4)  at  (3,0) {3};
\draw (1) -- (0.6,1); \draw (2) -- (0.6,1); \draw (3) -- (2.4,1); \draw (4) -- (2.4,1); \draw (0.6,1) -- (2.4,1);
\end{tikzpicture}\;,
\end{split}    
\end{equation}
which indeed yields the four-point amplitude upon using $\bar\phi_i=e^{ik_i\cdot x}$ and assuming $(k_i+k_j)^2\neq0$.

Having established the connection between the transferred brackets $\bar B_n$ and scattering amplitudes, we will now show that the $\bar B_n$ coincide with the brackets \eqref{on-shell Bs} obtained from the on-shell action. First of all, we evaluate the transferred brackets at coinciding inputs:
\begin{equation}
\bar B_n(\phi_0,\ldots,\phi_0)\;,\quad \B\phi_0=0\;,    
\end{equation}
where $\phi_0$ is given by linear combinations of on-shell plane waves. In doing so, all diagrams differing only by relabeling of the external legs sum up. For instance, from \eqref{bar bs homtransf} one obtains
\begin{equation}
\begin{split}
\bar B_2(\phi_0,\phi_0)&=P_{\rm h}B_2(\phi_0,\phi_0)\;,\\
\bar B_3(\phi_0,\phi_0,\phi_0)&=3\,P_{\rm h}B_2\big(hB_2(\phi_0,\phi_0),\phi_0\big)\;,\\
\bar B_4(\phi_0,\phi_0,\phi_0,\phi_0)&=P_{\rm h}\Big\{12\,B_2\Big(hB_2\big(hB_2(\phi_0,\phi_0),\phi_0\big),\phi_0\Big)+3\,B_2\big(hB_2(\phi_0,\phi_0),hB_2(\phi_0,\phi_0)\big)\Big\}\;.   
\end{split}    
\end{equation}
In general, the prefactor of a given diagram $\Gamma$ will be the number $c_{\Gamma}$ of inequivalent ways one can distribute the $n$ external legs. This is given by the ratio $c_\Gamma=\frac{n!}{S_\Gamma}$ of all $n!$ permutations of the external legs, divided by the number of permutations that leave the diagram unchanged which, by definition, is the symmetry factor $S_\Gamma$. We thus conclude that $\frac{1}{n!}\,\bar B_n(\phi_0,\ldots,\phi_0)$ is given by the sum of all rooted diagrams weighted by their symmetry factor.
This diagrammatic expansion coincides with the one obtained from the recursion \eqref{recursion with h} for the nonlinear field $\phi_{\rm n.l.}(\phi_0)$. More precisely, we have the following relation between $\phi_*(\phi_0)$ and $\bar B_n(\phi_0,\ldots,\phi_0)$:
\begin{equation}\label{phiBs relations}
\begin{split}
\phi_*(\phi_0)&=\phi_0+\sum_{n=1}^\infty \phi_n(\phi_0) \;,\quad\hspace{8mm}\cJ_n(\phi_0):=\frac{1}{2}\sum_{k+l=n}B_2\big(\phi_k,\phi_l\big)\\
\phi_n(\phi_0)&=h\,\cJ_{n-1}(\phi_0)\;,\quad\frac{1}{n!}\,\bar B_n(\phi_0,\ldots,\phi_0)=P_{\rm h}\,\cJ_{n-2}(\phi_0)\;,
\end{split}    
\end{equation}
where we have absorbed the factor $g^n$ into $\phi_n$ in order to have cleaner expressions, and defined the currents $\cJ_n(\phi_0)$. With this normalization (recall that $B_2=\cO(g)$) we have $\phi_n=\cO(g^{n})$. 

With these relations at hand, we can now determine that the brackets $\bar B_n$ come from the variation of the on-shell action $\bar S[\phi_0]$. Summing all the brackets $\bar B_n$ we obtain
\begin{equation}
\sum_{n=2}^\infty\frac{1}{n!}\,\bar B_n(\phi_0,\ldots,\phi_0)=P_{\rm h}\sum_{k,l=0}^\infty\frac{1}{2}\,B_2(\phi_k,\phi_l)=\frac{1}{2}\,P_{\rm h}B_2(\phi_*,\phi_*)\;,       
\end{equation}
which is precisely the obstruction \eqref{obstruction} of $\phi_*$ to solve the field equations:
\begin{equation}\label{obstruction Bs}
B_1(\phi_*)+\frac{1}{2}\,B_2\big(\phi_*,\phi_*\big)=\frac{1}{2}\,P_{\rm h}\,B_2\big(\phi_*,\phi_*\big)\;.
\end{equation}
Finally, let us see that the above expression is the variation of $\bar S[\phi_0]$ with respect to $\phi_0$. Using $\bar S[\phi_0]=S[\phi_*(\phi_0)]$ we compute the variation as
\begin{equation}
\begin{split}
\delta\bar S[\phi_0]&=\Big\l\delta\phi_0,\frac{\delta\bar S}{\delta\phi_0}\Big\r=\Big\l\delta\phi_*(\phi_0),\frac{\delta S}{\delta\phi}\Big\rvert_{\phi=\phi_*(\phi_0)}\Big\r\\
&=\Big\l\delta\phi_*(\phi_0),B_1(\phi_*)+\tfrac{1}{2}\,B_2\big(\phi_*,\phi_*\big)\Big\r
=\frac{1}{2}\Big\l\delta\phi_0+\delta\phi_{\rm n.l.}(\phi_0),P_{\rm h}\,B_2\big(\phi_*,\phi_*\big)\Big\r\;,
\end{split}    
\end{equation}
upon writing $\phi_*(\phi_0)=\phi_0+\phi_{\rm n.l.}(\phi_0)$ and using \eqref{obstruction Bs}. Thanks to \eqref{phiBs relations}, we can write the nonlinear part of $\phi_*$ as $\phi_{\rm n.l.}(\phi_0)=h\sum_{n=0}^\infty\cJ_n(\phi_0)$. The side condition $P_{\rm h}h=0$ then ensures that $\delta\phi_{\rm n.l.}(\phi_0)$ above is orthogonal to $P_{\rm h}B_2(\phi_*,\phi_*)$, thus proving that the brackets obtained from homotopy transfer are generated by the variation of $\bar S$:
\begin{equation}
\delta\bar S[\phi_0]=\frac12\,\Big\l\delta\phi_0,P_{\rm h}\,B_2(\phi_*,\phi_*)\Big\r=\sum_{n=2}^\infty\frac{1}{n!}\,\Big\l\delta\phi_0,\bar B_n(\phi_0,\ldots,\phi_0)\Big\r\;.   \end{equation}

\section{Amplitudes in Yang-Mills theory}

In this section we will discuss the $L_\infty$ interpretation of tree-level scattering amplitudes for the more interesting case of Yang-Mills theory. We will start by briefly introducing the $L_\infty$ algebra of Yang-Mills theory as a tensor product of its color Lie algebra with a kinematic $C_\infty$ algebra (see \cite{Zeitlin:2008cc,Borsten:2021hua,Bonezzi:2022yuh,Bonezzi:2022bse} for full details). In particular, we will show how the generalized Jacobi identities of the off-shell $L_\infty$ algebra imply the linear Ward identities of scattering amplitudes upon homotopy transfer.

\subsection{$L_\infty$ and $C_\infty$ algebras of Yang-Mills}\label{sec:YMoff}

We will now present the $L_\infty$ algebra associated to Yang-Mills theory, written as a tensor product between ``color'' and ``kinematics''. In the following we consider the Yang-Mills action, explicitly expanded in powers of the gauge field:
\begin{equation}\label{YM action}
S_{\rm YM}=\int dx\, \Tr\Big\{\tfrac{1}{2}\, \cA_{\mu}\square \cA^{\mu}+\tfrac{1}{2}\, (\del\cdot \cA)^2-g\,\del_\mu \cA_\nu[\cA^\mu, \cA^\nu]
-\tfrac14\,g^2\,[\cA_\mu, \cA_\nu][\cA^\mu, \cA^\nu]\Big\}\;,
\end{equation}
where $\cA_\mu=A_\mu^a\,T_a$ is the Lie algebra-valued vector field, and we use the short-hand notation $\del\cdot \cA = \partial_{\mu}\cA^{\mu}$. 
The generators and Cartan-Killing form of the color Lie algebra are normalized as
\begin{equation}
[T_a,T_b]=f_{ab}^c\,T_c\;,\quad\Tr(T_aT_b)=\delta_{ab}\;.    
\end{equation}
The Yang-Mills action is gauge invariant  under $\delta \cA_\mu=D_\mu\boldsymbol{\Lambda}=\del_\mu\bLambda+g[\cA_\mu,\bLambda]$, for a Lie algebra-valued parameter $\bLambda=\Lambda^a\,T_a$. The consistency of the theory is encoded, order by order in $\cA_\mu$, by the $L_\infty$ structure over the graded vector space $\cX_{\rm YM}=\bigoplus_{i=-1}^2X_i$. This consists of the chain complex
\begin{equation}
\begin{tikzcd}[row sep=2mm]
X_{-1}\arrow{r}{B_1}&X_{0}\arrow{r}{B_1}&X_{1}\arrow{r}{B_1}&X_{2}\\
\bLambda& \cA_\mu&\cJ_\mu&\cN
\end{tikzcd}    \;,
\end{equation}
associated with gauge parameters, fields, equations of motion (or currents) and Noether identities, endowed with graded symmetric multilinear brackets $B_1$, $B_2$ and $B_3$ of degree $|B_n|=+1$ (see e.g. \cite{Hohm:2017pnh,Bonezzi:2022yuh} for more details). These brackets obey nontrivial quadratic relations, encoding the gauge structure and covariance of the interacting theory. For instance, the differential $B_1$ acts as
\begin{equation}\label{B1 YM}
B_1(\bLambda)=\del_\mu\bLambda\;,\quad B_1(\cA)=\B \cA_\mu-\del_\mu\del\cdot \cA\;,\quad B_1(\cJ)=\del^\mu\cJ_\mu\;,    
\end{equation}
and its nilpotency corresponds to gauge invariance of the free theory. While brackets between fields $B_2(\cA,\cA)$ and $B_3(\cA,\cA,\cA)$ encode the interactions, as for a scalar theory, the nonlinear part of the gauge transformation is described by $B_2(\cA,\bLambda)$, the gauge algebra by $B_2(\bLambda_1,\bLambda_2)$, while the other non-vanishing brackets are related to the nonlinear Noether identity.

Given that all elements $\Psi\in\cX_{\rm YM}$ take values in the color Lie algebra $\mathfrak{g}$: $\Psi=\psi^a(x)\,T_a$, the $L_\infty$ algebra itself can be factorized into the tensor product $\cX_{\rm YM}=\cK_{\rm YM}\otimes \mathfrak{g}$. The elements of $\cK_{\rm YM}$ are spacetime fields whose color index $a$ is to be viewed as just a label of a basis expansion of $\mathfrak{g}$.
Following \cite{Bonezzi:2022bse}, we take the color Lie algebra to be concentrated in degree $-1$, meaning that we assign degree $|T_a|=-1$ to the generators of $\mathfrak{g}$ and degree $+1$ to the Lie bracket $[\cdot,\cdot]$. The resulting color-stripped complex $\cK_{\rm YM}=\bigoplus_{i=0}^3K_i$ is given by
\begin{equation}
\begin{tikzcd}[row sep=2mm]
K_{0}\arrow{r}{m_1}&K_{1}\arrow{r}{m_1}&K_{2}\arrow{r}{m_1}&K_{3}\\
\Lambda& A_\mu&J_\mu&N
\end{tikzcd}    \;,
\end{equation}
where we use a different font for the color-stripped gauge parameters, fields etc. to remind that they are \emph{not} Lie algebra-valued.

Expanding arbitrary elements of $\cX_{\rm YM}$ as $\Psi_i=\psi_i^a\,T_a$, with $\psi_i^a\in\cK_{\rm YM}$ and $T_a\in\mathfrak{g}$, the brackets $B_n$ of the $L_\infty$ algebra are given in terms of multilinear products $m_1$, $m_2$ and $m_3$ acting on $\cK_{\rm YM}$ and of the color Lie bracket $[\cdot,\cdot]$ as
\begin{equation}\label{Bs from ms}
\begin{split}
B_1(\Psi)&=m_1(\psi^a)\,T_a \;,\\
B_2(\Psi_1,\Psi_2)&=(-1)^{\Psi_1}g\,m_2(\psi_1^a,\psi_2^b)\,[T_a,T_b]\;,\\
B_3(\Psi_1,\Psi_2,\Psi_3)&=g^2\Big((-1)^{\Psi_2}m_3(\psi_1^a,\psi_2^b,\psi_3^c)+(-1)^{\Psi_1(\Psi_2+1)}m_3(\psi_2^a,\psi_1^b,\psi_3^c)\Big)\,[T_a,[T_b,T_c]]\;,
\end{split}    
\end{equation}
where the products $m_n$ have intrinsic degree $|m_n|=2-n$. Since the generators $T_a$ are taken to have degree $-1$, the degrees in $\cX_{\rm YM}$ and $\cK_{\rm YM}$ are related by $|\Psi|_{\cX_{\rm YM}}=|\psi^a|_{\cK_{\rm YM}}-1$, for $\Psi=\psi^a\,T_a$.

For $(\cX_{\rm YM},\{B_n\})$ to form an $L_\infty$ algebra, i.e.~for the brackets $B_n$ to obey the $L_\infty$ quadratic relations, the color-stripped space $\cK_{\rm YM}$ equipped with the products $m_n$ has to be a $C_\infty$ algebra \cite{Zeitlin:2008cc,Borsten:2021hua,Bonezzi:2022yuh}. This is a homotopy generalization of a commutative associative algebra, in the same fashion as $L_\infty$ generalizes Lie algebras. In particular, for the case of Yang-Mills theory with non-vanishing $m_n$ for $n=1,2,3$, the $C_\infty$ quadratic relations read
\begin{equation}\label{Cinfty relations}
\begin{split}
m_1^2(\psi)&=0\;,\\
m_1\big(m_2(\psi_1,\psi_2)\big)&=m_2\big(m_1(\psi_1),\psi_2\big)+(-1)^{\psi_1}m_2\big(\psi_1,m_1(\psi_2)\big)\;,\\
m_2\big(m_2(\psi_1,\psi_2),\psi_3\big)-m_2\big(\psi_1,m_2(\psi_2,\psi_3)\big)&=m_1\big(m_3(\psi_1,\psi_2,\psi_3)\big)+m_3\big(m_1(\psi_1),\psi_2,\psi_3\big)\\
&+(-1)^{\psi_1}m_3\big(\psi_1,m_1(\psi_2),\psi_3\big)\\
&+(-1)^{\psi_1+\psi_2}m_3\big(\psi_1,\psi_2,m_1(\psi_3)\big)\;,
\end{split}    
\end{equation}
denoting generic elements of $\cK_{\rm YM}$ by $\psi_i$. The above relations state that $m_1$ is a differential that acts as a derivation on $m_2$, and that $m_2$ is associative up to homotopy. These relations, in conjunction with the Jacobi identity of the color Lie bracket, ensure the $L_\infty$ relations of the $B_n$ and thus consistency of Yang-Mills theory.

These quadratic relations are the same for $A_\infty$ algebras. $C_\infty$ algebras differ in that the products have nontrivial symmetry properties, as they generalize commutative algebras. This is needed in $\cK_{\rm YM}$ in order to ensure that the $L_\infty$ brackets \eqref{Bs from ms} are graded symmetric. More precisely, the products $m_n$ vanish on shuffle permutations of inputs. We recall that the shuffle $P\shuffle Q$ between the ordered sets of labels $P=i_1\cdots i_p$ and $Q=j_1\cdots j_q$ is given by the sum of all permutations of $i_1\cdots i_p\,j_1\cdots j_q$ preserving the order of the elements within $P$ and $Q$. For $m_2$ and $m_3$, the independent symmetry properties are given by
\begin{equation}\label{Coo sym}
\begin{split}
m_2(\psi_1,\psi_2)-(-1)^{\psi_1\psi_2}m_2(\psi_2,\psi_1)&=0\;,\\    
m_3(\psi_1,\psi_2,\psi_3)-(-1)^{\psi_1\psi_2}m_3(\psi_2,\psi_1,\psi_3)+(-1)^{\psi_1(\psi_2+\psi_3)}m_3(\psi_2,\psi_3,\psi_1)&=0\;,
\end{split}    
\end{equation}
in terms of the signed graded\footnote{This means that the corresponding sign is given by the parity of the permutation plus the Koszul sign given by permuting graded elements, see e.g. \cite{Borsten:2021hua}.} shuffles $1\shuffle2$ and $1\shuffle23$. These reduce to simple shuffles with all plus signs when restricting all $\psi_i$ to color-stripped fields $A^\mu_i$ in $K_1$.

In analogy with the scalar case, we restrict elements of $\cK_{\rm YM}$ to finite sums of plane waves. Consequently, elements of $\cX_{\rm YM}$ are given by finite sums of Lie algebra-valued plane waves. For instance, we consider gauge fields of the form
\begin{equation}
\cA^\mu(x)=\sum_i a^\mu_{i}\,e^{ik_i\cdot x}\,T_i\;,   
\end{equation}
where here $T_i=c_i^a\,T_a$ are arbitrary vectors of $\mathfrak{g}$, not necessarily basis elements. Using plane waves, it is convenient to define the action of the products $m_n$ directly in terms of the Fourier coefficients. Taking color-stripped fields given by single plane waves: $A^\mu_i(x)=a^\mu_i\,e^{ik_i\cdot x}$, the $C_\infty$ products on the $A_i$ can be defined via
\begin{equation} \label{mn As}
 m_n(A_1,A_2,\ldots, A_n)=m_n(a_1,a_2,\ldots, a_n)\,e^{i(k_1+k_2+\cdots+k_n)\cdot x}\;,
\end{equation} 
where the products acting on polarizations $a^\mu_i$ read 
\begin{equation}\label{mn Fourier}
\begin{split}
m_1(a)&=-k^2a^\mu+k^\mu k\cdot a\;,\\
m_2( a_1, a_2)&=i\,k_1\cdot a^\mu_1\, a_2+2i\,k_2\cdot  a_1\, a_2^\mu+i\,k_1^\mu\, a_1\cdot a_2-(1\leftrightarrow2)\;,\\
m_3( a_1, a_2, a_3)&= a_1\cdot a_2\, a_3^\mu+ a_2\cdot a_3\, a_1^\mu-2\, a_3\cdot a_1\, a_2^\mu\;,
\end{split}    
\end{equation}
and are extended to arbitrary linear combinations of plane waves. The $C_\infty$ symmetry properties \eqref{Coo sym} in this case reduce to
\begin{equation}\label{m2m3sym}
\begin{split}
m_2( a_1, a_2)=-m_2( a_2, a_1)\;,\qquad 
 m_3( a_1, a_2, a_3)&=m_3( a_3, a_2, a_1)\;,\\
m_3( a_1, a_2, a_3)+m_3( a_2, a_3, a_1)+m_3( a_3, a_1, a_2)&=0\;.
\end{split}    
\end{equation}
Notice that, since $A^\mu_i\in K_1$ and $|m_n|=2-n$, all products \eqref{mn As} take values in $K_2$, the space of color-stripped currents.
For the complete list of $m_n$ acting on all elements of $\cK_{\rm YM}$ we refer to \cite{Bonezzi:2022yuh}.

We conclude this part by discussing the cyclic structure of the $L_\infty$ and $C_\infty$ algebras of Yang-Mills theory. Since we assigned degree $-1$ to the Lie algebra generators $T_a$, the Cartan-Killing form has degree $|\Tr|=+2$. We now define a degree $-3$ inner product on $\cK_{\rm YM}$, 
 \begin{equation}
 \omega:\cK_{\rm YM}\otimes \cK_{\rm YM}\longrightarrow\mathbb{C}\;,\quad \omega(\psi_1,\psi_2)=(-1)^{\psi_1\psi_2}\omega(\psi_2,\psi_1)\;,
 \end{equation} 
 by giving its non-vanishing components:
\begin{equation}\label{omega}
\begin{split}
\omega(A,J)&=\omega(J,A):=\int dx\,A^\mu(x)\,J_\mu(x)\;,\\
\omega(\Lambda,N)&=\omega(N,\Lambda):=\int dx\,\Lambda(x)\,N(x)\;,
\end{split}    
\end{equation}
with other pairings vanishing by degree.
With the above inner product, the $C_\infty$ algebra $(\cK_{\rm YM}, \{m_n\})$ is cyclic, meaning that the products obey
\begin{equation}\label{cyc Coo}
\begin{split}
\omega\big(m_1( \psi_1), \psi_2\big)&=(-1)^{1+ \psi_1 \psi_2}\omega\big(m_1( \psi_2), \psi_1\big)\;,\\    
\omega\big(m_2( \psi_1, \psi_2),\psi_3\big)&=(-1)^{\psi_1(\psi_2+\psi_3)}\omega\big(m_2(\psi_2,\psi_3),\psi_1\big)\;,\\ 
\omega\big(m_3(\psi_1,\psi_2,\psi_3),\psi_4\big)&=(-1)^{1+\psi_1(\psi_2+\psi_3+\psi_4)}\omega\big(m_3(\psi_2,\psi_3,\psi_4),\psi_1\big)\;.
\end{split}    
\end{equation}
Using $\omega$ and $\Tr$ one can construct a degree $-1$ inner product on $\cX_{\rm YM}$ as $\l\cdot,\cdot\r=\omega\otimes\Tr$. More explicitly, given two elements $\Psi_i=\psi_i\,T_i$, with $\psi_i\in\cK_{\rm YM}$ and $T_i\in\mathfrak{g}$, the inner product 
$\big\l\cdot,\cdot\big\r: \cX_{\rm YM}\otimes \cX_{\rm YM}\longrightarrow\mathbb{C}$ is given by
\begin{equation}\label{inner YM}
\begin{split}
%\big\l\cdot,\cdot\big\r&: \cX_{\rm YM}\otimes \cX_{\rm YM}\longrightarrow\mathbb{C}\;,\\
\big\l\Psi_1,\Psi_2\big\r&:=(-1)^{\psi_2}\omega(\psi_1,\psi_2)\,\Tr(T_1T_2)\;,\\    
\big\l\Psi_1,\Psi_2\big\r&=(-1)^{1+\Psi_1\Psi_2}\big\l\Psi_2,\Psi_1\big\r\;,
\end{split}
\end{equation}
which is extended to arbitrary elements of $\cX_{\rm YM}$ by linearity.
The inner product $\l\cdot,\cdot\r$ so defined makes the $L_\infty$ algebra $(\cX_{\rm YM}, \{B_n\})$ cyclic, in that the brackets \eqref{Bs from ms} obey
\begin{equation}
\big\l B_n(\Psi_1,\Psi_2,\ldots,\Psi_n),\Psi_{n+1}\big\r=(-1)^{\Psi_1(\Psi_2+\cdots+\Psi_{n+1})}\big\l B_n(\Psi_2,\ldots,\Psi_n,\Psi_{n+1}),\Psi_{1}\big\r\;, 
\end{equation}
thanks to cyclicity of the $C_\infty$ algebra with $\omega$ and standard cyclicity of the Cartan-Killing form $\Tr$. Upon expanding an arbitrary element of $\cX_{\rm YM}$ in a basis $T_a$ of $\mathfrak{g}$ one recovers the standard pairings of gauge fields with  currents, and of gauge parameters with Noether identities:
\begin{equation}\label{YM inner}
\begin{split}
\big\l \cA,\cJ\big\r&=\int dx\,A^{\mu a}(x)\,J^b_\mu(x) \,\Tr(T_aT_b)=\int dx\,A^{\mu}_a(x)\,J^a_\mu(x)\;,\\
\big\l \bLambda,\cN\big\r&=-\int dx\,\Lambda^{ a}(x)\,N^b(x) \,\Tr(T_aT_b)=-\int dx\,\Lambda_a(x)\,N^a(x)\;,
\end{split}    
\end{equation}
where we lowered the color index with $\delta_{ab}=\Tr(T_aT_b)$. With this inner product the Yang-Mills action \eqref{YM action} can be recast in the generalized Maurer-Cartan form
\begin{equation}
S_{\rm YM}[\cA]=\tfrac12\,\big\l \cA,B_1(\cA)\big\r+\tfrac{1}{3!}\,\big\l \cA,B_2(\cA,\cA)\big\r+\tfrac{1}{4!}\,\big\l \cA,B_3(\cA,\cA,\cA)\big\r\;,
\end{equation}
similar to \eqref{MC action scalar}.

\subsection{Homotopy transfer and amplitudes}

Since we have already exemplified the connection between homotopy transfer and the on-shell action method for the scalar case, here we will discuss Yang-Mills tree amplitudes directly from the perspective of homotopy transfer.

Our goal is to transfer the Yang-Mills $L_\infty$ structure from $\cX_{\rm YM}$ to a space $\bar\cX_{\rm YM}$, where the fields $\bar \cA_\mu\in\bar X_0$ are on-shell and transverse:
\begin{equation}\label{on-shell transverse A}
\B\bar \cA_\mu=0\;,\quad \del\cdot\bar \cA=0\;,  
\end{equation}
which implies $\bar B_1(\bar \cA)=0$ for the differential on $\bar\cX_{\rm YM}$. The fields $\bar \cA_\mu$ are still subject to residual on-shell gauge transformations $\delta\bar \cA_\mu=\del_\mu\bar\bLambda$ for parameters obeying $\B\bar\bLambda=0$, meaning that the space $\bar X_{-1}$ consists of harmonic gauge parameters. 
Furthermore, the on-shell and transverse conditions on $\bar \cA_\mu$ suggest that the space of currents $\bar X_1$ contains on-shell vectors defined modulo gradients, while the space of Noether identities $\bar X_2$ consists of harmonic scalars:
\begin{equation}\label{Xbar conditions}
\B\bar \cJ_\mu=0\;,\;\;\text{with} \quad \bar \cJ_\mu\sim\bar \cJ_\mu+\del_\mu\Phi\;,\qquad\B\bar \cN=0\;.
\end{equation}
The equivalence class condition defining $\bar X_1$ ensures that the pairing between an on-shell and transverse field $\bar \cA_\mu$ with a current $\bar \cJ_\mu$ remains non-degenerate. Indeed, if $\int dx\,\Tr\big(\bar \cA^\mu\bar \cJ_\mu\big)=0$ for any $\bar \cA_\mu$ with $\B\bar \cJ_\mu=0$, then $\bar \cJ_\mu=\del_\mu\Phi\sim0$.

Restricting to sums of Lie algebra-valued plane waves, we expand all elements in $\cX_{\rm YM}$ as
\begin{equation}\label{plane waves}
\begin{split}
\bLambda(x)&=\sum_i \lambda_i\,e^{ik_i\cdot x}\,T_i\;,\quad \cA^\mu(x)=\sum_i a^\mu_i\,e^{ik_i\cdot x}\,T_i\;,\\
\cJ^\mu(x)&=\sum_i j^\mu_i\,e^{ik_i\cdot x}\,T_i\;,\quad \cN(x)=\sum_i n_i\,e^{ik_i\cdot x}\,T_i\;,
\end{split}    
\end{equation}
and we will assume that the time component of all $k_i^\mu$ is non-vanishing: $k_i^0\neq0$.
We will denote a generic Fourier coefficient as $\psi(k)=\big(\lambda(k),a^\mu(k), j^\mu(k), n(k)\big)$, as a bookkeeping device for the momentum of the associated plane wave.
Given the expansions \eqref{plane waves}, one can define the harmonic projector $P_{\rm h}$ as in the previous section via a Kronecker delta, acting directly on the Fourier coefficients:
\begin{equation}
P_{\rm h}\psi(k)=\delta_{k^2,0}\,\psi(k)\;,\quad P_{\rm h}\Psi(x)=\sum_i P_{\rm h}(\psi_i)\,e^{ik_i\cdot x}T_i \;,
\end{equation}
which is well-defined and obeys $\B P_{\rm h}=P_{\rm h}\B=0$ and $P_{\rm h}^2=P_{\rm h}$.
We also introduce a transverse projector acting on vector polarizations. Splitting the time component on vectors we define
\begin{equation}
P_\perp\big(a^\mu(k) \big):=\left(\frac{\vec k\cdot\vec a}{k^0}, \vec a\right)\;,   
\end{equation}
which is extended by linearity to fields of the form \eqref{plane waves}, thus obeying $\del_\mu\big(P_\perp \cA\big)^\mu=0$. Notice that if a polarization $a_\mu(k)$ is already transverse, with $k^0\neq0$, then $P_\perp(a_\mu)=a_\mu$.

With these ingredients, we look for projection and inclusion maps $p$ and $\iota$ to perform homotopy transfer from the chain complex $\cX_{\rm YM}$ to the on-shell complex $\bar\cX_{\rm YM}$:
\begin{equation}
\begin{tikzcd}[row sep=10mm]
X_{-1}\arrow{d}{p}\arrow{r}{B_1}&X_{0}\arrow{d}{p}\arrow{r}{B_1}&X_{1}\arrow{d}{p}\arrow{r}{B_1}&X_{2}\arrow{d}{p}\\
\bar X_{-1}\arrow[shift left=2mm]{u}{\iota}\arrow{r}{\bar B_1}&\bar X_{0}\arrow[shift left=2mm]{u}{\iota}\arrow{r}{0}&\bar X_{1}\arrow[shift left=2mm]{u}{\iota}\arrow{r}{\bar B_1}&\bar X_{2}\arrow[shift left=2mm]{u}{\iota}
\end{tikzcd}    \;,
\end{equation}
where the differential $\bar B_1$ is given by
\begin{equation}
\bar B_1(\bar\bLambda)=\del_\mu\bar\bLambda\;,\quad\bar B_1(\bar \cA)=0\;,\quad\bar B_1(\bar \cJ)=\del^\mu\bar \cJ_\mu\;.    
\end{equation}
All elements $\bar\Psi\in\bar\cX_{\rm YM}$ are on-shell: $\B\bar\Psi=0$. Furthermore, the fields $\bar\cA_\mu$ are transverse and the currents $\bar\cJ_\mu$ are defined as the equivalence class \eqref{Xbar conditions}.
As sums of Lie algebra-valued plane waves, the elements of $\bar\cX_{\rm YM}$ are given by 
\begin{equation}\label{plane waves on-shell}
\begin{split}
\bar\bLambda(x)&=\sum_i \bar\lambda_i\,e^{ik_i\cdot x}\,T_i\;,\quad \bar \cA^\mu(x)=\sum_i \epsilon^\mu_i\,e^{ik_i\cdot x}\,T_i\;,\\
\bar \cJ^\mu(x)&=\sum_i \bar j^\mu_i\,e^{ik_i\cdot x}\,T_i\;,\quad \bar \cN(x)=\sum_i \bar n_i\,e^{ik_i\cdot x}\,T_i\;,
\end{split}    
\end{equation}
with all $k_i^2=0$. Here we adopted the standard notation $\epsilon^\mu_i$ for the transverse polarization vectors, instead of $\bar a^\mu_i$. Since the on-shell currents are defined modulo gradients, their Fourier coefficients are defined as the equivalence class $[\bar j^\mu(k)]= [\bar j^\mu(k)+k^\mu\phi(k)]$.

The projector $p:\cX_{\rm YM}\rightarrow\bar \cX_{\rm YM}$ can be defined directly on the polarizations $\psi(k)$ by 
\begin{equation}\label{p ontran}
\begin{split}
p(\lambda)&:=P_{\rm h}(\lambda)\;,\quad p(a_\mu):=P_{\rm h}P_\perp(a_\mu)\;,\\
p(j_\mu)&:=[P_{\rm h}j_\mu]\;,\quad\hspace{1mm} p(n):=P_{\rm h}(n)\;,
\end{split}
\end{equation}
and extended to fields \eqref{plane waves}.
For the inclusion $\iota:\bar\cX_{\rm YM}\rightarrow\cX_{\rm YM}$ we take the trivial inclusion acting on $\bar X_{-1}$, $\bar X_0$ and $\bar X_2$. For the inclusion of currents $\iota:\bar X_1\rightarrow X_1$ we pick a representative of the equivalence class, by demanding that its time component vanishes: $\big(\iota[\bar j]\big)^0=0$. This is possible since $k^0$ is non-vanishing, hence one has $[\bar j^0(k)]=[k^0\,\bar j^0(k)/k^0]=[0]$. We thus define
\begin{equation}
\iota\big(\bar j^\mu(k)\big):=\bar j^\mu(k)-k^\mu\,\frac{\bar j^0(k)}{k^0}\;.    
\end{equation}
This is well-defined on the equivalence class, since $\iota(\bar j+k\,\phi)=\iota(\bar j)$.
The above projector and inclusion obey the chain map conditions
\begin{equation}
\bar B_1\, p=p\, B_1\;,\quad B_1\,\iota=\iota\,\bar B_1 \;,   
\end{equation}
as well as $p\,\iota=1_{\bar \cX}$.

We now want to find a homotopy $h:\cX_{\rm YM}\rightarrow\cX_{\rm YM}$ of degree $-1$ obeying
\begin{equation}\label{h defining}
B_1h+hB_1=\iota\, p-1_{\cX}  \;. 
\end{equation}
One can find a one-parameter family of homotopies $h_\xi$, which differ only in their action on currents. As for the projector and inclusion, we define it on the Fourier coefficients:
\begin{equation}\label{homotopy YM}
\begin{split}
h_\xi\big(a(k)\big)&:=\frac{i}{k^2}\,(1-P_{\rm h})\,k\cdot a\;,\\
h_\xi\big(j(k)\big)&:=\frac{1}{k^2}\,\left(\delta^\mu{}_\nu-\xi\,\frac{k^\mu k_\nu}{k^2}\right)\,(1-P_{\rm h})\,j^\nu+P_{\rm h}\left(\frac{j^0}{(k^0)^2},\vec0\right)\;,\\
h_\xi\big(n(k)\big)&:=\frac{i}{k^2}\,(1-P_{\rm h})\,k^\mu n\;,
\end{split}
\end{equation}
and extend it to linear combinations of plane waves. Notice that for an off-shell current, thus obeying $\cJ^\mu=(1-P_{\rm h})\cJ^\mu$, the homotopy reduces to the familiar propagator in $R_\xi-$gauge. The homotopy $h_\xi$ obeys so-called  side conditions 
\begin{equation}
p\, h_\xi=0\;,\quad h_\xi\, \iota=0 \;,   
\end{equation}
for any value of $\xi$. Its square, on the other hand, is given by
\begin{equation}
h_\xi^2\big(j(k)\big)=(1-\xi)\,\frac{i}{k^4}\,(1-P_{\rm h})\,k\cdot j\;,\quad h_\xi^2\big(n(k)\big)=(1-\xi)\,\frac{i}{k^4}\,(1-P_{\rm h})\,k_\mu\,n\;,   
\end{equation}
showing that $h_\xi^2=0$ only for $\xi=1$, which corresponds to the Lorenz gauge propagator. In this algebraic language, the freedom in choosing $\xi$ for the propagator is reflected by the fact that homotopies with different $\xi$ differ by a $B_1-$exact term:
\begin{equation}
h_{\xi_1}-h_{\xi_2}=(\xi_1-\xi_2)\,\big(B_1g-gB_1\big)\;,  
\end{equation}
where $g$ is a higher homotopy of degree $-2$, acting as $g(j)=\frac{i}{k^4}(1-P_{\rm h})k\cdot j$ on currents and zero on all other elements.

Let us comment about using this homotopy for computing scattering amplitudes. The above expressions show that $h_\xi(j)$, as well as $p(a)$, break manifest Lorentz invariance. In order to compute amplitudes, however, one needs the transferred brackets acting on fields $\bar \cA_\mu$, which are already transverse. Moreover, these brackets all take values in the space of currents $\bar X_1$, meaning that the projection appearing in $\bar B_n=p\,(\cdots)$ never involves the non covariant $P_\perp$. Finally, as we have previously discussed, when computing scattering amplitudes one is implicitly assuming that exchanged momenta in propagators are never on-shell. This implies that in $h_\xi(j)$ the homotopy effectively acts as a standard propagator in $R_\xi$-gauge.

Before presenting the transferred brackets and related amplitudes, let us discuss the properties of the homotopy data $(p,\iota,h_\xi)$ with respect to the inner product. First of all, one can define the inner product on the transferred space $\bar\cX_{\rm YM}$ via the original one \eqref{inner YM} and the inclusion map as 
\begin{equation}\label{innerbar}
\big\l\bar\Psi_1,\bar\Psi_2\big\r_{\bar\cX}:=\big\l\iota(\bar\Psi_1),\iota(\bar\Psi_2)\big\r_{\cX}\;.   
\end{equation}
Using the explicit form \eqref{YM inner} and the plane wave expansion \eqref{plane waves on-shell} the non-vanishing pairings on $\bar\cX_{\rm YM}$ are given by
\begin{equation}\label{innerbar plane}
\begin{split}
\big\l\bar \cA,\bar \cJ\big\r_{\bar\cX}&=\sum_{il}(2\pi)^D\delta^D(k_i+k_l)\,\epsilon_i(k_i)\cdot\bar j_{l}(k_l)\,\Tr(T_iT_j)\;,\\   
\big\l\bar \bLambda,\bar \cN\big\r_{\bar\cX}&=-\sum_{ij}(2\pi)^D\delta^D(k_i+k_j)\,\bar\lambda_i(k_i)\,\bar n_{j}(k_j)\,\Tr(T_iT_j)\;.
\end{split}    
\end{equation}
The projector $\iota\, p:\cX_{\rm YM}\rightarrow\cX_{\rm YM}$ is symmetric with respect to the inner product, meaning that it obeys
\begin{equation}
\big\l\iota\, p(\Psi_1),\Psi_2\big\r_{\cX}=\big\l\Psi_1,\iota\, p(\Psi_2)\big\r_{\cX}    \;.
\end{equation}
In the case that $\Psi_2$ is the inclusion of an element in $\bar\cX_{\rm YM}$, i.e. $\Psi_2=\iota(\bar\Psi_2)$, the above relation yields
\begin{equation}\label{piota in inner}
\big\l p(\Psi_1),\bar\Psi_2\big\r_{\bar\cX}=\big\l \Psi_1,\iota(\bar\Psi_2)\big\r_{\cX} \;,   
\end{equation}
upon using $p\,\iota=1_{\bar\cX}$ and the definition \eqref{innerbar}. Finally, the homotopy $h_\xi$ obeys
\begin{equation}\label{homotopy sym YM}
\big\l h_\xi(\Psi_1),\Psi_2\big\r_\cX=(-1)^{\Psi_1}\big\l \Psi_1,h_\xi(\Psi_2)\big\r_\cX \;,   
\end{equation}
consistently with the above property of $\iota\, p$ and the defining relation \eqref{h defining}.

We now come to discussing the transferred brackets between on-shell fields $\bar \cA_i$. Similarly to the scalar case, they can be computed diagrammatically as the sum of all distinct rooted tree diagrams with cubic and quartic vertices, corresponding to brackets $B_2$ and $B_3$, respectively. Inequivalent diagrams are counted taking into account that $B_2$ and $B_3$ are totally symmetric in their inputs. From top to bottom, the brackets take fields in $X_0$ as inputs and produce a current in $X_1$ as output. Every internal line is a homotopy $h_\xi$, which inserts a propagator and shifts the current it acts on to the space of fields. The bottom line at the root of the diagram denotes the output of the last bracket, which is then projected to the space $\bar X_1$ of on-shell currents by $p$.
As an explicit example of this diagrammatic construction, the first few brackets $\bar B_n$ are given by (here we omit the subscript $\xi$ on $h$): 
\begin{equation*}
\begin{split}
\bar B_2(\bar \cA_1,\bar \cA_2)&=p\,B_2( \cA_1, \cA_2)=\begin{tikzpicture}[baseline={([yshift=-.5ex]current bounding box.center)},level distance=7mm,sibling distance=10mm]
  \node[scale=0.5,circle, draw] {}
    child[grow=up] {
     child {node[scale=0.7] {2}}  child{node[scale=0.7] {1}}
    };
\end{tikzpicture}\;,\end{split}
\end{equation*}
\begin{equation*}
\begin{split}
\bar B_3(\bar \cA_1,\bar \cA_2,\bar \cA_3)&=p\,\Big\{B_3( \cA_1,\cA_2,\cA_3)+3\,B_2\big(hB_2( \cA_{(1},\cA_2), \cA_{3)}\big)\Big\}\\
&=\begin{tikzpicture}[baseline={([yshift=-.5ex]current bounding box.center)},level distance=10.5mm,sibling distance=10mm]
  \node[scale=0.5,circle, draw] {}
    child[grow=up] {
     child {node[scale=0.7] {3}} child {node[scale=0.7] {2}} child {node[scale=0.7] {1}}}
    ;
\end{tikzpicture}+\begin{tikzpicture}[baseline={([yshift=-.5ex]current bounding box.center)},level distance=7mm,sibling distance=10mm]
  \node[scale=0.5,circle, draw] {}
    child[grow=up] {
     child{child {node[scale=0.7] {3}} child[missing]}    child {child {node[scale=0.7] {2}} child {node[scale=0.7] {1}}}
    };
\end{tikzpicture}+\begin{tikzpicture}[baseline={([yshift=-.5ex]current bounding box.center)},level distance=7mm,sibling distance=10mm]
  \node[scale=0.5,circle, draw] {}
    child[grow=up] {
     child{child {node[scale=0.7] {2}} child[missing]}    child {child {node[scale=0.7] {3}} child {node[scale=0.7] {1}}}
    };
\end{tikzpicture}+\begin{tikzpicture}[baseline={([yshift=-.5ex]current bounding box.center)},level distance=7mm,sibling distance=10mm]
  \node[scale=0.5,circle, draw] {}
    child[grow=up] {
     child{child {node[scale=0.7] {1}} child[missing]}    child {child {node[scale=0.7] {3}} child {node[scale=0.7] {2}}}
    };
\end{tikzpicture}\;,\\
\end{split}    
\end{equation*}
\begin{equation}\label{bar bs homtransf YM}
\begin{split}
\bar B_4(\bar \cA_1,\bar \cA_2,\bar \cA_3,\bar \cA_4)&=p\,\Big\{12\,B_2\Big(hB_2\big(hB_2( \cA_{(1}, \cA_2), \cA_3\big), \cA_{4)}\Big)+3\,B_2\big(hB_2( \cA_{(1}, \cA_2),hB_2( \cA_3, \cA_{4)})\big)\\
&+4\,B_2\big(hB_3( \cA_{(1}, \cA_2, \cA_3), \cA_{4)}\big)+6\,B_3\big(hB_2( \cA_{(1}, \cA_2), \cA_3, \cA_{4)}\big)\Big\}\\
&=\begin{tikzpicture}[baseline={([yshift=-.5ex]current bounding box.center)},level distance=7mm,sibling distance=10mm]
  \node[scale=0.5,circle, draw] {}
    child[grow=up]{child{child{child{node[scale=0.7] {4}} child[missing]} child[missing]} child{
     child{child {node[scale=0.7] {3}} child[missing]}    child {child {node[scale=0.7] {2}} child {node[scale=0.7] {1}}}
    }};
\end{tikzpicture}+\text{11 more}
+\begin{tikzpicture}[baseline={([yshift=-.5ex]current bounding box.center)},level distance=10mm]
\tikzstyle{level 2}=[sibling distance=15mm]
  \tikzstyle{level 3}=[sibling distance=10mm]
  \node[scale=0.5,circle, draw]{}[grow=up] child{child{child{node[scale=0.7] {4}} child{node[scale=0.7] {3}}} child{child{node[scale=0.7] {2}} child{node[scale=0.7] {1}}}} ;
\end{tikzpicture}+\text{2 more}\\
&+\begin{tikzpicture}[baseline={([yshift=-.5ex]current bounding box.center)},level distance=7mm]
\tikzstyle{level 2}=[sibling distance=10mm]
  \tikzstyle{level 3}=[sibling distance=7mm]
  \node[scale=0.5,circle, draw] {}
    child[grow=up] {
     child{child {node[scale=0.7] {4}} child[missing]}    child {child {node[scale=0.7] {3}} child {node[scale=0.7] {2}} child {node[scale=0.7] {1}}}
    };
\end{tikzpicture}+\text{3 more}+
\begin{tikzpicture}[baseline={([yshift=-.5ex]current bounding box.center)},level distance=7mm]
\tikzstyle{level 2}=[sibling distance=7mm]
  \tikzstyle{level 3}=[sibling distance=7mm]
  \node[scale=0.5,circle, draw] {}
    child[grow=up] {child{child {node[scale=0.7] {4}} child[missing]} child{child {node[scale=0.7] {3}} } child{child {node[scale=0.7] {2}} child {node[scale=0.7] {1}}}
    };
\end{tikzpicture}+\text{5 more}\;,    
\end{split}    
\end{equation}
where we used the notation $\cA_i:=\iota(\bar \cA_i)$ for the trivial inclusion of fields, and round brackets $(12\cdots n)$ stand for symmetrization in the labels with strength one. The transported brackets can also be computed recursively in an analytic way. To this end, let us introduce ``multiparticle'' fields $\cA^\mu_{i_1\cdots i_n}\in X_0$ and currents $\cJ^\mu_{i_1\cdots i_n}\in X_1$, by which we mean that they are multilinear functions of free field inputs $(\bar \cA_{i_1},\ldots,\bar \cA_{i_n})$. Given the seed of the recursion $\cA^\mu_i:=\iota(\bar \cA^\mu_i)$, for $n\geq2$  the currents and transported brackets $\bar B_n$ are given by
\begin{equation}\label{Recursive YM Loo}
\begin{split}
\cJ^\mu_{i_1\cdots i_n}&\stackrel{(12\cdots n)}{=}\frac12\,\sum_{k+l=n}B_2\big(\cA_{i_1\cdots i_k}, \cA_{i_{k+1}\cdots i_n}\big)+\frac{1}{3!}\,\sum_{k+l+m=n}B_3\big(\cA_{i_1\cdots i_k}, \cA_{i_{k+1}\cdots i_{k+l}}, \cA_{i_{k+l+1}\cdots i_{n}}\big)\;,\\
\cA^\mu_{i_1\cdots i_n}&=h\,\cJ^\mu_{i_1\cdots i_n}\;,\quad \frac{1}{n!}\,\bar B_n(\bar \cA_{i_1},\ldots,\bar \cA_{i_n})=p\,\cJ^\mu_{i_1\cdots i_n}\;,
\end{split}    
\end{equation}
where in the first equation we symmetrize in $(12\cdots n)$ with strength one. This is a particular case of the general recursive formula for the transferred brackets derived in Appendix \ref{Sec:Appendix Loo}. Let us point out that the construction of all $\bar B_n$ between fields $\bar \cA_i$ only requires the brackets $B_2$ and $B_3$ between fields, since all $\cA^\mu_{i_1\cdots i_n}\in X_0$.
This is the multiparticle version of the recursion \eqref{phiBs relations}, adapted to the case of Yang-Mills theory, and is essentially the color-dressed version of the Berends-Giele recursion, which we will address in the next section. 

Given the transferred brackets between on-shell fields, one can obtain the $n-$gluon tree level amplitude by choosing single plane waves as inputs and taking the inner product
\begin{equation}\label{YM amplitude Bbar}
\delta(K)\,\mathcal{A}_n^{\rm tree}=\Big\l\bar \cA_n,\bar B_{n-1}\big(\bar \cA_1,\ldots,\bar \cA_{n-1}\big)\Big\r_{\bar \cX}\;,\quad\bar \cA^\mu_i=\epsilon^\mu_i\,e^{ik_i\cdot x}\,T_i\;,    
\end{equation}
where $\delta(K)=(2\pi)^D\delta^D(k_1+\ldots+k_n)$ and $k_i^2=k_i\cdot\epsilon_i=0$.
The amplitude is a function of the triplets $\{k_i^\mu,\epsilon_i^\mu,T_i\}$ associated to each external gluon.
The symmetry \eqref{homotopy sym YM} of the homotopy, together with cyclicity of $B_2$ and $B_3$, ensures that the transferred brackets $\bar B_n$ between fields are cyclic, which amounts to total Bose symmetry of the amplitudes. 
To see how the general formula \eqref{YM amplitude Bbar} matches the standard results, we shall compute explicitly the four-point amplitude with this formalism. 

Taking the expression for $\bar B_3(\bar \cA_1,\bar \cA_2,\bar \cA_3)$ from \eqref{bar bs homtransf YM}, the four-point amplitude is given by
\begin{equation}\label{A4 abstract}
\begin{split}
\delta(K)\,\mathcal{A}_4^{\rm tree}&=\Big\l\bar \cA_4,\bar B_{3}\big(\bar \cA_1,\bar \cA_2,\bar \cA_3\big)\Big\r_{\bar \cX}\\
&=\Big\l \cA_4,B_2\big(hB_2( \cA_{1},\cA_2), \cA_{3}\big)+(231)+(312)+B_3( \cA_1,\cA_2,\cA_3)\big)\Big\r_\cX\;,
\end{split}    
\end{equation}
where we used the property \eqref{piota in inner}, denoted the trivial inclusion of the fields by $\iota(\bar \cA_i)=\cA_i$  and symmetrized the exchange terms $B_2hB_2$ over the labels $(123)$\footnote{Thanks to the symmetry of $B_2$, total symmetrization reduces to the sum over cyclic permutations.}. Using the factorized form \eqref{Bs from ms} for fields $\cA_i^\mu=\epsilon^\mu_i\,e^{ik_i\cdot x}\,T_i\in X_0$, the brackets read
\begin{equation}
\begin{split}
B^\mu_2(\cA_i,\cA_j)&=g\,m^\mu_2(\epsilon_i,\epsilon_j)\,[T_i,T_j]\,e^{i(k_i+k_j)\cdot x}\;,\quad i,j=1,2,3\;,\\
B^\mu_3(\cA_1,\cA_2,\cA_3)&=g^2\Big\{m^\mu_3(\epsilon_1,\epsilon_2,\epsilon_3)\,[T_1,[T_2,T_3]]+m^\mu_3(\epsilon_2,\epsilon_1,\epsilon_3)\,[T_2,[T_1,T_3]]\Big\}e^{i(k_1+k_2+k_3)\cdot x}\;,
\end{split}    
\end{equation}
where the $m_n$ acting on polarizations are given in \eqref{mn Fourier}.
In the above expressions we kept the vector index $\mu$ explicit, to remind us that the outputs take values in the space of currents $X_1$.

To proceed further, we assume that the exchanged momenta are not on-shell: $(k_i+k_j)^2\neq0$. In this case the homotopy acts on currents $B_2(\cA_i,\cA_j)$ as the propagator in $R_\xi-$gauge. In order to display the pole structure of the amplitude, we find it convenient to split the homotopy into a $\frac{1}{k^2}$ factor times a degree $-1$ operator $b$, which takes care of the degree shift and of the propagator numerator. Namely, for an off-shell current polarization $j^\mu(k)$ we define
\begin{equation}
b\big(j(k)\big):=j^\mu-\xi\,\frac{k^\mu k\cdot j}{k^2}\;,\quad k^2\neq0\quad\longrightarrow\quad h\big(j(k)\big)=\frac{1}{k^2}\,b\big(j(k)\big)\;,
\end{equation}
which is extended to plane waves as usual.
More generally, the $b$ operator plays a central role in understanding the so-called kinematic algebra of gauge theories \cite{Monteiro:2011pc,Reiterer:2019dys,Ben-Shahar:2021zww,Borsten:2022vtg,Bonezzi:2022bse,Bonezzi:2023pox}, but we will not discuss this further here.

With this notation at hand we evaluate the inner product in \eqref{A4 abstract} using \eqref{innerbar plane}, thus obtaining
\begin{equation}\label{A4 rough}
\begin{split}
\mathcal{A}_4^{\rm tree}&=\frac{g^2}{s}\,\epsilon_{4}\cdot m_2\big(bm_2(\epsilon_1,\epsilon_2),\epsilon_3\big)\,c_s+\frac{g^2}{t}\,\epsilon_{4}\cdot m_2\big(bm_2(\epsilon_2,\epsilon_3),\epsilon_1\big)\,c_t\\&+\frac{g^2}{u}\,\epsilon_{4}\cdot m_2\big(bm_2(\epsilon_3,\epsilon_1),\epsilon_2\big)\,c_u+g^2\Big[\epsilon_4\cdot m_3(\epsilon_2,\epsilon_1,\epsilon_3)\,c_u-\epsilon_4\cdot m_3(\epsilon_1,\epsilon_2,\epsilon_3)\,c_t\Big]\;,
\end{split}    
\end{equation}
where we introduced the Mandelstam variables and color factors, defined as
\begin{equation}
\begin{split}
s:=(k_1+k_2)^2\;,\quad c_s&:=\Tr\big(T_4[[T_1,T_2],T_3]\big)\;,\\
t:=(k_2+k_3)^2\;,\quad c_t&:=\Tr\big(T_4[[T_2,T_3],T_1]\big)\;,\\
u:=(k_3+k_1)^2\;,\quad c_u&:=\Tr\big(T_4[[T_3,T_1],T_2]\big)\;,
\end{split}
\end{equation}
with the color traces obeying $c_s+c_t+c_u=0$ thanks to the Jacobi identity of $\mathfrak{g}$. 
Notice that in \eqref{A4 rough} $bm_2(\epsilon_i,\epsilon_j)$ is a field polarization, since $|b|=-1$. This implies that the outermost $m_2$ is still the same map given in \eqref{mn Fourier}.
Since the exchange terms arising from the cubic vertices display manifest symmetry between the $s$, $t$
 and $u$ channels, one can use the symmetry \eqref{m2m3sym} of $m_3$,
together with the Jacobi identity of the $c_i$,
to assign the quartic contributions to the channels of the exchange diagrams. One then arrives at a symmetric form of the amplitude:
\begin{equation}\label{A4 BCJ}
\mathcal{A}^{\rm tree}_4=g^2\,\Big\{\frac{n_s\,c_s}{s}+\frac{n_t\,c_t}{t}+\frac{n_u\,c_u}{u}\Big\}\;,  
\end{equation}
written as a sum of formally cubic graphs. Here the kinematic numerators are given by
\begin{equation}\label{kinnum}
\begin{split}
n_s&:=\epsilon_4\cdot m_2\big(bm_2(\epsilon_1,\epsilon_2),\epsilon_3\big)+\frac{s}{3}\,\big(m_3(\epsilon_1,\epsilon_2,\epsilon_3)-m_3(\epsilon_2,\epsilon_1,\epsilon_3)\big)\;,\\
n_t&:=\epsilon_4\cdot m_2\big(bm_2(\epsilon_2,\epsilon_3),\epsilon_1\big)+\frac{t}{3}\,\big(m_3(\epsilon_2,\epsilon_3,\epsilon_1)-m_3(\epsilon_3,\epsilon_2,\epsilon_1)\big)\;,\\
n_u&:=\epsilon_4\cdot m_2\big(bm_2(\epsilon_3,\epsilon_1),\epsilon_2\big)+\frac{u}{3}\,\big(m_3(\epsilon_3,\epsilon_1,\epsilon_2)-m_3(\epsilon_1,\epsilon_3,\epsilon_2)\big)\;,
\end{split}    
\end{equation}
and play a central role in the double copy program pioneered in \cite{Bern:2008qj}.
Alternatively, one may want to decompose the amplitude in independent color structures, which in this case are only two out of $(c_s,c_t,c_u)$, due to the Jacobi identity. In the next section we will take this route and discuss its relation with $C_\infty$ algebras.

\subsection{Ward identities and on-shell gauge algebra}\label{sec:Wardfull}

In the remainder of this section, we will first prove that the amplitudes \eqref{YM amplitude Bbar} obey linear Ward identities, if the momenta are suitable
for a scattering process. We will then show that, allowing for generic momentum configurations, the on-shell gauge algebra is not abelian and in fact not even a Lie algebra.  
To do so, we start by recalling the quadratic relations of $L_\infty$ algebras. 

Consider a generic $L_\infty$ algebra $(\cX,\{B_n\})$, with elements $\Psi_i\in\cX$ of arbitrary degree. The brackets $B_n$ obey a (possibly infinite) set of generalized Jacobi identities.  Denoting by $\{12\cdots n\}$ total graded symmetrization\footnote{This means to sum over all permutations with strength one and the appropriate Koszul signs. For instance, $\Psi_{\{1}\Psi_{2\}}=\frac12\,(\Psi_{1}\Psi_{2}+(-1)^{\Psi_1 \Psi_2}\Psi_{2}\Psi_{1})$.} in the arguments $\Psi_1,\ldots,\Psi_n$, the identity with $n$ inputs of arbitrary degree is given by
\begin{equation}\label{GenGenJac}
\sum_{l=1}^n\binom{n}{l}\, B_{n+1-l}\big( B_l(\Psi_{\{1},\ldots,\Psi_l),\Psi_{l+1},\ldots,\Psi_{n\}}\big)=0\;, \quad n\geq1\;,   
\end{equation}
where $\binom{n}{l}$ is the binomial coefficient.
The Yang-Mills $L_\infty$ algebra $\cX_{\rm YM}$ has no higher brackets than $B_3$. Its nontrivial identities then reduce to the following:
\begin{equation}\label{Loo relations YM}
\begin{split}
B_1^2\big(\Psi\big)&=0\;,\\
B_1\big(B_2(\Psi_1,\Psi_2)\big)+B_2\big(B_1(\Psi_1),\Psi_2\big)+(-1)^{\Psi_1}B_2\big(\Psi_1,B_1(\Psi_2)\big)&=0\;,\\
B_2\big(B_2(\Psi_1,\Psi_2),\Psi_3\big)+(-1)^{\Psi_1(\Psi_2+\Psi_3)}B_2\big(B_2(\Psi_2,\Psi_3),\Psi_1\big)\\
+(-1)^{\Psi_3(\Psi_1+\Psi_2)}B_2\big(B_2(\Psi_3,\Psi_1),\Psi_2\big)
+B_1\big(B_3(\Psi_1,\Psi_2,\Psi_3)\big)+B_3\big(B_1(\Psi_1),\Psi_2,\Psi_3\big)\\
+(-1)^{\Psi_1}B_3\big(\Psi_1,B_1(\Psi_2),\Psi_3\big)+(-1)^{\Psi_1+\Psi_2}B_3\big(\Psi_1,\Psi_2,B_1(\Psi_3)\big)&=0\;,
\end{split}    
\end{equation}
plus the $n=4$ identity between $B_2$ and $B_3$ with vanishing $B_4$. In Yang-Mills theory, the above identities descend from the $C_\infty$ relations \eqref{Cinfty relations} of the $m_n$ products, combined with the Jacobi identity of the color Lie bracket. 
Although the original theory has only non-vanishing maps $B_1$, $B_2$ and $B_3$, homotopy transfer generates infinitely many brackets $\bar B_n$ on $\bar \cX_{\rm YM}$ and, as we have discussed, the transferred brackets between on-shell fields encode tree-level scattering amplitudes. The homotopy transfer theorem guarantees that the $\bar B_n$ obey the infinite set of generalized Jacobi identities \eqref{GenGenJac}, as we also prove explicitly in Appendix \ref{Sec:Appendix Loo}.

Given the general relations obeyed by the transferred brackets, we can now discuss Ward identities for amplitudes. Usually, the gauge invariance of a Yang-Mills amplitude is stated
as its invariance under shifts of any polarization vector by the corresponding momentum: $\delta\epsilon^\mu_i=ik^\mu_i$.
In terms of the Lie algebra-valued plane waves entering the amplitude \eqref{YM amplitude Bbar}, this corresponds to the linearized gauge transformation
\begin{equation}
\begin{split}
\delta\bar \cA_i^\mu&=\del^\mu\bar \bLambda_i=\bar B_1(\bar\bLambda_i)\;,\quad{\rm where}\\
\bar \cA^\mu_i&=\epsilon^\mu_i\,e^{ik_i\cdot x}\,T_i\,\in\bar X_0\;,\quad \bar\bLambda_i=\,e^{ik_i\cdot x}\,T_i\,\in\bar X_{-1}\;.        
\end{split}
\end{equation}
Thanks to total Bose symmetry, it does not matter which external leg we shift. We choose the $n-$th particle and vary the amplitude \eqref{YM amplitude Bbar} under $\delta\bar \cA_n=\bar B_1(\bar\bLambda_n)$:
\begin{equation}
\begin{split}
\delta(K)\,\delta\mathcal{A}_n^{\rm tree}&=\Big\l \bar B_1(\bar\bLambda_n),\bar B_{n-1}(\bar \cA_1,\ldots,\bar \cA_{n-1})\Big\r_{\bar \cX}\\
&=\Big\l \bar\bLambda_n,\bar B_1\big(\bar B_{n-1}(\bar \cA_1,\ldots,\bar \cA_{n-1})\big)\Big\r_{\bar \cX}\;,
\end{split}    
\end{equation}
where we used cyclicity in the form $\l\bar B_1(\bar\Psi_1),\bar\Psi_2\r_{\bar\cX}=(-1)^{\bar\Psi_1+1}\l\bar \Psi_1,\bar B_1(\bar\Psi_2)\r_{\bar\cX}$.
Since the inner product is non-degenerate, gauge invariance of the amplitude is equivalent to the linear Ward identity
\begin{equation}\label{Ward}
\bar B_1\big(\bar B_{n-1}(\bar \cA_1,\ldots,\bar \cA_{n-1})\big)=\del_\mu \bar B^\mu_{n-1}(\bar \cA_1,\ldots,\bar \cA_{n-1})=0\;.   
\end{equation}

In order to prove that \eqref{Ward} holds, we examine the generalized Jacobi identity \eqref{GenGenJac} with $n-1$ fields as inputs, separating the terms involving $\bar B_1$:   
\begin{equation}\label{GenGenJac fields}
\begin{split}
&\bar B_{1}\big(\bar B_{n-1}(\bar \cA_{1},\ldots,\bar \cA_{n-1})\big)+(n-1)\,\bar B_{n-1}\big(\bar B_1(\bar \cA_{(1}),\ldots,\bar \cA_{n-1)}\big)\\&+\sum_{l=2}^{n-2}\binom{n-1}{l}\,\bar B_{n-l}\big(\bar B_l(\bar \cA_{(1},\ldots,\bar \cA_l),\bar \cA_{l+1},\ldots,\bar \cA_{n-1)}\big)=0\;.   \end{split}
\end{equation}
The graded symmetrization here reduces to simple symmetrization over the $n-1$ legs, since the fields are in degree zero. First of all, since the $\bar \cA_i^\mu$ are on-shell and transverse the second term above vanishes, as $\bar B_1(\bar \cA_i)=0$. For the second sum we have to look at the recursive definition \eqref{Recursive YM Loo}, which gives $\bar B_l(\bar \cA_{i_1},\ldots,\bar \cA_{i_l})=l!\,p\cJ^\mu_{i_1\cdots i_l}$, for $l=2,\ldots,n-2$. As we have remarked multiple times, in an actual amplitude all momenta exchanged in propagators are off-shell. In the case at hand, the momenta involved in \eqref{GenGenJac fields} are $\{k^\mu_i\}_{i=1}^{n-1}$, and obey
\begin{equation}
\begin{split}
(k_{i_1}+k_{i_2}+\ldots+k_{i_l})^2&\neq0\;,\quad l=2,\ldots,n-2\;,\\
(k_{1}+k_{2}+\ldots+k_{n-1})^2&=0\;,\quad l=n-1\;.
\end{split}    
\end{equation}
The reason for this is that, in order to have total symmetry in the external legs, any combination of at most $n-2$ momenta is present in a propagator at least once. The sum of any $n-1$ momenta, instead, is on-shell by momentum conservation.
Looking back at \eqref{Recursive YM Loo}, this implies that the currents $\cJ^\mu_{i_1\cdots i_l}$ obey
\begin{equation}
h\cJ_{i_1\cdots i_l}=\frac{b\cJ_{i_1\cdots i_l}}{(k_{i_1}+k_{i_2}+\ldots+k_{i_l})^2}=\cA_{i_1\cdots i_l} \;,\quad p\cJ_{i_1\cdots i_l}=0\;,\quad l=2,\ldots,n-2\;.    
\end{equation}
This proves that $\bar B_l(\bar \cA_{i_1},\ldots,\bar \cA_{i_l})=0$ for all $l=2,\ldots,n-2$, thus implying that the generalized Jacobi identity \eqref{GenGenJac fields} reduces to the linear Ward identity \eqref{Ward} for amplitudes\footnote{Notice that the Ward identities of Yang-Mills theory are linear \emph{only} if the external gluons are transverse, which is the case here.}.

The fact that the Ward identities are linear suggests that the corresponding gauge algebra becomes abelian. We will now show that, if one allows for special configurations of momenta, the on-shell gauge algebra is not only non-abelian, but not even a strict Lie algebra. Let us start from the original gauge algebra of the off-shell theory. This is computed by taking the commutator of two successive gauge variations of the gauge field, leading to the two-bracket
\begin{equation}
B_2(\bLambda_1,\bLambda_2)=\bLambda_{12}=-\Lambda_1^a\,\Lambda_2^b\,[T_a,T_b]\;,    
\end{equation}
upon expanding the $\bLambda_i$ in a basis of $\mathfrak{g}$. Notice that the above bracket involves the pointwise product of the two gauge parameters, which is associative.

On the on-shell complex $\bar\cX_{\rm YM}$, the gauge parameters are harmonic: $\B\bar\bLambda_i=0$. Their gauge algebra is encoded in the transferred bracket $\bar B_2$, which reads
\begin{equation}\label{bar B2 Lambdas}
\bar B_2\big(\bar\bLambda_1,\bar\bLambda_2\big)=p\,B_2\big(\iota(\bar\bLambda_1),\iota(\bar\bLambda_2)\big)=-P_{\rm h}(\bar\Lambda_1^a\,\bar\Lambda_2^b)\,[T_a,T_b]\;.    
\end{equation}
Due to the harmonic projector, the above product is not associative, which is the reason the on-shell gauge algebra is not strict. By the homotopy transfer theorem, the above $\bar B_2$ has to obey the $n=3$ generalized Jacobi identity \eqref{GenGenJac}:
\begin{equation}\label{GenGenJac Lambdas}
\begin{split}
&\bar B_2\big(\bar B_2(\bar\bLambda_1,\bar\bLambda_2),\bar\bLambda_3\big)+\bar B_2\big(\bar B_2(\bar\bLambda_2,\bar\bLambda_3),\bar\bLambda_1\big)+\bar B_2\big(\bar B_2(\bar\bLambda_3,\bar\bLambda_1),\bar\bLambda_2\big)\\
&+\bar B_3\big(\bar B_1(\bar\bLambda_1),\bar\bLambda_2,\bar\bLambda_3\big)+\bar B_3\big(\bar B_1(\bar\bLambda_2),\bar\bLambda_3,\bar\bLambda_1\big)+\bar B_3\big(\bar B_1(\bar\bLambda_3),\bar\bLambda_1,\bar\bLambda_2\big)
=0\;,
\end{split}    
\end{equation}
with $\bar B_3(\bar\bLambda_1,\bar\bLambda_2,\bar\bLambda_3)
$ vanishing by degree. The above three-bracket $\bar B_3(\bar \cA,\bar\bLambda_1,\bar\bLambda_2)$ is given by homotopy transfer as
\begin{equation}
\bar B_3(\bar \cA,\bar\bLambda_1,\bar\bLambda_2)=p\,B_2\big(hB_2(\iota(\bar \cA),\iota(\bar\bLambda_1)),\iota(\bar\bLambda_2)\big)-p\,B_2\big(hB_2(\iota(\bar \cA),\iota(\bar\bLambda_2)),\iota(\bar\bLambda_1)\big)\;,    
\end{equation}
where we used the fact that $B_3$ is non-vanishing only on three fields, and that $h$ gives zero by degree on a gauge parameter.
For a field and gauge parameters given by single plane waves: $\bar \cA^\mu=\epsilon^\mu\,e^{ip\cdot x}\,T$ and $\bar\bLambda_i=\bar\lambda_i\,e^{ik_i\cdot x}\,T_i$, the above three-bracket is given by
\begin{equation}\label{bar B3 Lambdas}
\bar B_3(\bar \cA,\bar\bLambda_i,\bar\bLambda_j)=
-\frac{ik_i\cdot\epsilon}{(p+k_i)^2}\,\delta_{(p+k_i+k_j)^2,0}\,\big(1-\delta_{(p+k_i)^2,0}\big)\,\bar\lambda_i\,\bar\lambda_j\,e^{i(p+k_i+k_j)\cdot x}\,[[T,T_i],T_j]-(i\leftrightarrow j) \;.   
\end{equation}

For generic on-shell momenta obeying $(k_i+k_j)^2=2\,k_i\cdot k_j\neq0$ the gauge algebra is abelian, as expected, since the two bracket \eqref{bar B2 Lambdas} vanishes in this case. We will now present a simple example where the momentum assignment instead generates a non-trivial $L_\infty$ algebra. For this, we choose the gauge parameters $\bar\bLambda_1$, $\bar\bLambda_2$ and $\bar\bLambda_3$ to be single plane waves: $\bar\bLambda_i=e^{ik_i\cdot x}\,T_i$ with on-shell momenta chosen so that
\begin{equation}
(k_1+k_2+k_3)^2=0\;,\quad (k_1+k_2)^2=0\;,\quad (k_2+k_3)^2=-(k_1+k_3)^2=t\neq 0\;.
\end{equation}
The only non vanishing bracket between parameters is thus $\bar B_2(\bar\bLambda_1,\bar\bLambda_2)$, yielding a non-vanishing Jacobiator:
\begin{equation}\label{JacLambdas}
\begin{split}
&\bar B_2\big(\bar B_2(\bar\bLambda_1,\bar\bLambda_2),\bar\bLambda_3\big)+\bar B_2\big(\bar B_2(\bar\bLambda_2,\bar\bLambda_3),\bar\bLambda_1\big)+\bar B_2\big(\bar B_2(\bar\bLambda_3,\bar\bLambda_1),\bar\bLambda_2\big)\\
&=\bar B_2\big(\bar B_2(\bar\bLambda_1,\bar\bLambda_2),\bar\bLambda_3\big)=e^{i(k_1+k_2+k_3)\cdot x}\,[[T_1,T_2],T_3]\;.    
\end{split}    
\end{equation}
This is cancelled by adding the three-bracket contribution. Using \eqref{bar B3 Lambdas} with $\bar \cA=\bar B_1(\bar\bLambda_i)$ and the above assignment of momenta we obtain
\begin{equation}
\begin{split}
\bar B_3\big(\bar B_1(\bar\bLambda_1),\bar\bLambda_2,\bar\bLambda_3\big)&=\frac{k_3\cdot k_1}{t}\,e^{i(k_1+k_2+k_3)\cdot x}\,[[T_1,T_3],T_2]=\frac12\,e^{i(k_1+k_2+k_3)\cdot x}\,[[T_3,T_1],T_2]\;,\\
\bar B_3\big(\bar B_1(\bar\bLambda_2),\bar\bLambda_3,\bar\bLambda_1\big)&=\frac{k_3\cdot k_2}{t}\,e^{i(k_1+k_2+k_3)\cdot x}\,[[T_2,T_3],T_1]=\frac12\,e^{i(k_1+k_2+k_3)\cdot x}\,[[T_2,T_3],T_1]\;,\\
\bar B_3\big(\bar B_1(\bar\bLambda_3),\bar\bLambda_1,\bar\bLambda_2\big)&=-\Big(\frac{k_1\cdot k_3}{t}\,[[T_3,T_1],T_2]+\frac{k_2\cdot k_3}{t}\,[[T_3,T_2],T_1]\Big)\,e^{i(k_1+k_2+k_3)\cdot x}\\
&=\frac12\,e^{i(k_1+k_2+k_3)\cdot x}\,\Big([[T_3,T_1],T_2]+[[T_2,T_3],T_1]\Big)\;.
\end{split}    
\end{equation}
These contributions vanish when summed with \eqref{JacLambdas}, thus providing a nontrivial example of the generalized Jacobi identity \eqref{GenGenJac Lambdas}.

\section{Color-ordered amplitudes}\label{Sec:Coo}

In this section we introduce color-ordered Yang-Mills scattering amplitudes and discuss their homotopy-algebraic interpretation. More precisely, we show that the algebraic structure underlying tree-level color-ordered amplitudes is a $C_{\infty}$ algebra, similarly to how $L_{\infty}$ algebras underlie the amplitudes of the color-dressed theory (see also \cite{Borsten:2021hua} for a related discussion in the minimal model). This follows from the factorization  of the $L_{\infty}$ algebra of Yang-Mills into the color Lie algebra $\mathfrak{g}$ and the kinematic $C_{\infty}$ algebra $\cK_{\rm{YM}}$ discussed in section \ref{sec:YMoff}. Furthermore, we prove that the Ward identities obeyed by the color-ordered amplitudes are a consequence of the $C_{\infty}$ relations, after performing homotopy transfer to on-shell fields. 

\subsection{Color-ordering and $C_\infty$ algebras}

Scattering amplitudes in Yang-Mills theory display a factorization  between color and kinematics. The color information of the amplitudes is encoded in generators and structure constants of the Lie algebra $\mathfrak{g}$. In contrast, their kinematic information is encoded in the momenta of the external particles and their polarization. Formally, Yang-Mills amplitudes can be written in \textit{color-decomposed} forms which manifests such a factorization. In this paper, we will use a color decomposition in the so-called DDM (Del Duca, Dixon, Maltoni) basis \cite{DelDuca:1999rs,Bandiera:2020aqn}, which for $n$-point tree-level amplitudes explicitly reads
\begin{equation}\label{DDMdecomp}
\mathcal{A}^{\rm{tree}}_{n}=g^{n-2}\, \sum_{\sigma\in S_{n-2}}\Tr\big\{[\cdots[T_1,T_{\sigma(2)}],T_{\sigma(3)}],\, ...\,],T_{\sigma(n-1)}]T_{n}\big\}\, A[1,\sigma(2),\sigma(3),...,\sigma(n-1),n]\;.
\end{equation}
The sum runs over the permutation of $n-2$ indices while keeping the first and last indices fixed. The $(n-2)!$ coefficients $A[1,\sigma, n]$ are called \textit{color-ordered} or \textit{partial amplitudes}, and they carry all the kinematic information of the color-dressed amplitude $\mathcal{A}^{\rm{tree}}_{n}$. As an example, a 4-point tree-level Yang-Mills amplitude can be decomposed as the sum of two terms, namely
\begin{equation}
\mathcal{A}^{\rm tree}_4=g^{2}\, \Tr\big([[T_1,T_2],T_3]T_4\big)\,A_4[1,2,3,4]+g^{2}\,\Tr\big([[T_1,T_3],T_2]T_4\big)\,A_4[1,3,2,4]\; ,
\end{equation}
while a 5-point amplitude would have six terms and so on.

The partial amplitudes $A[1,...,n]$ can be computed using \textit{color-ordered Feynman rules}, only considering planar diagrams that preserve a particular order of the external legs determined by the labels of the generators $T_{i}$. These Feynman rules are typically derived by isolating independent color structures in the Feynman rules in the color-dressed theory and reading off the kinematic information accompanying each color structure. In the following, we state the relation between the color-ordered amplitudes/Feynman rules, and the kinematic products $m_{n}$ that follow from the algebraic color decomposition in section \ref{sec:YMoff}. To that end, let us first consider a 4-point tree-level amplitude which, following the discussion of the previous section, can be written in terms of the color factors $c_{i}$ and the kinematic numerators $n_{i}$
\begin{equation}\label{A4 BCJ2}
\mathcal{A}^{\rm tree}_4=g^2\,\Big\{\frac{n_s\,c_s}{s}+\frac{n_t\,c_t}{t}+\frac{n_u\,c_u}{u}\Big\}\; ,  
\end{equation}
where we recall that the color factors $c_{i}$ are defined as
\begin{equation}
\begin{split}
c_s&:=\Tr\big(T_4[[T_1,T_2],T_3]\big)\;,\\
c_t&:=\Tr\big(T_4[[T_2,T_3],T_1]\big)\;,\\
c_u&:=\Tr\big(T_4[[T_3,T_1],T_2]\big)\;,
\end{split}
\end{equation}
and the kinematic numerators $n_{i}$ are determined by the kinematic products $m_{n}$ as in equation \eqref{kinnum}.

The decomposition of the amplitude in equation \eqref{A4 BCJ2} is overcomplete, meaning that not all color structures are independent due to the Jacobi identity of the Lie algebra, which in this notation reads $c_{s}+c_{t}+c_{u}=0$. For this reason, one may want to decompose the amplitude into independent color structures. Given the Jacobi identity, we choose $c_s=\Tr\big([[T_1,T_2],T_3]T_4\big)$ and $-c_u=\Tr\big([[T_1,T_3],T_2]T_4\big)$ as a basis. Then, solving for $c_t$ one can rewrite \eqref{A4 BCJ2} in the DDM basis as 
\begin{equation}
\begin{split}
\mathcal{A}^{\rm tree}_4&=g^2\,\Tr\big([[T_1,T_2],T_3]T_4\big)\,A_4[1,2,3,4]+g^2\,\Tr\big([[T_1,T_3],T_2]T_4\big)\,A_4[1,3,2,4]\;,
\end{split}
\end{equation}
where the partial amplitudes are written in terms of the kinematic numerators as 
\begin{equation}
A_4[1,2,3,4]=\frac{n_s}{s}-\frac{n_t}{t}\;,\quad A_4[1,3,2,4]=\frac{n_t}{t}-\frac{n_u}{u}\;.
\end{equation}
Using the definition of the kinematic numerators in \eqref{kinnum}, we can write the partial amplitudes in terms of the kinematic products $m_{n}$, namely
\begin{equation}\label{colorfromDDM}
\begin{split}
A_4[1,2,3,4]&=\epsilon_4\cdot m_3(\epsilon_1,\epsilon_2,\epsilon_3)+\epsilon_4\cdot m_2\big(hm_2(\epsilon_1,\epsilon_2),\epsilon_3\big)+\epsilon_4\cdot m_2\big(\epsilon_1,hm_2(\epsilon_2,\epsilon_3)\big)\;,\\
%&=\epsilon_{4}\cdot \bar m_{3}(\epsilon_{1},\epsilon_{2},\epsilon_{3})\; ,\\
A_4[1,3,2,4]&=\epsilon_4\cdot m_3(\epsilon_1,\epsilon_3,\epsilon_2)+\epsilon_4\cdot m_2\big(hm_2(\epsilon_1,\epsilon_3),\epsilon_2\big)+\epsilon_4\cdot m_2\big(\epsilon_1,hm_2(\epsilon_3,\epsilon_2)\big)\;,
%&= \epsilon_{4}\cdot \bar m_{3}(\epsilon_{1},\epsilon_{3},\epsilon_{2})\; ,
\end{split}
\end{equation}
and one immediately infers that the color-ordered Feynman rules, namely the vertices that make up the color-ordered amplitudes, are encoded in the kinematic products $m_{n}$. The polarization vectors above should be thought of as being acted upon by the trivial inclusion map, so that the actions of the $m_{n}$ are well defined. The above expressions highlight the planar nature of the color-ordered amplitudes, meaning that the vertices $m_2$ and $m_3$ only join adjacent inputs, whose order is kept fixed. 

The above color-ordered amplitudes can be obtained by performing homotopy transfer of the kinematic $C_{\infty}$ algebra $\cK_{\rm{YM}}$ of Yang-Mills theory, in complete analogy to the color-dressed case where one performs homotopy transfer of the $L_{\infty}$ algebra of Yang-Mills. We now turn to constructing color-ordered amplitudes by transferring the $C_{\infty}$ structure on $\cK_{\rm{YM}}$ to the on-shell kinematic space $\bar{\cK}_{\rm{YM}}$. Let us start with the kinematic chain complex
\begin{equation}
\begin{tikzcd}[row sep=2mm]
K_{0}\arrow{r}{m_1}&K_{1}\arrow{r}{m_1}&K_{2}\arrow{r}{m_1}&K_{3}\\
\Lambda& A_\mu&J_\mu&N
\end{tikzcd}    \;,
\end{equation}
which follows from stripping off color as presented in section \ref{sec:YMoff}. As in the color-dressed case, we restrict our attention to sums of plane waves, namely
\begin{equation}\label{plane waves kin}
\begin{split}
\Lambda(x)&=\sum_i \lambda_i\,e^{ik_i\cdot x}\;,\quad A^\mu(x)=\sum_i a^\mu_i\,e^{ik_i\cdot x}\;,\\
J^\mu(x)&=\sum_i j^\mu_i\,e^{ik_i\cdot x}\;,\quad N(x)=\sum_i n_i\,e^{ik_i\cdot x}\;.
\end{split}    
\end{equation}
Our goal now is to perform homotopy transfer to the on-shell chain complex defined by the on-shell space $\bar{\cK}_{\rm{YM}}=\bigoplus_{i=0}^{3} \bar K_{i}$ and the projected differential $\bar m_{1}:\bar{\cK}_{YM}\to \bar{\cK}_{\rm{YM}}$
\begin{equation}
\begin{tikzcd}[row sep=2mm]
\bar K_{0}\arrow{r}{\bar m_1}&\bar K_{1}\arrow{r}{\bar m_1}&\bar K_{2}\arrow{r}{\bar m_1}&\bar K_{3}\\
\bar \Lambda& \bar A_\mu& \bar J_\mu&\bar N
\end{tikzcd}    \;.
\end{equation}
The elements of this on-shell space are sums of harmonic plane waves
\begin{equation}\label{plane waves kin on}
\begin{split}
\bar \Lambda(x)&=\sum_i \bar \lambda_i\,e^{ik_i\cdot x}\;,\quad \bar A^\mu(x)=\sum_i \epsilon^\mu_i\,e^{ik_i\cdot x}\;,\\
\bar J^\mu(x)&=\sum_i \bar j^\mu_i\,e^{ik_i\cdot x}\;,\quad \bar N(x)=\sum_i \bar n_i\,e^{ik_i\cdot x}\;,
\end{split}    
\end{equation}
where the polarization vectors are transverse, i.e $k_{i}\cdot \epsilon_{i}=0$, and the Fourier coefficients of the color-ordered currents $j^{\mu}(k)$ belong to the equivalence class $[\bar j^{\mu}(k)]=[\bar j^{\mu}(k)+k^{\mu}\phi(k)]$. In order to perform homotopy transfer to the on-shell space, we first need to define a projector $p:\cK_{\rm{YM}}\to \bar{\cK}_{\rm{YM}}$ and an inclusion map $\iota:\bar{\cK}_{\rm{YM}}\to \cK_{\rm{YM}}$ which diagrammatically act as
\begin{equation}
\begin{tikzcd}[row sep=10mm]
K_{0}\arrow{d}{p}\arrow{r}{m_1}&K_{1}\arrow{d}{p}\arrow{r}{m_1}&K_{2}\arrow{d}{p}\arrow{r}{m_1}&K_{3}\arrow{d}{p}\\
\bar K_{0}\arrow[shift left=2mm]{u}{\iota}\arrow{r}{\bar m_1}&\bar K_{1}\arrow[shift left=2mm]{u}{\iota}\arrow{r}{0}&\bar K_{2}\arrow[shift left=2mm]{u}{\iota}\arrow{r}{\bar m_1}&\bar K_{3}\arrow[shift left=2mm]{u}{\iota}
\end{tikzcd}    \;.
\end{equation}
To that end, we use the same projector and inclusion maps as we did in the $L_{\infty}$ case, which act on the Fourier coefficients as
\begin{equation}\label{p ontran2}
\begin{split}
p(\lambda)&:=P_{\rm h}(\lambda)\;,\quad p(a_\mu):=P_{\rm h}P_\perp(a_\mu)\;,\\
p(j_\mu)&:=[P_{\rm h}j_\mu]\;,\quad\hspace{1mm} p(n):=P_{\rm h}(n)\;,
\end{split}
\end{equation}
and the inclusion is trivial, except when acting on $K_{2}$, where it picks the representative of the equivalence class with vanishing time component:
\begin{equation}
\iota\big(\bar j^{\mu}(k)\big):=\bar j^{\mu}-k^{\mu}\frac{\bar j^{0}(k)}{k^{0}}\; .
\end{equation}
Notice that we can choose the same maps as in the $L_{\infty}$ case because in the color-dressed theory all these maps act only non-trivially on the Fourier coefficients with no color dependence. Similarly, the homotopy map can be chosen to be the same as in the previous section
\begin{equation}
\begin{split}
h_\xi\big(a(k)\big)&:=\frac{i}{k^2}\,(1-P_{\rm h})\,k\cdot a\;,\\
h_\xi\big(j(k)\big)&:=\frac{1}{k^2}\,\left(\delta^\mu{}_\nu-\xi\,\frac{k^\mu k_\nu}{k^2}\right)\,(1-P_{\rm h})\,j^\nu+P_{\rm h}\left(\frac{j^0}{(k^0)^2},\vec0\right)\;,\\
h_\xi\big(n(k)\big)&:=\frac{i}{k^2}\,(1-P_{\rm h})\,k^\mu n\;,
\end{split}
\end{equation}
which thus obeys the homotopy relation
\begin{equation}
    m_{1}h+hm_{1}=\iota \, p-1_{\cK}\; .
\end{equation}

As in the $L_{\infty}$ case and as we shall prove in the following section, performing homotopy transfer of the $C_{\infty}$ algebra of Yang-Mills to the on-shell space $\bar{\cK}_{\rm{YM}}$ generates an infinite number of higher transferred products $\bar m_{n}$ that define a $C_{\infty}$ algebra. These transferred products are defined in terms of the projector, inclusion and homotopy maps, as well as kinematic products $m_{n}$ of the off-shell $C_{\infty}$ algebra. In the interest of computing color-ordered amplitudes, it is convenient to define the action of the transferred products $\bar m_{n}$ on the Fourier coefficients of the elements of $\bar{\cK}_{\rm{YM}}$. Taking fields given by a single plane wave $\bar A^{\mu}_{i}(x)=\epsilon^{\mu}_{i}\, e^{ik_{i}\cdot x}$, the $\bar m_{n}$ act as
\begin{equation}
\bar m_{n}(\bar A_{1},\bar A_{2},\ldots,\bar A_{n})=\bar m_{n}(\epsilon_{1},\epsilon_{2},\ldots,\epsilon_{n})\, e^{i(k_{1}+k_{2}+\ldots+k_{n})\cdot x}\;,
\end{equation}
which extends by linearity to sums of plane waves. The products $\bar m_{n}$ acting on the polarization vectors $\epsilon^{\mu}_{i}$ define color-ordered currents which subsequently we will use to construct color-ordered amplitudes. Similarly to the diagrammatic formulation of the transported brackets in the $L_{\infty}$ case, we can represent the transported $C_{\infty}$ products using tree diagrams. However, in contrast to the $L_{\infty}$ case, here we shall keep the order of the inputs fixed. For instance, the transferred product $\bar m_{3}$ acting on three polarization vectors can be represented as
\begin{equation}\label{barm3}
\begin{split}
\bar m_3(\epsilon_{1},\epsilon_{2},\epsilon_{3})&=p\,\Big\{m_3( a_1,a_2,a_3)+m_{2}(h m_{2}(a_{1},a_{2}),a_3)+m_{2}(a_{1},hm_{2}(a_{2},a_{3}))\Big\}\\
&=\begin{tikzpicture}[baseline={([yshift=-.5ex]current bounding box.center)},level distance=10.5mm,sibling distance=10mm]
  \node[scale=0.5,circle, draw] {}
    child[grow=up] {
     child {node[scale=0.7] {3}} child {node[scale=0.7] {2}} child {node[scale=0.7] {1}}}
    ;
\end{tikzpicture}+\begin{tikzpicture}[baseline={([yshift=-.5ex]current bounding box.center)},level distance=7mm,sibling distance=10mm]
  \node[scale=0.5,circle, draw] {}
    child[grow=up] {
     child{child {node[scale=0.7] {3}} child[missing]}    child {child {node[scale=0.7] {2}} child {node[scale=0.7] {1}}}
    };
\end{tikzpicture}+\begin{tikzpicture}[baseline={([yshift=-.5ex]current bounding box.center)},level distance=7mm,sibling distance=10mm]
  \node[scale=0.5,circle, draw] {}
    child[grow=up] {
         child {child {node[scale=0.7] {3}} child {node[scale=0.7] {2}}}
         child{child[missing] child {node[scale=0.7] {1}} }
    };
\end{tikzpicture}\;,\\
\end{split}    
\end{equation}
with $a_{i}=\iota(\epsilon_{i})$ with the trivial inclusion. Similarly, the 4-product $\bar m_4(\epsilon_{1},\epsilon_{2},\epsilon_{3},\epsilon_{4})$ reads
\begin{equation}\label{bar ms homtransf}
\begin{split}
\bar m_4(\epsilon_{1},\epsilon_{2},\epsilon_{3},\epsilon_{4})=p \,\Big\{& m_{2}(hm_{2}(a_{1},a_{2}),hm_{2}(a_{3},a_{4})))+m_{2}(h m_{2}(hm_{2}(a_{1},a_{2}),a_{3}),a_{4})\\
+ & m_{2}(a_{1},hm_{2}(a_{2},hm_{2}(a_{3},a_{4})))+m_{2}(hm_{2}(an_{1},hm_{2}(a_{2},a_{3})),a_{4})\\
+& m_{2}(a_{1},hm_{2}(hm_{2}(a_{2},a_{3}),a_{4}))+m_{2}(hm_{3}(a_{1},a_{2},a_{3}),a_{4})\\
+ & m_{2}(a_{1},hm_{3}(a_{2},a_{3},a_{4}))+m_{3}(hm_{2}(a_{1},a_{2}),a_{3},a_{4})\\
+& m_{3}(a_{1},hm_{2}(a_{2},a_{3}),a_{4})+m_{3}(a_{1},a_{2},hm_{2}(a_{3},a_{4}))\Big\}\\
=&\begin{tikzpicture}[baseline={([yshift=-.5ex]current bounding box.center)},level distance=10mm]
\tikzstyle{level 2}=[sibling distance=15mm]
  \tikzstyle{level 3}=[sibling distance=10mm]
  \node[scale=0.5,circle, draw]{}[grow=up] child{child{child{node[scale=0.7] {4}} child{node[scale=0.7] {3}}} child{child{node[scale=0.7] {2}} child{node[scale=0.7] {1}}}} ;
\end{tikzpicture}+\begin{tikzpicture}[baseline={([yshift=-.5ex]current bounding box.center)},level distance=7mm,sibling distance=10mm]
  \node[scale=0.5,circle, draw] {}
    child[grow=up]{child{child{child{node[scale=0.7] {4}} child[missing]} child[missing]} child{
     child{child {node[scale=0.7] {3}} child[missing]}    child {child {node[scale=0.7] {2}} child {node[scale=0.7] {1}}}
    }};
\end{tikzpicture}+\begin{tikzpicture}[baseline={([yshift=-.5ex]current bounding box.center)},level distance=7mm,sibling distance=10mm]
  \node[scale=0.5,circle, draw] {}
    child[grow=up]{child{child{child{node[scale=0.7] {4}} child{node[scale=0.7]{3}}} child{child[missing] child{node[scale=0.7] {2}}}} child{child[missing]child{child[missing] child{node[scale=0.7] {1}}}}
    };
\end{tikzpicture}\\
+& \begin{tikzpicture}[baseline={([yshift=-.5ex]current bounding box.center)},level distance=7mm,sibling distance=10mm]
  \node[scale=0.5,circle, draw] {}
    child[grow=up]{child{child{child{node[scale=0.7] {4}}child[missing]}child[missing]
    }child{child[missing]child{child{node[scale=0.7] {3}}child{node[scale=0.7] {2}}}child{child[missing]child{node[scale=0.7]{1}}}child[missing]}
    };
\end{tikzpicture}+\begin{tikzpicture}[baseline={([yshift=-.5ex]current bounding box.center)},level distance=7mm,sibling distance=10mm]
  \node[scale=0.5,circle, draw] {}
    child[grow=up]{child{child{child{node[scale=0.7] {4}}child[missing] }child{child{node[scale=0.7] {3}}child{node[scale=0.7] {2}}}} child{child[missing] child{child[missing] child{node[scale=0.7] {1}}}}
    };
\end{tikzpicture}+\begin{tikzpicture}[baseline={([yshift=-.5ex]current bounding box.center)},level distance=7mm,sibling distance=8mm]
  \node[scale=0.5,circle, draw] {}
    child[grow=up]{child{child{child{node[scale=0.7] {4}}child[missing]}child[missing]
    }child{child[missing]child{child{node[scale=0.7] {3}}child{node[scale=0.7] {2}}child{node[scale=0.7]{1}}}}
    };
\end{tikzpicture}\\
+& \begin{tikzpicture}[baseline={([yshift=-.5ex]current bounding box.center)},level distance=7mm,sibling distance=8mm]
  \node[scale=0.5,circle, draw] {}
    child[grow=up]{child{child{child{node[scale=0.7] {4}}child{node[scale=0.7] {3}}child{node[scale=0.7] {2}}}
    child[missing]}child{child[missing]child{child[missing]child{node[scale=0.7]{1}}}}
    };
\end{tikzpicture}+\begin{tikzpicture}[baseline={([yshift=-.5ex]current bounding box.center)},level distance=7mm,sibling distance=8mm]
  \node[scale=0.5,circle, draw] {}
    child[grow=up]{child{child{child{node[scale=0.7]{4}}child[missing]child[missing]}child{child{node[scale=0.7]{3}}}child{child{node[scale=0.7]{2}}child{node[scale=0.7]{1}}}}
    };
\end{tikzpicture}+\begin{tikzpicture}[baseline={([yshift=-.5ex]current bounding box.center)},level distance=7mm,sibling distance=8mm]
  \node[scale=0.5,circle, draw] {}
    child[grow=up]{child{child{child{node[scale=0.7]{4}}child[missing]child[missing]}child{child{node[scale=0.7]{3}}child{node[scale=0.7]{2}}}child{child[missing]child[missing]child{node[scale=0.7]{1}}}}
    };
\end{tikzpicture}\\
+&\begin{tikzpicture}[baseline={([yshift=-.5ex]current bounding box.center)},level distance=7mm,sibling distance=8mm]
  \node[scale=0.5,circle, draw] {}
    child[grow=up]{child{child{child{node[scale=0.7]{4}} child{node[scale=0.7]{3}}}child{child{node[scale=0.7]{2}}}child{child[missing]child[missing]child{node[scale=0.7]{1}}}}
    };
\end{tikzpicture}\;.
\end{split}    
\end{equation}

In the previous section we developed recursion relations for the currents $\cJ^{\mu}_{i_{1}\ldots i_{n}}$, and showed their relation to the transported $L_{\infty}$ brackets $\bar B_{n}$ (see equation \eqref{Recursive YM Loo}). Here we can proceed similarly and relate the so-called \textit{Berends-Giele currents} \cite{Berends:1987me,Mafra:2015vca,Macrelli:2019afx} to the transported $C_{\infty}$ products $\bar m_{n}$. Starting with the recursion seed $a^\mu_i=\iota(\epsilon^\mu_{i})$, for $n\geq 2$, the Berends-Giele currents $j^{\mu}_{i_{1}\ldots i_{n}}$ and transported products $\bar m_{n}$ are given by
\begin{equation}\label{Berends-Giele}
\begin{split}
j^{\mu}_{i_{1}\ldots i_{n}}&=\sum_{k+l=n}m_{2}(a_{i_{1} \ldots i_{k}},a_{i_{k+1}\ldots i_{n}})+\sum_{k+l+m=n} m_{3}(a_{i_{1}\ldots i_{k}},a_{i_{k+1}\ldots i_{k+l}},a_{i_{k+l+1}\ldots i_{n}})\;,\\
a^{\mu}_{i_{i}\ldots i_{n}}&=h j^{\mu}_{i_{1}\ldots i_{n}}\;,\quad \bar m_{n}(\epsilon_{i},\ldots, \epsilon_{i_{n}})=p\, j^{\mu}_{i_{1}\ldots i_{n}}\; .
\end{split}
\end{equation}

To obtain the amplitudes from the above currents, it is necessary to introduce an inner product. To this end, let us recall that the cyclic inner product of the $L_{\infty}$ algebra of Yang-Mills, $\langle\cdot,\cdot\rangle_{\cX}$, factorizes into the trace over color indices and a cyclic kinematic pairing $\omega$ (see equation \eqref{inner YM}). Similarly, the pairing transferred to the on-shell space $\bar{\cX}$, $\langle\cdot,\cdot\rangle_{\bar{\cX}}$, factorizes into the trace over color indices and a kinematic pairing $\bar{\omega}$, defined via the inclusion map as
\begin{equation}
\bar\omega\big(\bar\psi_1,\bar\psi_2\big):=\omega\big(\iota(\bar\psi_1),\iota(\bar\psi_2)\big)\;,\quad\bar\psi_i\in\bar\cK_{\rm YM}\;.    
\end{equation}
Restricting the inputs to color-stripped fields $\bar A_i$, ciclicity of $\bar\omega$ is ensured by the homotopy obeying
\begin{equation}
    \omega(h_{\xi} J_{1},J_{2})=\omega(J_{1},h_{\xi} J_{2})\; ,
\end{equation}
where $J_{1},\,J_{2}\in K_{2}$. The color-ordered amplitudes are then given by
\begin{equation}\label{cycliccolor}
\begin{split}
\delta\big(K\big)\, A[1,2,\ldots,n-1,n]&=\bar\omega\big(\bar{A}_{n},\bar m_{n-1}(\bar{A}_{1},\bar{A}_{2},\ldots,\bar{A}_{n-2},\bar{A}_{n-1})\big)\\
&=\delta\big(K\big)\, \epsilon_{n}\cdot \bar m_{n-1}(\epsilon_{1},\epsilon_{2},\ldots,\epsilon_{n-2},\epsilon_{n-1})\;,
\end{split}
\end{equation}
where all inputs $\bar A^{\mu}_{i}$ are single plane waves $\bar A^{\mu}_{i}=\epsilon_{i}^{\mu}\, e^{ik_{i}\cdot x}$. Using this pairing, the 4-point and 5-point amplitudes are
\begin{equation}
\begin{split}
A[1,2,3,4]=\epsilon_{4}\cdot \bar m_{3}(\epsilon_{1},\epsilon_{2},\epsilon_{3})\;,\\
A[1,2,3,4,5]=\epsilon_{5}\cdot \bar m_{4}(\epsilon_{1},\epsilon_{2},\epsilon_{3},\epsilon_{4})\; .
\end{split}
\end{equation}
Using the explicit expression of $\bar m_{3}(\epsilon_{1},\epsilon_{2},\epsilon_{3})$ in equation \eqref{barm3} in the color-ordered amplitude above, one immediately sees that it coincides with the first color-ordered amplitude in equation \eqref{colorfromDDM} which we derived from color decomposition.

Color-ordered amplitudes obey cyclicity, reflection, a set of relations called \textit{Kleiss-Kuijf relations} \cite{KLEISS1989616}, the photon decoupling identity, and the so-called BCJ relations \cite{Bern:2008qj}. All these properties and identities imply that not all color-ordered amplitudes are independent of each other. In this paper, however, we focus on the relations that are a consequence of the symmetry properties of the kinematic products $\bar m_{n}$ and the kinematic pairing $\bar \omega$. Given that the maps $\bar m_{n}$ form a $C_{\infty}$ algebra, these products have the corresponding symmetry properties, namely they vanish upon taking the sums of graded shuffle permutations of the inputs as discussed in the previous section. The relevant relations for this paper are cyclicity, reflection, and the Kleiss-Kuijf relations, which we discuss in the following. 

%\medskip

\vspace{0.5ex} 

\noindent 
\textbf{Cyclicity} \\[0.5ex] 
The cyclicity of the color-ordered amplitudes follows from the cyclicity of the kinematic pairing $\bar\omega$. Indeed, the kinematic pairing of the on-shell algebra $\bar \omega$ inherits the cyclic property of the off-shell pairing $\omega$ (recall equation \eqref{cyc Coo}). Cyclicity of $\bar\omega$ implies
\begin{equation}
\epsilon_{n}\cdot \bar m_{n-1}(\epsilon_{1},\epsilon_{2},\ldots,\epsilon_{n-2},\epsilon_{n-1})=\epsilon_{n-1}\cdot \bar m_{n-1}(\epsilon_{n},\epsilon_{1},\ldots,\epsilon_{n-3},\epsilon_{n-2})\; ,
\end{equation}
which translates into the cyclic property of color-ordered amplitudes
\begin{equation}
    A[1,2,\ldots,n-1,n]=A[n,1,\ldots,n-2,n-1]\; .
\end{equation}

\medskip

\noindent 
\textbf{Reflection}\\[1ex] 
Color-ordered amplitudes obey the following reflection property:
\begin{equation}
A[1,2,\ldots,n-1,n]=(-1)^{n}\, A[n,n-1,\ldots,2,1]\;.
\end{equation}
This relation follows from the symmetry property of the kinematic products $m_{n}$, namely vanishing on signed shuffles, in combination with cyclicity of the inner product. In order to motivate the more general discussion of the following section, let us consider the 4-point case and show that $A[1,2,3,4]=A[4,3,2,1]$. Indeed, the original amplitude in terms of the transferred products and polarization vectors reads
\begin{equation}
A[1,2,3,4]=\epsilon_{4}\cdot \bar m_{3}(\epsilon_{1},\epsilon_{2},\epsilon_{3})\;.
\end{equation}
Next, in the above amplitudes we can use the symmetry property of $\bar m_{3}$
\begin{equation}
\bar m_{3}(\epsilon_{1},\epsilon_{2},\epsilon_{3})=\bar m_{3}(\epsilon_{3},\epsilon_{2},\epsilon_{1})\; ,
\end{equation}
which yields the following color-ordered amplitude
\begin{equation}
    A[3,2,1,4]=\epsilon_{4}\cdot \bar m_{3}(\epsilon_{3},\epsilon_{2},\epsilon_{1})\;.
\end{equation}
Finally, in order to arrive to prove the reflection property, we use cyclicity of the partial amplitude, leading to the desired result
\begin{equation}
    A[4,3,2,1]=\epsilon_{1}\cdot m_{3}(\epsilon_{4},\epsilon_{3},\epsilon_{2})\; .
\end{equation}

\medskip

\noindent 
\textbf{Kleiss-Kuijf relations} \\[0.5ex] 
These Kleiss-Kuijf relations can be expressed in the following form \cite{Garozzo:2018uzj}:
\begin{equation}
\sum_{\sigma \in P\shuffle Q}A[\sigma,n]=0\; ,
\end{equation}
where the sum runs over the shuffle $P\shuffle Q$ of the non-empty ordered sets of particle labels $P$ and $Q$. These relations follow immediately from the vanishing on shuffles of the kinematic products $\bar m_{n}$, namely
\begin{equation}
\sum_{\sigma \in P\shuffle Q}A[\sigma,n]= \sum_{\sigma \in P\shuffle Q} \epsilon_{n}\cdot \bar m_{n-1}(\epsilon_{\sigma})=0\; .
\end{equation}

\subsection{Color-ordered Ward identities}
In this section, we turn to gauge invariance of color-ordered amplitudes and their Ward identities. To this end, let us perform color decomposition of a full gluon tree amplitude in the so-called \textit{trace basis} where, in contrast to the DDM basis, one uses traces of products of the generators of the gauge algebra instead of commutators, namely
\begin{equation}
    \mathcal{A}^{\rm{tree}}_{n}=g^{n-2}\sum_{\sigma \in S_{n-1}}\text{Tr}\Big\{T_{1}T_{\sigma(2)}...T_{\sigma(n)}\Big\}\, A[1,\sigma(2),\ldots,\sigma(n)]\; .
\end{equation}
Starting from the color decomposition in the DDM basis, one can arrive at this form of the amplitude by writing out the commutators of the generators in equation \eqref{DDMdecomp} and using the Kleiss-Kuijf relations. The sum above runs through the independent $(n-1)!$ orderings of the generators inside of the trace. Consequently, to ensure gauge invariance of the full amplitude, each color-ordered amplitude must be gauge invariant on its own, implying the existence of color-ordered Ward identities for every color-ordered amplitude. In the following, we give an algebraic interpretation of these identities using the $C_{\infty}$ relations, in the same spirit of section \eqref{sec:Wardfull} where we presented an analogous discussion for the full Yang-Mills amplitudes. 

Gauge invariance of a color-ordered amplitude is stated as its invariance under the shift of the polarization vector of one of the external particles $\delta \epsilon^{\mu}_{i}=i\, k^{\mu}_{i}\bar \lambda_{i}=\bar m_{1}(\bar \lambda_{i})$. Algebraically, such a shift is encoded in the following variation of the color-ordered amplitude\footnote{Notice that by cyclicity of the color-ordered amplitude, we could have shifted any other leg.}
\begin{equation}
\begin{split}
    \delta(K)\, \delta A[1,\ldots,n]&=\bar\omega\big( \bar m_{1}(\bar\lambda_{n}),\bar m_{n-1}(\epsilon_{1},\ldots ,\epsilon_{n-1}) \big)\\
    &=\bar \omega\big( \bar\lambda_{n},\bar m_{1}(\bar m_{n-1}(\epsilon_{1},\ldots ,\epsilon_{n-1}) \big)\;,
\end{split}
\end{equation}
whose vanishing requires the linear color-ordered Ward identity
\begin{equation}
\bar m_{1}(\bar m_{n-1}(\epsilon_{1},\ldots,\epsilon_{n-1}))=0\;.
\end{equation}
Our task now is to prove that the above linear Ward identity holds. Let us recall that the transferred brackets $\bar m_{n}$ form a $C_{\infty}$ algebra, which means that they obey the following homotopy associativity relations
\begin{equation}
    \sum_{i+j+k=n-1}(-1)^{k}\, \bar m_{i+k+1}\big(\epsilon_{1},\ldots,\epsilon_{i},\bar m_{j}(\epsilon_{i+1},\ldots,\epsilon_{i+j}),\epsilon_{i+j+1},\ldots, \epsilon_{i+j+k}\big)=0\; ,
\end{equation}
where $i\geq 0\, ,\, j\geq 1 \, ,\, k\geq 0$ and we considered $n-1$ polarization vectors as inputs because these are the relevant inputs for Ward identities of $n$-point amplitudes. Recall that the action of the differential $\bar m_{1}$ on polarization vectors is trivial, i.e $\bar m_{1}(\epsilon_{i})=0$, and hence the terms in the above $C_{\infty}$ relations when $j=1$ vanish. Thus, the non-trivial contributions to the homotopy associativity relations are those where $j\geq 2$. In the interest of proving that the color-ordered Ward identities are linear, it will be convenient to rewrite the above equation by singling out the case when $i=k=0$ as
\begin{equation}\label{nonlinward}
\begin{split}
\bar m_{1}(\bar m_{n-1}(\epsilon_{1},\ldots,\epsilon_{n-1}))+\sum_{l+r=n-1}\, (-1)^{r}\, \bar m_{r+1}(\bar m_{l}(\epsilon_{1},\ldots,\epsilon_{l}),\epsilon_{l+1},\ldots, \epsilon_{l+r})\\
+\sum_{i+j+k=n-1}(-1)^{k}\, \bar m_{i+k+1}(\epsilon_{1},\ldots,\epsilon_{i},\bar m_{j}(\epsilon_{i+1},\ldots,\epsilon_{i+j}),\epsilon_{i+j+1},\ldots, \epsilon_{i+j+k})=0\; ,
\end{split}
\end{equation}
with $r\geq 1\, ,\, l\geq 2\, , \, i\geq 1\, ,\, j\geq 2 \, ,\, k\geq 0$. As we will show next, the second and last terms vanish because $\bar m_{l}$ and $\bar m_{j}$ contain off-shell combinations of momenta, while the first term, which encodes linear color-ordered Ward identities, remains. To see this, we recall that $\bar m_{l}$ (as well as $\bar m_{j}$) above can be written in terms of Berends-Giele currents as $\bar m_{l}(\epsilon_{i_{1}},\ldots,\epsilon_{i_{l}})=p\, j^{\mu}_{i_{1}\ldots i_{l}}$ with $2\leq l\leq n-2$. The currents obey
\begin{equation}
    h\, j^{\mu}_{i_{1}\ldots i_{l}}=\frac{b\, j^{\mu}_{i_{1}\ldots i_{l}}}{(k_{i_{1}}+\ldots +k_{i_{l}})^{2}}=a^\mu_{i_{1}\ldots i_{l}}\;,
\end{equation}
where the sum in the denominator is over the momenta of $l$ adjacent particles. The sum of $l\leq n-2$ adjacent momenta has to be off-shell so that all the momenta exchanged in the propagators are off-shell and, as a consequence, the on-shell projector acting on such currents vanishes, implying the vanishing of $\bar m_{l}$ with $l\leq n-2$. For this reason, the only non-trivial contribution to the $C_{\infty}$ relations in \eqref{nonlinward} is the linear Ward identity
\begin{equation}
    \bar m_{1}(\bar m_{n-1}(\epsilon_{1},\ldots,\epsilon_{n-1}))=0\; .
\end{equation}

We now turn to a concrete example. Let us consider the 4-point color-ordered amplitude $A[1,2,3,4]$ written in terms of the kinematic numerators 
\begin{equation}\label{partialex}
    \delta(K)\, A[1,2,3,4]=\bar \omega\big(\epsilon_{4}, \bar m_{3}(\epsilon_{1},\epsilon_{2},\epsilon_{3})\big)=\delta(K)\Big(\frac{n_{s}}{s}-\frac{n_{t}}{t}\Big)\;.
\end{equation}
The two terms above are related to each other by exchanging particle labels. Indeed, starting with the first term written in terms of the kinematic products
\begin{equation}
    \delta(K)\, \frac{n_s}{s}:=\bar\omega\Big(\epsilon_4,p\,m_2\big(hm_2(\epsilon_1,\epsilon_2),\epsilon_3\big)+\frac{1}{3}\,p\big(m_3(\epsilon_1,\epsilon_2,\epsilon_3)-m_3(\epsilon_2,\epsilon_1,\epsilon_3)\big)\Big)\;,
\end{equation}
one can exchange the labels $1\to 2$ and $2\to 3$ to recover the second term $-\delta(K)\frac{n_{t}}{t}$. We remind the reader that we are omitting the inclusion maps that act on the inputs of the brackets $m_{n}$ to avoid cluttering the equations. As we pointed out before, a gauge variation of $A[1,2,3,4]$ is given by the shift of one of the polarization vectors. In this case, we choose the fourth external particle so that the gauge variation of the amplitude is generated by the shift $\delta \epsilon_{4}=\bar m_{1}(\bar \lambda)$. Such a shift leads to
\begin{equation}
\begin{split}
    \delta(K)\, \delta\Big(\frac{n_{s}}{s}\Big)&=\bar \omega\Big( \bar m_{1}(\bar \lambda), p\, m_2\big(hm_2(\epsilon_1,\epsilon_2),\epsilon_3\big)+\frac{1}{3}\,\big(p\, m_3(\epsilon_1,\epsilon_2,\epsilon_3)-p\, m_3(\epsilon_2,\epsilon_1,\epsilon_3)\big) \Big)\\
    &=\bar \omega\Big(\bar \lambda, \bar m_{1}\,\big(p\, m_2\big(hm_2(\epsilon_1,\epsilon_2),\epsilon_3\big)\big)+\frac{1}{3}\, \bar m_{1}\, \big(p\, m_3(\epsilon_1,\epsilon_2,\epsilon_3)-p\, m_3(\epsilon_2,\epsilon_1,\epsilon_3)\big) \Big)\;.
\end{split}
\end{equation}
%where in the second equality we dropped the bar on the differential because it is projected to the on-shell space by the pairing with the projected gauge parameter $\bar \lambda$. 
In the following we prove that this variation is invariant under cyclic permutations of the particle labels $(1,2,3)$ and, as a consequence, it cancels against the variation of the term involving $n_{t}$. To this end, we shall use the $C_{\infty}$ relations in the second entry of the above pairing, namely
\begin{equation}\label{wardexaaa}
\bar m_{1}\big(p\, m_2\big(hm_2(\epsilon_1,\epsilon_2),\epsilon_3\big)\big)+\frac{1}{3}\,\bar m_{1}\big(p\, m_3(\epsilon_1,\epsilon_2,\epsilon_3)-p\, m_3(\epsilon_2,\epsilon_1,\epsilon_3)\big)\;.
\end{equation}
Let us begin with the first term. Here, we can use the fact that $p$ is a chain map ($\bar m_{1}p=p\,m_{1}$) as well as the Leibniz rule of the differential with respect to the two-product $m_{2}$ to obtain
\begin{equation}
    \bar m_{1}\big(p\, m_2\big(hm_2(\epsilon_1,\epsilon_2),\epsilon_3\big)\big)=p\, m_2\big(m_{1}\big(hm_2(\epsilon_1,\epsilon_2)\big),\epsilon_3\big)\big)\;,
\end{equation}
where we used $m_{1}(\epsilon_{i})=0$ for on-shell and transverse polarization vectors. We can further manipulate the above expression by using the homotopy transfer relation (again, omitting the inclusion map for simplicity) $m_{1}h+hm_{1}= p-1$, which yields
\begin{equation}\label{partial1}
\begin{split}
    p\, m_2\big(m_{1}\big(hm_2(\epsilon_1,\epsilon_2)\big),\epsilon_3\big)\big)=&-p\, m_{2}\big(hm_{1}m_{2}\big(\epsilon_{1},\epsilon_{2}\big),\epsilon_{3}\big)- p\, m_{2}\big((1-p)\, m_{2}\big(\epsilon_{1},\epsilon_{2}\big),\epsilon_{3}\big)\\
    =&-p\, m_{2}\big(m_{2}\big(\epsilon_{1},\epsilon_{2}\big),\epsilon_{3}\big)\;,
\end{split}
\end{equation}
where to get to the last line we used the fact that the sum of two momenta in the 4-point amplitude is off-shell ($s\neq0$), namely $p\, m_{2}\big((1-p)\, m_{2}\big(\epsilon_{1},\epsilon_{2}\big),\epsilon_{3}\big)=p\, m_{2}\big(m_{2}\big(\epsilon_{1},\epsilon_{2}\big),\epsilon_{3}\big)$. The last two terms in \eqref{wardexaaa} which involve $m_{3}$ can be manipulated by using homotopy associativity 
\begin{equation}\label{partial2}
\begin{split}
\frac{1}{3}\,\bar m_{1}\big(p\, m_3(\epsilon_1,\epsilon_2,\epsilon_3)-p\, m_3(\epsilon_2,\epsilon_1,\epsilon_3)\big)=\frac{1}{3}\, p\, \Big(2\, m_{2}\big(m_{2}\big(\epsilon_{1},\epsilon_{2}\big),\epsilon_{3}\big)&- m_{2}\big(m_{2}\big(\epsilon_{2},\epsilon_{3}\big),\epsilon_{1}\big)\\
&-m_{2}\big(m_{2}\big(\epsilon_{3},\epsilon_{1}\big),\epsilon_{2}\big)\Big)\;,
\end{split}
\end{equation}
where we used the symmetry properties of $m_{2}$. Summing equations \eqref{partial1} and \eqref{partial2} yields the following variation of the first term of the color-ordered amplitude
\begin{equation}
    \delta(K)\, \delta\Big(\frac{n_{s}}{s}\Big)=\bar \omega\big(\bar \lambda, \alpha\big)\;,
\end{equation}
with $\alpha$ the following combination of terms that is invariant under cyclic permutations of the labels $(1,2,3)$ 
\begin{equation}
    \alpha:=-p\, m_{2}\big(m_{2}\big(\epsilon_{[1},\epsilon_{2}\big),\epsilon_{3]}\big)\;.
\end{equation}
Thus, due to all Mandelstam variables $s_{ij}$ and kinematic numerators $n_{s_{ij}}$ being related by cyclic permutations of the particle labels, the variation of term involving $n_{t}$ in the color-ordered amplitude $A[1,2,3,4]$ yields
\begin{equation}
    \delta(K)\, \delta\Big(-\frac{n_{t}}{t}\Big)=-\bar \omega\big(\bar \lambda, \alpha\big)\;,
\end{equation}
and the variation of the amplitude vanishes. 

\section{Homotopy commutative algebras}

In this section we develop a self-contained mathematical framework for the color-stripped amplitudes described in the previous section. 
We first give a definition of homotopy associative ($A_\infty$) algebras in terms of differential graded algebras. This definition can be found for example in \cite{Kajiura:2003ax}, where also the homotopy transfer is described. We then describe homotopy commutative ($C_\infty$) algebras as a special case of the associative algebras \cite{CHENG20082535}. The structure of homotopy algebras, without any reference to an underlying vector space, can be described as operads \cite{AlgebraicOperads}. This reference also describes the bar construction for a wide class of homotopy algebras.

\subsection{Coalgebras}

A powerful tool to describe homotopy algebras is the language of coalgebras, due to the bar construction we will describe below. In practice, coalgebras can be obtained from algebras by going to the dual vector space, at least in finite dimensions.\footnote{An interesting fact is that one can obtain algebras from coalgebras by going to the dual without the finite dimensionality assumption.} Even without this construction, the axioms of coalgebras can be obtained by reading the axioms of algebras backwards. Let us explain what we mean by that. An algebra is defined via a product, that is a bilinear map
\begin{equation}\label{aproduct}
\mu_2: V \otimes V \rightarrow V    
\end{equation}
on some vector space $V$. This vector space can be graded, in which case we require $\mu_2$ to have degree zero. The fundamental object defining a coalgebra is obtained from reversing the arrow in \eqref{aproduct}, that is a linear map
\begin{equation}
    \Delta: V \rightarrow V \otimes V\;.
\end{equation}
Instead of merging (or multiplying) two elements like a product, $\Delta$ splits an element into two. 

We usually put additional axioms on products. For example, when describing associative algebras, we want the product to satisfy the associativity condition
\begin{equation}
    \mu_2(\mu_2(a,b),c) = \mu_2(a,\mu_2(b,c)) \, .
\end{equation}
To get the corresponding axiom on coalgebras, we first get rid of the inputs. In its input-free form, associativity reads
\begin{equation}
    \mu_2\circ (\mu_2 \otimes 1) = \mu_2\circ(1 \otimes \mu_2) \, ,
\end{equation}
where $1: V \rightarrow V$ is the identity map. Reversing the order of the maps, we get the coassociativity condition
\begin{equation}
    (\Delta \otimes 1) \circ \Delta = (1 \otimes \Delta) \circ \Delta \, .
\end{equation}

Sometimes (but not always) we work with algebras having a unit. This is an element $e \in V$, such that
\begin{equation}\label{aunit}
    \mu_2(a,e) = \mu_2(e,a) = a
\end{equation}
for all $a \in V$. To get the dual relation, we have to do something that may seem akward at first. Instead of thinking of a unit as an element $e \in V$, we define it to be a linear map $\eta: \mathbb{C} \rightarrow V$. By linearity we have $\eta(\lambda) = \lambda\, \eta(1)$ for all $\lambda \in \mathbb{C}$. Therefore, $\eta$ is fully described by the element $e := \eta(1)$. One can check that \eqref{aunit} boils down to
\begin{equation}
    \mu_2 \circ (1 \otimes \eta) =\mu_2 \circ (\eta \otimes 1) = 1 \, .
\end{equation}
Here, strictly speaking the maps act on different spaces. The first map acts on $V \otimes \mathbb{C}$, the second acts on $\mathbb{C} \otimes V$ and the last one acts on $V$. However, all these spaces can be identified using
\begin{equation}\label{natiso1}
    a \otimes \lambda \cong \lambda \otimes a \cong \lambda\, a \, ,
\end{equation}
where $a \in V$ and $\lambda \in \mathbb{C}$. We will always take such identifications as implicit whenever they are necessary.

With this set up, the dual concept to a unit $\eta: \mathbb{C} \rightarrow V$ is a counit $\epsilon: V \rightarrow \mathbb{C}$. It is required to satisfy
\begin{equation}
    (1 \otimes \epsilon) \circ \Delta = (\epsilon \otimes 1) \circ \Delta = 1 \, .
\end{equation}

There are many more axioms for algebras that have a corresponding axiom for coalgebras, like that for an algebra (homo-)morphism or for a derivation. The way to obtain these from the often more familiar definitions appearing in the context of algebras should be more or less clear now. We will give all of them below.

\subsubsection*{General definition}
\vspace{-1.5ex} 
In this section, we will state the axioms of coalgebras given above in a compact form. We will also give further definitions if necessary.

A coalgebra $(C,\Delta,\epsilon)$ is a graded vector space $C$, equipped with a coproduct
\begin{equation}
\Delta: C \longrightarrow C \otimes C
\end{equation}
and a counit
\begin{equation}
\epsilon: C \longrightarrow \mathbb{C} \, ,
\end{equation}
such that $\Delta$ is coassociative and $\epsilon$ is a counit of $\Delta$. This means that the diagrams
\begin{equation*}
\begin{tikzcd}
C \arrow[r,"\Delta"]  \arrow[d,"\Delta"] & C \otimes C \arrow[d,"\Delta \otimes 1"] \\
C \otimes C \arrow[r,"1 \otimes \Delta"] & C \otimes C \otimes C
\end{tikzcd} \ , \qquad
\begin{tikzcd}
C \arrow[r,"\Delta"] \arrow[dr,"\cong"] & C \otimes C \arrow[d,"\epsilon \otimes 1"] \\
& \mathbb{C} \otimes C
\end{tikzcd}\ , \qquad
\begin{tikzcd}
C \arrow[r,"\Delta"] \arrow[dr,"\cong"] & C \otimes C \arrow[d,"1 \otimes \epsilon"] \\
& C \otimes \mathbb C
\end{tikzcd}
\end{equation*}
commute. Here, $1$ denotes the identity on $C$ and $\cong$ stands for the natural isomorphisms given in \eqref{natiso1}.

A morphism $f: (C,\Delta_C, \epsilon_C) \rightarrow (D,\Delta_D, \epsilon_D)$ of coalgebras is a linear map $f: C \rightarrow D$, such that
\begin{equation}
\Delta_D \circ f = (f \otimes f) \circ \Delta_C \, , \quad \epsilon_D \circ f = \epsilon_C \, .
\end{equation}

\subsubsection*{Conilpotent coalgebras}
\vspace{-1.5ex} 

In what follows, we need to restrict to a special class of coalgebras. The first simplification comes by considering coalgebras with a rather trivial counit. Given a coalgebra $C$, we now consider those we can write as $C = \mathbb{C} \oplus \bar C$, where $\bar{C}$ comes equipped with a coassociative coproduct $\overline{\Delta}: \bar C \rightarrow \bar C \otimes \bar C$, so that the full coproduct is given by
\begin{equation}
\Delta(1) = 1 \otimes 1 \, , \quad \Delta(x) = \bar{\Delta}(x) + x \otimes 1 + 1 \otimes x \, ,
\end{equation}
where $x \in \bar C$ and $1 \in \mathbb{C}$. The counit is given by the projection $\epsilon: \mathbb{C} \oplus \bar C \rightarrow \mathbb{C}$. We also have the inclusion $\eta: \mathbb{C} \rightarrow \mathbb{C} \oplus \bar C$, which we call the augmentation map. Coalgebras with such a structure are called augmented. Also, $(\bar{C},\overline{\Delta})$ itself is a coalgebra, however without a counit. Given an augmented coalgebra $C$, we call $\bar{C}$ the reduced coalgebra.

We denote augmented coalgebras by a tuple $(C,\Delta,\epsilon,\eta)$. A morphism $f: (C,\Delta_C,\epsilon_C,\eta_C) \rightarrow (D,\Delta_D,\epsilon_D,\eta_D)$ is a morphism of coalgebras, with the additional property that
\begin{equation}
f \circ \eta_C = \eta_D \, .
\end{equation}

We also want to put a certain finiteness assumption on our coalgebras. Recall that we can think of a coproduct $\Delta$ as a certain split of elements in $C$. We can ask that elements become in some sense smaller under repeated applications of the coproduct. To make this precise, we take the reduced coproduct $\overline \Delta$ and define the $n$-split recursively as
\begin{equation}
\overline \Delta_{n+1} = (1 \otimes \overline \Delta_n) \circ \overline \Delta \ , \quad \overline \Delta_1 = \text{id}_{\bar{C}}\, . 
\end{equation}
Note that by coassociativity, we could have also used $\overline \Delta_n \otimes 1$ instead of $1 \otimes \overline \Delta_n$. The condition we want to put on elements $x \in \bar{C}$ is that $\overline \Delta_{n}(x) = 0$ for large enough $n$. This means that after splitting $x$ enough times using $\overline{\Delta}$, it cannot be split further and therefore gives zero. We define $F_n C := \ker \overline{\Delta}_n$. Note that $\ker \overline \Delta_{n} \subseteq \ker \overline{\Delta}_{n+1}$. Assuming that all $x$ live in some $F_n C$ then means that
\begin{equation}
\overline C = \bigcup_{n \ge 1} F_n C \, .
\end{equation}
Augmented coalgebras with this property are called conilpotent.

\subsubsection*{The tensor coalgebra}
\vspace{-1.5ex} 
For  a graded vector space $V$ one can define the tensor coalgebra
\begin{equation}
T^c(V) := \bigoplus_{k \ge 0} V^{\otimes k} \, .
\end{equation}
We denote the elements in $V^{\otimes n}$ by $x_1\cdots x_n$ with $x_i \in V$. A general element in $T^c(V)$ is just a finite linear combination of such elements. The vector space $T^c(V)$ comes equipped with a coproduct $\Delta: T^c(V) \rightarrow T^c(V) \otimes T^c(V)$ given by 
\begin{equation}
\Delta(x_1 \cdots x_n) := \sum_{k = 0}^n (x_1 \cdots x_{k}) \otimes (x_{k+1} \cdots x_n) \, .
\end{equation}
The first and the last term in the sum are given by $1\otimes x_1 \cdots x_n$ and $x_1 \cdots x_n \otimes 1$, respectively, where $1 \in V^{\otimes 0} \cong \mathbb{C}$. The counit $\epsilon: T^c(V) \rightarrow \mathbb{C}$ is given by the projection onto the zeroth tensor power. The tensor coalgebra is both augmented and conilpotent, with the augmentation map $\eta: \mathbb{C} \rightarrow T^c(V)$ being the inclusion and $F_p T^c(V) = \bigoplus_{k \le p} V^{\otimes k}$. We denote the reduced tensor coalgebra by $\bar T^c(V)$.

The tensor coalgebra has a certain universal property, namely that it is cofree among all conilpotent coalgebras. This means that given any conilpotent coalgebra $(C,\Delta_C,\epsilon_C,\eta_C)$ and a linear map $f: C \rightarrow V$ with $f \circ \eta_C = 0$, there is a unique coalgebra morphism
\begin{equation}
T^c(f): C \longrightarrow T^c(V) \, ,
\end{equation}
such that $\pi_1 \circ T^c(f) = f$. Here, $\pi_1: T^c V \rightarrow V$ is the projection to the first tensor power. The map $T^c(f)$ is constructed as
\begin{equation}\label{CoalgebraMorLift}
T^c(f) = \sum_{n = 0}^{\infty} f^{\otimes n}\circ {{\Delta}_C}_n \, ,
\end{equation}
where 
\begin{equation}
    {{\Delta}_C}_n =({{\Delta}_C}_{n-1} \otimes 1) \circ \Delta \, , \quad {{\Delta}_C}_0 = \epsilon \, , \quad {{\Delta}_C}_1 = \text{id}_C \, .
\end{equation}
Note that the $k$th term in the above sum is the part of $T^c(f)$ that takes values in $V^{\otimes k}$. Conilpotency of $C$ together with $f \circ \eta_C = 0$ guarantees that the sum is actually finite when evaluated on any $x \in C$. 

The universal property of the tensor coalgebras makes it very easy to describe morphisms between them. Below we will only encounter tensor coalgebras. In this case, we know that any morphism $F: T^c(V) \rightarrow T^c(W)$ is equivalent to a linear map $f: T^c(V) \rightarrow W$.

\subsubsection*{Differential graded coalgebras}
\vspace{-1.5ex} 
A differential graded coalgebra is a coalgebra $(C,\Delta,\epsilon)$ equipped with a linear map $M: C \rightarrow C$ of degree one and such that $M^2 = 0$. We further demand that $M$ is a coderivation, i.e. satisfies the co-Leibniz rule
\begin{equation}
(1 \otimes M + M \otimes 1) \circ \Delta = \Delta \circ M \, .
\end{equation}
Coderivations of augmented coalgebras are required to satisfy $M \circ \eta = 0$.

A morphism $f: (C,\Delta_C,\epsilon_C,M) \rightarrow (D,\Delta_D,\epsilon_D,N)$ of differential graded coalgebras with differentials $M$ and $N$ is a morphism of coalgebras, such that
\begin{equation}
f \circ M = N \circ f \, .
\end{equation}

In the case when $C = T^c(V)$, any coderivation $M: T^c(V) \rightarrow T^c(V)$ is uniquely determined by $m :=\pi_1 M: T^c(V) \rightarrow V$. Given such an $m$, we can reconstruct $T^c(m) := M$ via
\begin{equation}
T^c(m) = \nabla_3(1 \otimes m \otimes 1)\Delta_3 \, ,
\end{equation}
where $\nabla_3: T^c(V) \otimes T^c(V) \otimes T^c(V) \rightarrow T^c(V)$ acts as
\begin{equation}
    \nabla_3(x_1\cdots x_k \otimes y_1 \cdots y_l \otimes z_1 \cdots z_m) = x_1\cdots x_k y_1 \cdots y_l z_1 \cdots z_m \, .
\end{equation} 
Note that we can write $m = \sum_{k \ge 1} m_k$, where $m_k: V^{\otimes n} \rightarrow V$ is extended to $T^c(V)$ via $T^c(V) \rightarrow V^{\otimes k}$. The coderivation associated to $m_k$ is then given by
\begin{equation}
T^c(m_k)(x_1\cdots x_n) = \sum_{i = 0}^{n-k} x_1 \cdots x_{i}\, m_k(x_{i+1},...,x_{i+k})\,x_{i+k+1} \cdots x_n \, .
\end{equation}
Using this formula, a general linear map $m: T^c(V) \rightarrow V$ can then be extended by linearity, i.e.
\begin{equation}
T^c(m) = \sum_{k \ge 1} T^c(m_k) \, .
\end{equation} 

Just like derivations of algebras, coderivations are closed under taking commutators, meaning that if $M$ and $N$ is a coderivation, so is $[M,N] = M \circ N -(-)^{MN}N\circ M$. In particular, if $M$ is of degree one, we have that
\begin{equation}
M^2 = \frac{1}{2}[M,M] \, ,
\end{equation}
so $M^2$ is also a coderivation. Now if $M$ is a coderivation on $T^c(V)$, the condition $M^2 = 0$ is equivalent to $\pi_1 \circ M^2 = 0$, since $M^2$ as a coderivation is determined by $\pi_1 M^2$. Writing $m = \sum_{k \ge 1} m_k$, we obtain an infinite set of relations among the $m_k$. The first three are given by
\begin{equation}\label{Firstainfinites}
\begin{split}
m_1^2(x) &= 0 \, , \\
 m_1 m_2(x_1,x_2) &= - m_2 (m_1 (x_1),x_2) - (-)^{x_1} m_2(x_1,m_1 (x_2)) \, , \\
[m_1,m_3](x_1,x_2,x_3) &= -m_2(m_2(x_1,x_2),x_3) - (-)^{x_1}m_2(x_1,m_2(x_2,x_3)) \, ,
\end{split}
\end{equation}
where
\begin{align*}
[m_1,m_3](x_1,x_2,x_3) 	&= m_1m_3(x_1,x_2,x_3) + m_3(m_1(x_1),x_2,x_3) \\
						 & \quad + (-)^{x_1}m_3(x_1,m_1(x_2),x_3) + (-)^{x_1 + x_2} m_3(x_1,x_2,m_1(x_3)) \, .
\end{align*}
Note that, in particular, the first relation  says  that $(V,m_1)$ is a differential graded vector space.

\subsection{Differential graded associative algebras}

\subsubsection*{Associative algebras}
\vspace{-1.5ex} 
An associative algebra $(A,\mu_2)$ consists of a potentially graded $\mathbb{C}$-vector space $A$, together with a degree zero associative product $\mu_2: A \otimes A \rightarrow A$. This means that the following diagram commutes:
\begin{equation*}
\begin{tikzcd}
A\otimes A \otimes A \arrow[d,"1 \otimes \mu_2"] \arrow[r,"\mu_2 \otimes 1"] & A \otimes A \arrow[d,"\mu_2"] \\
A \otimes A \arrow[r,"\mu_2"]& A
\end{tikzcd}
\ , \quad
\end{equation*}
A morphism $\varphi: (A,{\mu_2}) \rightarrow (B,\nu_2)$ of associative algebras is a linear map $\varphi: A \rightarrow B$ of degree zero, such that
\begin{equation}
\nu_2 \circ (\varphi \otimes \varphi) = \varphi \circ \mu_2 \, . 
\end{equation}

We consider associative algebras without a unit. The reason is that the bar construction explained below works only either for non-unital algebras or for augmented algebras, which is the concept dual to augmented coalgebras. Augmented algebras are equivalent to non-unital algebras in every respect. 
This is because any augmented algebra can be obtained from a non-unital algebra by adding a single vector $e$ in degree zero and defining
\begin{equation}
\mu_2(a,e) = \mu_2(e,a) = a \, .
\end{equation}
For this reason, and in order not to overcomplicate things, we will always work with non-unital algebras.

\subsubsection*{Differential graded algebras}
\vspace{-1.5ex} 
A differential graded algebra $(A,\text d,\mu_2)$ is a an algebra $(A,\mu_2)$, such that $\text d: A \rightarrow A$ is a linear map of degree one with the property that $\text d^2 = 0$. We further demand that it satisfies the Leibniz rule with respect to $\mu_2$, i.e.
\begin{equation}
\text d \mu_2(a,b) = \mu_2 (\text d a,b) + (-)^a \mu_2(a, \text d b) \, .
\end{equation}
A morphism $\varphi: (A,\text d_A,\mu_2) \rightarrow (B,\text d_B,\nu_2)$ of differential graded algebras is required to be a chain map, i.e.
\begin{equation}
\varphi\circ \text d_A = \text d_B \circ \varphi \, .
\end{equation}

We recall that the cohomology $H(V)$ of a differential graded vector space $(V,\text d)$ is defined to be the quotient
\begin{equation}
H(V) = \frac{\ker \text d}{\text{Im}\, \text d} \, .
\end{equation}
In case of a differential graded algebra, the Leibniz rule implies that the product $\mu_2$ restricts to an associative product
\begin{equation}
H(\mu_2): H(A) \otimes H(A) \rightarrow H(A) \, . 
\end{equation}
Similarly, the chain map property of an algebra morphism $\varphi: (A,\text d_A,\mu_2) \rightarrow (B,\text d_B,\nu_2)$ says that it descends to an algebra morphism
\begin{equation}
H(\varphi): H(A) \rightarrow H(B)
\end{equation}
on cohomology. We also have $H(\varphi_2 \circ \varphi_1) = H(\varphi_2) \circ H(\varphi_1)$.

When working over chain complexes, we take  two differential algebras to be equivalent if an algebra morphism $\varphi: (A,\text d_A,\mu_2) \rightarrow (B,\text d_B,\nu_2)$ becomes an isomorphism on cohomology. In this case, we call $\varphi$ a \emph{quasi-isomorphism}, and we say that two 
differential graded algebras $(A,\text d_A,\mu_2) $ and $ (B,\text d_B,\nu_2)$ are quasi-isomorphic 
if there is either a quasi-isomorphism from $A$ to $B$ or from $B$ to $A$.
Note that we explicitly require that such a morphism only has to exist in one direction. This is because even if there is a quasi-isomorphism $\varphi: (A,\text d_A,\mu_2) \rightarrow (B,\text d_B,\nu_2)$, there is in general no quasi-isomorphism in the other direction, although it necessarily exists as an isomorphism at the level of cohomology. In order to have a good equivalence class of differential graded algebras, we therefore require that it only exists in one direction. Requiring it to exist in both directions would be  too restrictive. Therefore, we call the differential graded algebras $A$ and $B$ equivalent, if there is a \emph{zig-zag} of quasi-isomorphisms
\begin{equation}\label{zigzag}
    A = A_0 \leftarrow A_1 \rightarrow A_2 \leftarrow ... \rightarrow A_n = B \, ,
\end{equation}
where each arrow stands for some quasi-isomorphism.

Apart from non-invertibility of quasi-isomorphisms, there is another issue with differential graded algebras. In general, the differential graded algebras $(A,\text d,\mu_2)$ and $(H(A),0,H(\mu_2))$ are not quasi-isomorphic. Even worse, there is in general no sequence of quasi-isomorphisms of the form \eqref{zigzag} that connects the two. Therefore, when going to cohomology, some information about the equivalence of a differential graded algebra is lost. The solution to this problem lies in the concept of homotopy algebras, where we also allow for higher products $\mu_k: A^{\otimes k} \rightarrow A$. In this framework, even when starting from a differential graded algebra, its cohomology, while still defining an associative algebra,  also comes equipped with higher maps $H(\mu_k): H(A)^{\otimes k} \rightarrow A$. This homotopy algebra is called the \emph{minimal model} of $A$. The higher products contain all the information that would be lost when considering $H(A)$ as a plain associative algebra. Indeed, one can show that two differential graded algebras are equivalent, possibly via a sequence of quasi-isomorphisms as in \eqref{zigzag}, if and only if they have isomorphic minimal models.

When homotopy algebras describe perturbative field theories, for example as $L_\infty, A_\infty$ or $ C_\infty$ algebras, the story boils down to the fact that the physical information we are interested in is contained in the scattering amplitudes. Even if the theory only has a cubic interaction, in general we have $n$-point amplitudes for arbitrary  $n$. The maps $H(\mu_n)$ of the minimal model are exactly  the scattering amplitudes, although in their fully gauge fixed form without the residual on-shell gauge freedom considered in this work. Obviously, physical information about the theory will be lost if we only consider the three-point amplitude, given in the algebraic language by $H(\mu_2)$.

We will introduce homotopy associative algebras by first giving the bar construction for strict associative algebras. Here one describes a differential graded algebra equivalently as a differential graded coalgebra. When formulated in this way, the generalization to homotopy algebras is almost automatic.

\subsubsection*{The bar construction}

The bar construction associates to a differential graded algebra $(A,\text d,\mu)$ a certain differential graded coalgebra $\pazocal B(A)$. One  first performs 
a shift in degree, denoted by $A \rightarrow s A$. The degree of the $k$ component $(sA)^k$ of $s A$ is given by
\begin{equation}
(sA)^{k} = A^{k+1} \, , 
\end{equation} 
i.e. the degree of each element gets lowered by one. We then define
\begin{equation}
m_1(a) := -\text d a \,, \qquad m_2(a,b) = (-)^{a+1}\mu_2(a,b)  \, ,
\end{equation}
where on the right hand side we use the degree of $s A$ in the exponent. Obviously we still have $m_1^2 = 0$. The Leibniz rule and associativity then induce the following relations on $m_1$ and $m_2$:
\begin{equation}
\begin{split}
m_1 m_2(a,b) &= - m_2(m_1(a),b) - (-)^a m_2(a,m_1(b)) \, , \\  0 &= m_2(m_2(a,b),c) + (-)^a m_2(a,m_2(b,c)) \, .
\end{split}
\end{equation}

We observe that the above relations are exactly the ones we found in \eqref{Firstainfinites} in the case $m_3 = 0$. Therefore, we see that $T^c(m_1 + m_2)$ defines a coderivation on the coalgebra $T^c(s A)$. This is called the bar construction for differential graded algebras. It associates to $(A,\text d,\mu)$ the coalgebra $\pazocal B(A) = (T^c(sA),T^c(m_1 + m_2),\Delta,\epsilon)$. Further, a morphism $\varphi (A,\text d_A,\mu_2) \rightarrow (B,\text d_B,\nu_2)$ of differential graded algebras defines a degree zero linear map 
\begin{equation}
\begin{split}
f: sA \longrightarrow sB \,, \qquad 
	a \longmapsto \varphi(a) \, ,  
\end{split}
\end{equation}
which, when extended as $f\circ \pi_1: T^c(s A) \rightarrow sB$, defines a morphism
\begin{equation}
\pazocal{B}(\varphi) := T^c(f\circ \pi_1): \pazocal{B}(A) \rightarrow \pazocal{B}(B)
\end{equation}
using the lift \eqref{CoalgebraMorLift} on $f\circ \pi_1$. By the property of the tensor coalgebra, it is automatically a morphism of coalgebras. On the other hand, the fact that it commutes with the differential $M:= T^c(m_1 + m_2)$ uses both
\begin{equation}
\varphi \circ \text d_A = \text d_B \circ \varphi \, , \quad \varphi\circ \mu_2 = \nu_2\circ(\varphi\otimes \varphi)\, .
\end{equation}

To summarize, the bar construction associates to a (non-unital) differential graded algebra $A$ the conilpotent differential graded coalgebra $\pazocal{B}(A)$, and to any morphism $\varphi$ of differential graded algebras a morphism $\pazocal{B}(\varphi)$ of conilpotent differential graded coalgebras.\footnote{In mathematics, an object like $\pazocal{B}(-)$ is called a functor. A functor further satisfies $\pazocal{B}(\text{id}) = \text{id}$ and $\pazocal{B}(f \circ g) = \pazocal{B}(f) \circ \pazocal{B}(g)$. It is easy to check that $\pazocal{B}(-)$ has these properties.}

\subsection{Homotopy associative algebras}

\subsubsection*{Motivation and first definition}
\vspace{-1.5ex} 
Homotopy associative algebras are obtained from differential graded algebras by relaxing associativity \emph{up to homotopy}. To check associativity of a differential graded algebra $(A,\text d, \mu_2)$ we need to compute the associator
\begin{equation}\label{associator}
\text{Ass}(a,b,c) = \mu_2(\mu_2(a,b),c) - \mu_2(a,\mu_2(b,c)) \, .
\end{equation}
Note that the associator is a degree zero map $\text{Ass}: A^{\otimes 3} \rightarrow A$ of differential graded algebras. We ask for $\text{Ass}$ to be homotopy trivial, 
i.e.~that it can be written as
\begin{equation}\label{end3diff}
\text{Ass} = - \partial \mu_3 := - \text d \circ \mu_3 - \mu_3 \circ \text{d}_{A^{\otimes 3}} 
\end{equation}
for some degree $-1$ map $\mu_3: A^{\otimes 3} \rightarrow A$. Here, the differential on tensor products is given by
\begin{equation}
\text{d}_{A^{\otimes n}} = \sum_{k = 1}^n 1^{\otimes (k-1)} \otimes \text d \otimes 1^{\otimes (n-k)} \, .
\end{equation}
We already encountered \eqref{end3diff} in \eqref{Cinfty relations} with a different, but equivalent sign rule. The sign rule made here will be justified once we introduce the bar construction for homotopy algebras. 
The fact that $\text{Ass}$ is homotopy trivial implies that $\mu_2$ becomes associative on cohomology, since the map $\partial \mu_3$ is trivial on cohomology.

Having introduced $\mu_3$, new relations among $\mu_2$ and $\mu_3$ appear. One can check that
\begin{equation}\label{Higherhomotopy}
\mu_2 (\mu_3 \otimes 1) + \mu_3 (1 \otimes \mu_2 \otimes 1) + \mu_2 (1 \otimes \mu_3) - \mu_3(\mu_2 \otimes 1 \otimes 1) - \mu_3( 1 \otimes 1 \otimes \mu_2) \, ,
\end{equation}
as a map $A^{\otimes 4} \rightarrow A$, is $\partial$-closed. In an $A_\infty$ algebra, we require that this is actually $\partial$-exact. Therefore, it is equal, up to a sign, to $\partial \mu_4$, where $\mu_4: A^{\otimes 4} \rightarrow A$ is a degree $-2$ map and
\begin{equation}\label{end4diff}
\partial \mu_4 = \text d \circ \mu_4 - \mu_4 \circ \text d_{A^{\otimes 4}} \, .
\end{equation}

Let us elaborate on this observation. Note that the spaces of multilinear maps
\begin{equation}
\text{End}_n^\bullet(A) := \text{Hom}(A^{\otimes n},A)
\end{equation}
can be endowed with the differential
\begin{equation}
\partial{M} := \text d_A \circ M - (-)^M M \circ \text d_{A^{\otimes n}} \, ,
\end{equation}
which generalizes the differentials in \eqref{end3diff} and \eqref{end4diff}. The bullet in $\text{End}_n^\bullet(A)$ labels the degree of the $n$-linear map.
We have $\mu_2 \in \text{End}^0_2(A)$. The compositions $\mu_2\circ(\mu_2 \otimes 1) $ and $\mu_2\circ(1 \otimes \mu_2)$ live in $\text{End}^0_3(A)$. The fact that these are equal up to homotopy is described by $\mu_3 \in \text{End}^{-1}_3(A)$. Pictorially, we draw this as
\begin{equation}
\mu_2 \circ (\mu_2 \otimes 1) \longrightarrow \mu_2 \circ (1 \otimes \mu_2) \, ,
\end{equation}
where the arrow means that there is a homotopy (path) from one to the other.

We can also take more compositions of the $\mu_2$. If we take three of the $\mu_2$, there are in total five combinations. Let us write $ab := \mu_2(a,b)$. The five combinations are
\begin{equation}\label{pentagonvertices}
((ab)c)d \ , \quad (a(bc))d \ , \quad a((bc)d) \ , \quad a(b(cd)) \ , \quad (ab)(cd) \, .
\end{equation}
All of them define elements in $\text{End}_4^0(A)$. The fact that we have the homotopy $\mu_3$ implies that all of the objects in \eqref{pentagonvertices} are homotopic to each other. The homotopies can be constructed by combining $\mu_3$ with $\mu_2$, producing elements in $\text{End}_4^{-1}(A)$. But there is something new happening here. There are actually two distinct ways to show that each pair in \eqref{pentagonvertices} are homotopic to each other. Take for example the cases $((ab)c)d$ and $a(b(cd))$. We have
\begin{equation}\label{associahedra}
\begin{tikzcd}
((ab)c)d \arrow[r] \arrow[dr] & (a(bc))d \arrow[r] & a((bc)d) \arrow[d] \\
& (ab)(cd)  \arrow[r] & a(b(cd)) \, .
\end{tikzcd}
\end{equation}
Each arrow represents a movement of a single bracket $(ab)c \rightarrow a(bc)$. Let $h_1$ and $h_2$ be two homotopies connecting $((ab)c)d$ with $a(b(cd))$, i.e.
\begin{equation}
\partial h_1(a,b,c,d) = \partial h_2(a,b,c,d) = a(b(cd)) - ((ab)c)d \, .
\end{equation}
It follows that
\begin{equation}
\partial (h_1 - h_2) = 0 \, .
\end{equation}
It turns out that $h_1 - h_2$ is given by \eqref{Higherhomotopy}. This is the origin of this relation. The product $\mu_4 \in \text{End}^{-2}_4(A)$ can therefore be interpreted as a \emph{second order} homotopy between the homotopies $h_1$ and $h_2$. 

The upshot is that the homotopy $\mu_3$ makes all ways to compose the $\mu_2$ to get maps in $\text{End}_3^0(A)$ and $\text{End}_4^0(A)$ equivalent up to homotopy. It is easy to convince oneself that this continues to higher orders, i.e. no matter how we compose the $\mu_2$ to get an element in $\text{End}_n^0(A)$, all of them will be equivalent up to homotopy. In the same fashion, there are multiple ways to identify two elements in $\text{End}_n^0(A)$, described by distinct homotopies $h_i \in \text{End}_n^{-1}(A)$. The differences $h_i - h_j$ are then $\partial$-closed. This closedness is made exact by compositions of $\mu_4$ with an appropriate number of $\mu_2$ and/or by a composition of two $\mu_3$ with an appropriate number of $\mu_2$ (in general, such a homotopy will be a superposition of some using one $\mu_4$ and some using two $\mu_3$). The upshot is that all closed elements in $\text{End}_n^{-1}(A)$ can be made exact with second order homotopies in $\text{End}^{-2}_n(A)$.

Each introduction of a $\mu_n \in \text{End}_n^{2-n}(A)$ gives rise to new $\partial$-closed terms. These terms then are required to be exact with the help of a new product $\mu_{n+1}$. In this way, an infinite set $\{\mu_n\}_{n \ge 2}$ arises.

The relations among the $\{\mu_k\}_{k \ge 2}$ also has a nice geometric interpretation in terms of the Stasheff polytopes or associahedra \cite{Stasheff1963HomotopyAO,LodayPolytopes}. To each independent element in $\text{End}_k^{-l}(A)$ coming from compositions of $\{\mu_{k}\}_{k \ge 2}$ one associates a polytope (which are the analog of polygons in arbitrary dimensions) of dimension $l$.
The operation $\partial$ then computes the boundary of that polytope. The exactness condition then boils down to the fact that all closed surfaces made from the polytopes are the boundary of another polytope, and therefore can be contracted to a point. This contraction encodes the fact that all ways to compose the $\mu_2$ are equivalent up to homotopies, and that all homotopies and their higher analogs are equivalent. The diagram in \eqref{associahedra} represents the situation in two dimensions, where the polytope is a pentagon.

We are now ready to give a definition of $A_\infty$-algebras. A homotopy associative $(A_\infty)$ algebra $(A,\text d,\{\mu_k\}_{k \ge 2})$ consists of differential graded vector space $(A,\text d)$ and a set of degree $2-k$ multilinear maps $\mu_k: A^{\otimes k} \rightarrow A$, such that
\begin{itemize}
\item $\text{d}$ satisfies Leibniz rule with respect to $\mu_2$.
\item All ways to multiply $n$ elements with $\mu_2$ are homotopic in $\text{End}_n^0(A)$, with homotopy $\mu_3$. In other words, $\mu_2$ is associative up to homotopy $\mu_3$.
\item All $\partial$-closed relations made out of the $\{\mu_k\}_{k \ge 2}$ in $\text{End}_n^{k}(A)$ are $\partial$-exact for negative $k$.
\end{itemize}

We will not prove that this definition actually makes sense. For example, it is a non-trivial fact that it is sufficient to have a single map $\mu_k: A^{\otimes k} \rightarrow A$ for each integer $k$ to make all $\partial$-closed relations exact. It is also a drawback that this definition is inductive. Having all relations involving up to some $\mu_k$, we can find the relation involving $\mu_{k+1}$ by looking for $\partial$-closed relations in degree $2-k$. On the other hand, it gives a very clear picture of what an $A_\infty$ algebra is and how it arises from differential graded associative algebras. By requiring the condition of associativity to hold only up to homotopy, we get a seed for the relation involving $\mu_3$ as a homotopy. All higher homotopies are found from that by requiring that closed relations are exact. 

Below we will give a compact, non-inductive definition of $A_\infty$ algebras by generalizing the bar construction. But before doing so, we first have to discuss 
morphisms of $A_\infty$ algebras.

The conditions on morphisms of $A_\infty$ algebras are described very similarly to the conditions on the products $\mu_k$. Recall that a morphism $\varphi_1$ of differential graded algebras $\varphi_1 : (A,\text d_A,\mu_2) \rightarrow (B,\text d_B, \nu_2)$ is required to be both a chain map and an algebra morphism. We will not touch the chain map condition, since we want to preserve the notion of differential graded vector spaces. On the other hand, we will relax the condition that $\varphi_1$ is an algebra morphism. Like with associativity of $\mu_2$, we want $\varphi_1$ to be an algebra morphism up to some homotopy $\varphi_2$. Therefore, we require that
\begin{equation}\label{moruptohom}
\varphi_1 \circ \mu_2 - \nu_2 \circ (\varphi_1 \otimes \varphi_1) = -\partial \varphi_2 \, .
\end{equation}
Here, $\varphi_2: A\otimes A \rightarrow B$ is a degree $-1$ map and
\begin{equation}
\partial \varphi_2 = \text d_B \circ \varphi_2 + \varphi_2 \circ \text d_{A^{\otimes 2}}
\end{equation}
is the canonical differential on this map.

From the experience we have gathered so far, we now know how to continue. The introduction of $\varphi_2$ will make certain combinations of $\varphi_2$ with the product $\mu_2$ $\partial$-closed. We require it to be $\partial$-exact, hence we introduce another $\varphi_3: A^{\otimes 3} \rightarrow B$ of degree $-2$ as a homotopy. In the most general case, this continues forever and therefore we obtain a whole family of $\varphi_k: A^{\otimes k} \rightarrow B$ of degree $1-k$, ensuring that all $\partial$-closed relations in negative degree are exact.

Morphisms of differential $A_\infty$ algebras are called quasi-isomorphisms, if the linear part $\varphi_1: A \rightarrow B$ is a quasi-isomorphism of differential graded vector spaces. A crucial property of $A_\infty$-algebras is that given a quasi-isomorphism $\{\varphi_k\}_{k \ge 1}$ from $A$ to $B$, there is always a quasi-isomorphism $\{\varsigma_k\}_{k \ge 1}$ from $B$ to $A$. We explained that this is not true for differential graded algebras. The existence of a quasi-isomorphism $\varphi_1: (A,\text d_A,\mu_2) \rightarrow (B,\text d_B,\nu_2)$ does not guarantee that there is a quasi-isomorphism $\varsigma_1: (B,\text d_B,\nu_2) \rightarrow (A,\text d_A,\mu_2)$ in the other direction. Only if we allow ourselves to think of them as certain $A_\infty$-algebras, $\varphi_1$ has a quasi-inverse $\{\varsigma_k\}_{k \ge 1}$. Allowing $\varsigma_1$ to be an algebra morphism only up to homotopy $\varsigma_2$ makes all the difference. 

The quasi-invertibility of quasi-isomorphisms allows us to call $ A_\infty$ algebras equivalent, if there is a quasi-isomorphism between them. We no longer need to rely on zig-zags of quasi-isomorphisms as in the case of differential graded algebras. Every quasi-isomorphisms in \eqref{zigzag} has a quasi-inverse and since compositions of quasi-isomorphisms are again quasi-isomorphisms (the composition of morphisms of $A_\infty$ algebras will be defined below when introducing the bar construction for $A_\infty)$, there is always a single quasi-isomorphism connecting any equivalent $A_\infty$ algebras.

\subsubsection*{The bar construction for $ A_\infty$ algebras}
\vspace{-1.5ex} 
Given a differential graded algebra $A$, we saw that there is a description in terms of a differential graded coalgebra $\pazocal{B}(A)$. This construction extends to $A_\infty$-algebras. 

The bar construction of $A_\infty$ algebras should include the bar construction of differential graded algebras. From this it is clear that we want to work on the coalgebra $T^c(sA)$. The differential $\text d : A \rightarrow A$ and products $\mu_{k}: A^{\otimes k} \rightarrow A$ induce maps $m_k : (s A)^k \rightarrow s A$ of degree one by defining
\begin{equation}
m_1(a) = - \text d a \, , \qquad m_k(a_1,...,a_k) = (-)^{k + \sum_{j = 1}^{k}(k-j) (a_j + 1)} \mu_k(a_1,...,a_k) \, ,
\end{equation}
where the degree in the exponent on the right hand side is with respect to $s A$. The choice of signs here is not unique. We follow the conventions given in \cite{MappingCone}. The maps $\{m_k\}_{ k\ge 1}$
combine into a degree one map $m = \sum_{k \ge 1} m_k: T^c(s A) \rightarrow s A$ and therefore can be lifted to a coderivation $Q := T^c(m)$. It can be shown that the $A_\infty$ relations among $\text d$ and the $\mu_k$ are equivalent to $Q^2 = 0$. Since we never explicitly stated the $A_\infty$ relations, we can take this also as a definition. An $A_\infty$ algebra is equivalent to the differential graded coalgebra $T^c(s A)$ with differential $Q$. We denote this coalgebra by $\pazocal{B}(A)$. As in the case of differential graded algebras, we will still call it the bar construction.

For morphisms, we can give a similar construction. Let $\varphi_k: A^{\otimes k} \rightarrow B$ be a family of degree $1-k$ maps such that they define a morphism of $A_\infty$ algebras. These give a family of degree zero maps $f_k: (s A)^{\otimes k} \rightarrow s B$ via
\begin{equation}
f_k(a_1,...,a_k) = (-)^{\sum_{j = 1}^k (k-j)(a_j+1)}\varphi_k(a_1,...,a_k) \, .
\end{equation}
The degree in the exponent are with respect to $sA$. The $f_k$ then combine into a degree zero map $f = \sum_{k \ge 1} f_k: T^c(s A) \rightarrow s B$ which in turn gives a coalgebra morphism $F := T^c(f): T^c(s A) \rightarrow T^c(s B)$. The fact that the $\varphi_k$ are morphisms of $A_\infty$ algebras then implies that $F$ commutes with the differentials obtained from the bar construction. Since we did not state the conditions on the $\varphi_k$ explicitly in the previous section, we can also take this as a definition. In this language, compositions of $A_\infty$-algebra morphisms are given by their composition as coalgebra morphisms. 

The upshot of this section is that $A_\infty$ algebras can be fully described by a differential $Q$ on the coalgebra $T^c(sA)$. Further, this is the most general way to obtain such a differential, any such  differential describes an $A_\infty$ algebra. These two notions are therefore equivalent. Similarly, morphisms of $A_\infty$ algebras are equivalent to morphisms of the tensor coalgebra, equipped with a coderivation.

\subsubsection*{Homotopy Transfer}
\vspace{-1.5ex}
Here, we discuss the homotopy transfer of $A_\infty$ algebras in full generality. The homotopy transfer theorem \cite{vallette2012algebrahomotopyoperad,AlgebraicOperads} states that given an $A_\infty$ algebra $(A,\text d_A, \{\mu_k\}_{k \ge 2})$ together with a chain complex $(B,\text d_B)$ that is homotopy equivalent to $(A,\text d_A)$ as a chain complex, there is an $A_\infty$ structure on $B$ quasi-isomorphic to that on $A$.

We recall that, given chain complexes $(A,\text d_A)$ and $(B,\text d_B)$, a chain map $\pi_1: (A,\text d_A) \rightarrow (B,\text d_B)$ is a homotopy equivalence, if there is another chain map $\iota_1: (B,\text d_B) \rightarrow (A,\text d_A)$ together with a degree minus one map $h: (A,\text d_A) \rightarrow (A, \text d_A)$ so that 
\begin{equation}\label{genhomotopy}
\iota_1 \pi_1 - 1 = \text d_A h + h \text d_A\;. 
\end{equation}
This relation implies that $\iota_1 \pi_1$ is the identity on cohomology. Similarly, one demands that $\pi_1 \iota_1$ is the identity on cohomology. Therefore,  $\iota_1$ and $\pi_1$ are inverse to each other on cohomology. The datum $(\iota_1,\pi_1,h)$ defines the homotopy equivalence. Note that this generalizes the assumption made in \eqref{homotopy relation}, where we demanded that $\pi_1 \iota_1 = 1$ on the full chain complex.

Often there are more conditions on $\iota_1$ and $\pi_1$, so-called side conditions, which we do not want to impose right now. The above is sufficient for what follows. We now assume that we have an $A_\infty$-algebra $(A,\text d, \{\mu_k\}_{k \ge 2})$ together with another chain complex $(B,\text d_B)$ and a homotopy equivalence $(\pi_1,\iota_1,h)$ from $(A,\text d_A)$ to $(B,\text d_B)$. The homotopy transfer theorem then states that there is another $A_\infty$ structure on $(B,\text d_B)$ with maps $\{\nu_k\}_{k \ge 2}$, such that the $A_\infty$ algebras $(A,\text d_A,\{\mu_k\}_{k \ge 2})$ and $(B,\text d_B,\{\nu_k\}_{k \ge 2})$ are quasi-isomorphic. Further, there is a formula for the quasi-isomorphism 
\begin{equation}
\iota: (B,\text d_B,\{\nu_k\}_{k \ge 2}) \longrightarrow (A,\text d_A,\{\mu_k\}_{k \ge 2})
\end{equation}
given by a family of maps $\{\iota_k\}_{k \ge 1}$, such that $\iota_1$ is the chain map that is part of the homotopy equivalence.

The formulas for the $\{\nu_k\}_{k \ge 2}$ and $\{\iota_k\}_{k \ge 1}$ are best stated in terms of rooted and planar tree diagrams. We first introduce the primitive rooted trees $P_k$ for $k \ge 1$:
\begin{equation}
P_k = \begin{tikzpicture}[baseline=(current bounding box.center), line width = 1]
\draw 	(-.5,.5) -- (0,0) -- (0,-.60) 
		(.5,.5) -- (0,0)
		(-.3,.5) -- (0,0);
\path 	(-.3,.5) -- (.5,.5) node[midway] {$\ldots$}
		(-.5,.65) -- (.5,.65) node[midway,above] {\scriptsize $k$ times};
\draw 	[decorate, decoration = {brace, raise=3}] (-.5,.5) --  (.5,.5);
\end{tikzpicture} \ .
\end{equation}
In particular, $P_1$ is depicted as a straight line:
\begin{equation}
P_1 = \quad  \begin{tikzpicture}[baseline=(current bounding box.center), line width = 1]
\draw 	(0,.5) -- (0,0) -- (0,-.5);
\end{tikzpicture} \ .
\end{equation}
The primitive rooted trees can be composed. For example, we can form a tree with five leaves composing a $P_4$ and a $P_2$:
\begin{equation}
\begin{tikzpicture}[baseline=(current bounding box.center), line width = 1]
\draw (.25,.5) -- (0,0) -- (-.25,.5) (0,0) -- (.25,-.5) -- (.5,0) (.25,-.5) -- (1,0) (.25,-1.15) -- (.25,-.5) -- (-.5,0); \ .
\end{tikzpicture}
\end{equation}
The primitive tree $P_1$ serves as an identity for this operation. Any tree $T$ composed with $P_1$ in any fashion gives back $T$. In the same fashion, 
any trees, not necessarily primitive, can be composed. In this way, any rooted tree with an arbitrary number of vertices can be inductively built from primitive trees.

We consider the spaces of all formal superpositions of rooted trees with $n$ leaves, which we call $\mathbb{Z}[\pazocal{T}_n]$. This means that trees $T_1$ and $T_2$ can be formally added. Furthermore, we demand that the composition of trees is distributive, in the sense that
\begin{equation}
\begin{tikzpicture}[baseline=(current bounding box.center), line width = 1]
\node[rectangle,draw] (1) at (0,0) {\scriptsize $A + B$};
\draw	(1) -- (-.5,.65) (1) -- (.5,.65) (1) -- (0,-1) -- (1,0) (0,-1) -- (1.8,0) (0,-1) -- (-1,0) (0,-2) -- (0,-1) -- (-1.8,0);
\path	(-.5,.65) -- (.5,.65) node[midway] {$\ldots$}
		(1,0) -- (1.8,0) node[midway] {$\hspace{-1mm}\ldots$}
		(-1,0) -- (-1.8,0) node[midway] {$\hspace{1mm}\ldots$};
\end{tikzpicture}
\quad =
\quad
\begin{tikzpicture}[baseline=(current bounding box.center), line width = 1]
\node[rectangle,draw] (1) at (0,0) {\scriptsize $A$};
\draw	(1) -- (-.5,.65) (1) -- (.5,.65) (1) -- (0,-1) -- (.5,0) (0,-1) -- (1.3,0) (0,-1) -- (-.5,0) (0,-2) -- (0,-1) -- (-1.3,0);
\path	(-.5,.65) -- (.5,.65) node[midway] {$\ldots$}
		(.5,0) -- (1.3,0) node[midway] {$\hspace{-1mm}\ldots$}
		(-.5,0) -- (-1.3,0) node[midway] {$\hspace{1mm}\ldots$};
\end{tikzpicture}
\quad + \quad
\begin{tikzpicture}[baseline=(current bounding box.center), line width = 1]
\node[rectangle,draw] (1) at (0,0) {\scriptsize $B$};
\draw	(1) -- (-.5,.65) (1) -- (.5,.65) (1) -- (0,-1) -- (.5,0) (0,-1) -- (1.3,0) (0,-1) -- (-.5,0) (0,-2) -- (0,-1) -- (-1.3,0);
\path	(-.5,.65) -- (.5,.65) node[midway] {$\ldots$}
		(.5,0) -- (1.3,0) node[midway] {$\hspace{-1mm}\ldots$}
		(-.5,0) -- (-1.3,0) node[midway] {$\hspace{1mm}\ldots$};
\end{tikzpicture} \ .
\end{equation}
The box with letter $A$ or $B$ is a placeholder for any superposition of trees, not necessarily primitive. For example, we could have
\begin{equation}
\begin{tikzpicture}[baseline=(current bounding box.center), line width = 1]
\node[rectangle,draw] (1) at (0,0) {\scriptsize $A$};
\draw (1) -- (0,.7) (1) -- (.5,.7) (1) -- (1,.7) (1) -- (-1,.7) (1) -- (-.5,.7) (1) -- (0,-.8);
\end{tikzpicture}
\quad
=
\quad
\begin{tikzpicture}[baseline=(current bounding box.center), line width = 1]
\draw (.25,.5) -- (0,0) -- (-.25,.5) (0,0) -- (.25,-.5) -- (.5,0) (.25,-.5) -- (1,0) (.25,-1.15) -- (.25,-.5) -- (-.5,0); \ .
\end{tikzpicture}
\end{equation}

Of particular interest for us is the sum of all rooted trees with fixed number of leaves. The sum of trees with $n$ leaves is denoted by $\Sigma_n$. They can be built inductively as follows. For $n = 1$, we set $\Sigma_1 = P_1$. The higher $\Sigma_n$ can then be inductively built as
\begin{equation}\label{Treesum}
\Sigma_n = \sum_{k = 2}^n \sum_{[i_1,...,i_k | n]}
\begin{tikzpicture}[baseline=(current bounding box.center), line width = 1]
\node[rectangle,draw] (1) at (-1.25,.75) {\scriptsize $\Sigma_{i_1}$};
\node[rectangle,draw] (2) at (-.25,.75) {\scriptsize $\Sigma_{i_2}$};
\node[rectangle,draw] (3) at (1.25,.75) {\scriptsize $\Sigma_{i_k}$};
\draw	(1) -- (0,-.4) -- (0,-1) (2) -- (0,-.4) (3) -- (0,-.4)
		(1) -- (-1.65,1.4) (1) -- (-.85,1.4)
		(2) -- (-.65,1.4) (2) -- (.15,1.4)
		(3) -- (.85,1.4) (3) -- (1.65,1.4);
\path	(2) -- (3) node[midway] {$\ldots$}
		(-1.65,1.4) -- (-.85,1.4) node[midway] {$\ldots$}
		(-.65,1.4) -- (.15,1.4) node[midway] {$\ldots$}
		(.85,1.4) -- (1.65,1.4) node[midway] {$\ldots$};
\end{tikzpicture} \ ,
\end{equation}
where $[i_1,...,i_k|n]$ is the set of all tuples $(i_1,...,i_k)$ of positive integers, such that $\sum_{j}i_j = n$. This sum ensures that $\Sigma_n$ consists only of superpositions of trees with $n$ leaves.

The tree sums $\Sigma_n$ can be used to describe both $\nu_n$ and $\iota_n$. To avoid any complications with signs, we consider the maps on the suspended complexes $sA$ and $s B$. We denote the maps on $sA$, resp. $sB$ by $\{m_k\}_{k \ge 1}$ (resp. $\{n_k\}_{k \ge 1}$) and the morphism by $\{i_k\}_{k \ge 1}$. We also write $p_1$ for the suspended projection $\pi_1$. Given any tree $T$, we associate to a vertex $P_k$ with $k \ge 2$ the product map $m_k$, for each internal line connecting two vertices we associate the homotopy $h$ and to each leaf we associate the map $\iota_1$. These maps are composed as indicated by the tree $T$. In this way, we obtain a linear map
\begin{equation}
\begin{split}
    \pazocal{C}:\mathbb{Z}[\pazocal{T}_n] &\longrightarrow \text{End}_n(B,A) \, , \\
    T &\longmapsto \pazocal C(T) \, ,
\end{split}
\end{equation}
where $\text{End}_n(B,A)$ is the space of linear maps $B^{\otimes n} \rightarrow A$. For example,
\begin{equation}
\begin{tikzpicture}[baseline=(current bounding box.center), line width = 1]
\draw (.25,.5) -- (0,0) -- (-.25,.5) (0,0) -- (.25,-.5) -- (.5,0) (.25,-.5) -- (1,0) (.25,-1.15) -- (.25,-.5) -- (-.5,0); \ .
\end{tikzpicture}
 \quad \longmapsto \quad \mu_4 (\iota_1 \otimes h\mu_2(\iota_1 \otimes \iota_1) \otimes \iota_1 \otimes \iota_1) \, .
\end{equation}
Using this map, we have
\begin{equation}
n_k = p_1 \circ \pazocal{C}(\Sigma_k) \, , \quad i_k = h \circ \pazocal{C}(\Sigma_k) \, .
\end{equation}

The maps $f_k = \pazocal{C}(\Sigma_k)$ define a generalization of the Berends-Giele currents \eqref{Berends-Giele} for any theory that can be described in terms of $A_\infty$ or $C_\infty$ algebras. In general, at the level of maps it can be inductively defined as
\begin{equation}\label{Treerecursion}
f_n = \sum_{k = 2}^{n} \sum_{[i_1,...,i_k|n]} m_n(i_{i_1},...,i_{i_k}) \, ,
\end{equation}
which immediately follows from the recursive formula \eqref{Treesum} and the definition of $\pazocal{C}: \mathbb{Z}[\pazocal{T}] \rightarrow \text{End}_n(B,A)$.

\subsubsection*{Proof of the homotopy transfer theorem}

In this section, we will prove the homotopy transfer theorem using the recursive formula. We will also show how to apply this proof to the $L_\infty$ case, where it is essentially the same. We are well aware that homotopy transfer theorem is well established in the language of operads \cite{AlgebraicOperads}, especially in the case of the $A_\infty$ and $L_\infty$ operad, and therefore immediately translates to $A_\infty$ and $L_\infty$ algebras as representations of these operads.  For illustration, we nevertheless think it is fruitful to rederive this theorem directly for the algebras itself, in particular to establish that the formulas \eqref{ainfinityinduction} and \eqref{linfinityinduction} are correct.

The following proof relies on the formulation of homotopy algebras in terms of differential graded coalgebras. We will also give a fundamental proof in the appendix for the $L_\infty$ case. The two proofs illustrate the power of the coalgebraic formulation, since using this language, the proof is much shorter.

Recall that for $A_\infty$-algebras $(A,\text d_A, \{\mu_k\}_{k \ge 2})$ and $(B, \text d_B, \{\nu_k\}_{k \ge 2})$, a morphism is a degree zero linear map $f: T^c(sA) \rightarrow sB$, such that its lift $T^c(f)$ commutes with the coderivation. We write $f = \sum_{k \ge 1} f_k$, where $f_k: (sA)^{\otimes k} \rightarrow sB$ is its $k$-linear part. Given two morphisms $f$ and $g$, their composition in components is given by
\begin{equation}
(g \circ f)_n = \sum_{k = 1}^n \sum_{[i_1,...,i_k|n]} g_k \circ(f_{i_1} \otimes \cdots \otimes f_{i_k}) \, . 
\end{equation}
This formula is derived by considering $g \circ T^c(f) = \pi_1 \circ T^c(g) \circ T^c(f)$ and then restricting to $n$ inputs.

The above formula shows that the recursive formula given in \eqref{Treerecursion} can be compactly written as
\begin{equation}\label{ainfinityinduction}
f = m_{\ge 2} \circ T^c(i) \, , \quad i = i_1 + h \circ  f \, , \quad n = n_1 + p_1 \circ f \, ,
\end{equation}
where $m_{\ge 2} = \sum_{k \ge 2} m_k$. The fact that this works as an inductive definition relies on the fact that $m_{\ge 2}$ starts at quadratic order. This ensures that, when evaluated on some $x_1 \cdots x_n \in A^{\otimes n}$, $f$ depends only on the $i_k$ for $k < n$. Therefore, also each $i_n$ is defined recursively in terms of the $i_k$ with $k < n$. 

In order to prove the homotopy transfer theorem, we need to show two things. First, $i: T^c(sA) \rightarrow sB$ is an $A_\infty$ morphism from $(A,\text d_A,\{\mu_k\}_{k \ge 2})$ to $(B,\text d_B,\{\nu\}_{k \ge 2})$, i.e. that it satisfies
\begin{equation}
T^c(i) \circ T^c(n) = T^c(m) \circ T^c(i) \, .
\end{equation}
This is equivalent to showing that
\begin{equation}\label{iisachainmap}
i \circ T^c(n) = m \circ T^c(i) \, .
\end{equation}
Second, we need to show that $(B,\text d_B,\{\nu\}_{k \ge 2})$ is an actual $A_\infty$ algebra. This means that we need to show that
\begin{equation}
T^c(n) \circ T^c(n) = 0 \, ,
\end{equation}
which is equivalent to
\begin{equation}
n \circ T^c(n) = 0 \, .
\end{equation}

\paragraph*{$i$ is an $A_\infty$ morphism:}

We prove the chain map condition \eqref{iisachainmap} by induction over the number of inputs. In other words, the induction step assumes that we proved \eqref{iisachainmap} on $\bigoplus_{k < n}{B}^{\otimes n}$ and then prove from there that it is true on elements in $B^{\otimes n}$.

Let us assume that we have a homotopy equivalence data, i.e. that there are chain maps $i_1: (sB,n_1) \rightarrow (sA,m_1)$, $p_1: (sA,m_1) \rightarrow (sB,n_1)$ satsifying
\begin{equation}
h m_1 + m_1 h = i_1\circ p_1 - 1 \, ,
\end{equation} 
where $h$ is some map $h: sA \rightarrow sA$ of degree $-1$. We begin by proving \eqref{iisachainmap} on $x \in V$. In this case, we have
\begin{equation}
m \circ T^c(i)(x) = m_1 \circ i(x) = i \circ n_1(x) = i \circ T^c(n)(x) \, .
\end{equation}
Here we used the assumption that $i_1$ is a chain map. 

Having proved the initial case, now lets assume that $i = \sum_{k \ge 1} i_k$ satisfies \eqref{iisachainmap} up to $n-1$ inputs. Therefore, we assume that
\begin{equation}
i \circ T^c(n) = m \circ T^c(i) + r_n \, , 
\end{equation}
where $r_n: T^c(sB) \rightarrow sA$ is some linear map of degree one vanishing on $\bigoplus_{k < n} (sB)^{\otimes k}$. It follows that also the lift 
\begin{equation}
T^c(r_n) := T^c(i) \circ T^c(n) - T^c(m) \circ T^c(i) 
\end{equation} 
vanishes on $k < n$ inputs. Also, we necessarily have that
\begin{equation}
T^c(r_n)(x_1,...,x_n) \in sA \, ,    
\end{equation}
i.e. there are no terms in $\bigoplus_{k \ge 2} (sA)^{\otimes k}$. With this set up, we find
\begin{equation}\label{chainmap}
\begin{split}
i \circ T^c(n) &= i_1 \circ T^c(n) + h \circ m_{\ge 2} \circ T^c(i) \circ T^c(n) \\
&= i_1 \circ n + h \circ m_{\ge 2} \circ T^c(m) \circ T^c(i)
+ h \circ m_{\ge 2} \circ T^c(r_n)\, .
\end{split}
\end{equation}
When evaluated on elements $(x_1,...,x_n) \in V^{\otimes n}$, we find that
\begin{equation}
T^c(r_n)(x_1,...,x_n) = r_n(x_1,...,x_n) \in sA \, .
\end{equation}
Since the output does not take values in $\bigoplus_{k \ge 2} (sA)^{\otimes k}$, the composition $m_{\ge 2} \circ T^c(r_n)$ vanishes when evaluated on $V^{\otimes n}$. From now on, we assume that \eqref{chainmap} is evaluated on $V^{\otimes n}$, so that we can set $m_{\ge 2} \circ T^c(r_n) = 0$ in \eqref{chainmap}.

The appearance of $m_{\ge 2} \circ T^c(m)$ in $\eqref{chainmap}$ suggests to use the $A_\infty$ relations. We have
\begin{equation}
0 = m \circ T^c(m) = (m_1 + m_{\ge 2}) \circ T^c(m)
\end{equation}
and therefore $m_{\ge 2} \circ T^c(m) = - m_1 \circ T^c(m) = -m_1 \circ m_{\ge 2}$. With this, we can continue the computation in \eqref{chainmap}.
\begin{equation}
\begin{split}
i_1 \circ n + h \circ m_{\ge 2} \circ T^c(m) \circ T^c(i) = i_1 \circ m - h \circ m_1 \circ m_{\ge 2} \circ T^c(i) \\
 = i_1 \circ m + (m_1 \circ h - i_1\circ p_1 + 1) \circ m_{\ge 2} \circ T^c(i) \\
 = i_1 \circ (n - p \circ m_{\ge 2} \circ T^c(i)) + m_1 \circ h \circ m_{\ge 2} \circ T^c(i) + m_{\ge 2} \circ T^c(i) \\
 = i_1 \circ n_1 + m_1 \circ i_{\ge 2} + m_{\ge 2} \circ T^c(i) = m_1 \circ (i + i_{\ge 2}) + m_{\ge 2} \circ T^c(i) \\
 = m_1 \circ i + m_{\ge 2} \circ T^c(i) = m \circ T^c(i) \, .
\end{split}
\end{equation}
Here, $i_{\ge 2} = \sum_{k \ge 2} i_k$. This proves the chain map property $i \circ T^c(n) = m \circ T^c(i)$ on $V^{\otimes n}$. This completes the induction.

\paragraph*{Transferred $A_\infty$ relations:} We also need to show that the transferred brackets define an $A_\infty$ algebra. In this case, we don't need to do any induction. We simply need to use that $T^c(i)\circ T^c(n) = T^c(m) \circ T^c(i)$, which we just proved. Further, we again use the $A_\infty$ relations of the $m_k$ in the form $m_{\ge 2} \circ T^c(m) = - m_1 \circ T^c(m) = -m_1 \circ m_{\ge 2}$. We have
\begin{equation}
\begin{split}
n \circ T^c(n) &= n_1 \circ n_{\ge 2} + n_{\ge 2} \circ T^c(\bar n) = n_1 \circ p_1 \circ b_{\ge 2} \circ T^c(i) + n_{\ge 2} \circ T^c(n) \\
&= p_1 \circ m_1 \circ m_{\ge 2} \circ T^c(i) + n_{\ge 2} \circ T^c(n) \\
&= - p_1 \circ m_{\ge 2} \circ T^c(m) \circ T^c(i) + n_{\ge 2} \circ T^c(n) \\
&= -p_1 \circ m_{\ge 2} \circ T^c(i) \circ T^c(n) + n_{\ge 2} \circ T^c(n) \\
&= - n_{\ge 2} \circ T^c(n)  + n_{\ge 2} \circ T^c(n) \\ 
&= 0 \, .
\end{split}
\end{equation}
This proves that $(B,\text d_B,\{\nu_k\}_{k \ge 2})$ defines an $A_\infty$ algebra. Assuming that $i_1$ is a quasi-isomorphism then implies that $(B,\text d_B,\{\nu_k\}_{k \ge 2})$ is quasi-isomorphic to $(A,\text d_A,\{\mu_k\}_{k \ge 2})$.

\paragraph{The $L_\infty$ case:} As advertised in the beginning of this section, the proof in the $L_\infty$ case works essentially the same. We recall that for $L_\infty$ algebras, we have to work with the symmetric coalgebra $S^c(sA)$, where, for any vector space $V$, we have
\begin{equation}
S^c(V) = \bigoplus_{n \ge 0} V^{\wedge n} \, .
\end{equation}
The coproduct $\Delta: S^c(V) \rightarrow S^c(V)$ is defined as
\begin{equation}
\Delta(x_1 \cdots x_n) = \sum_{k = 0}^n\sum_{\sigma \in \overline{\text{Sh}}(k,n)} (-)^{\epsilon(\sigma,x_1,...,x_n)} x_{\sigma(1)} \cdots x_{\sigma(k)} \otimes x_{\sigma(k+1)} \cdots x_{\sigma(n)} \, .
\end{equation}
The sum runs over all permutations $\sigma$ of $\{1,...,n\}$ elements, such that $\sigma(1) < ... < \sigma(k)$ and $\sigma(k+1) < ... < \sigma(n)$.

A degree zero linear map $f: S^c(V) \rightarrow W$ can be lifted to a coalgebra morphism $S^c(f)$ using the formula
\begin{equation}\label{SymmetricLift}
S^c(f) = \sum_{n \ge 0} \frac{1}{k!}  f^{\wedge n} \circ \Delta^n \, ,
\end{equation}
where $\Delta^n: S^c(V) \rightarrow S^c(V)^{\otimes n}$ is the iterated coproduct and $f^{\wedge n}: S^c(V)^{\otimes n} \rightarrow W^{\wedge n}$ is the $n$th tensor power $f^{\otimes n}$, composed with the quotient map $W^{\otimes n} \rightarrow W^{\wedge n}$, given by
\begin{equation}
x_1 \otimes \cdots \otimes x_n \longmapsto x_1 \wedge \cdots \wedge x_n \, .
\end{equation}
Note that\eqref{SymmetricLift} uses the same formula as in \eqref{CoalgebraMorLift}, up to the  symmetry factor $\frac{1}{k!}$. A linear map $b: S^c(V) \rightarrow V$ lifts to a coderivation $S^c(b)$ using the formula
\begin{equation}
S^c(b) = (1 \wedge b) \circ \Delta \, ,
\end{equation}
where $1 \wedge b$ is $1 \otimes b$, followed by the quotient $V \otimes S^c(V) \rightarrow V \wedge S^c(V) \subseteq S^c(V)$.

The proof of the homotopy transfer theorem in the $L_\infty$ case follows along the same line as in the $A_\infty$ case. One simply has to replace $T^c(-) \rightarrow S^c(-)$. In particular, we have that 
\begin{equation}\label{linfinityinduction}
f = b_{\ge 2} \circ S^c(i) \, .
\end{equation}

In the appendix, we will give a more explicit proof of the fact that the transferred $L_\infty$ brackets satisfy the $L_\infty$ relation. This proof does not rely on the formulation of $L_\infty$ algebras as coalgebras (which is the bar construction). The length of that proof in comparison to the one presented above really illustrates the power of the bar construction. In order to compare the two formulas, we will give a formula for the composition $b_{\ge 2} \circ S^c(i)$ when evaluated on $n$ vectors. To get rid of all symmetry factors, we make the following definition. Given $n$ vectors $x_1,...,x_n \in V$ and a set $I = \{i_1,...,i_k\} \subseteq \{1,...,n\} =: [n]$ with $i_j < i_{j+1}$ for $j = 1,...,k-1$, we define
\begin{equation}
x_I = x_{i_1} \cdots x_{i_k} \, .
\end{equation}
Further, we call a set $\{I_1,...,I_k\}$ of disjoint subsets of $[n]$ a $k$-partition of $[n]$, if
\begin{equation}
I_1 \sqcup ... \sqcup I_k = [n] \, .
\end{equation} 
We denote by $P_k(n)$ the set of all $k$ partitions of $n$.  With this formula, we find
\begin{equation}\label{NoOvercounting}
b_k \circ S^c(i)(x_1,...,x_n) = \sum_{k = 1}^n \sum_{\{I_1,...,I_k\} \in P_k(n)} \pm b_k(i_{|I_1|}(x_{I_1}),...,i_{|I_n|}(x_{I_k}))\, ,
\end{equation}
where $|I_j|$ is the number of elements of $I_j$. The sign is the usual Koszul sign obtained from reordering $x_1\cdots x_n$ to $x_{I_1}\cdots x_{I_k}$.

\subsection{Homotopy commutative algebras}

\subsubsection*{Shuffle product}
\vspace{-1.5ex}
Recall that, given a graded vector space $V$, there is the tensor coalgebra $T^c(V)$. We want to give $T^c(V)$ the structure of a bialgebra. A bialgebra can be thought of a coalgebra $(C,\Delta,\epsilon)$ together with an associative product $\nabla: C \otimes C \rightarrow C$ and unit $\eta: \mathbb{C} \rightarrow C$, with the additional property that $\eta$ and $\nabla$ are coalgebra morphisms. Here, the coalgebra structure on $C \otimes C$ is given by $\Delta_{T^c(V)\otimes T^c(V)} = (1 \otimes \tau \otimes 1)(\Delta\otimes \Delta$) with
\begin{equation}\label{NatIso}
\begin{split}
\tau: C \otimes C &\longrightarrow C \otimes C \, , \\
     a \otimes b &\longmapsto (-)^{ab} b \otimes a \, ,
\end{split}
\end{equation}
while the coalgebra structure on $\mathbb{C}$ is simply the isomorphism $\mathbb{C} \cong \mathbb{C} \otimes \mathbb{C}$.

To construct a compatible product on $T^c(V)$, we can use the fact that any coalgebra morphism $\nabla: T^c(V) \otimes T^c(V) \rightarrow T^c(V)$ is uniquely determined by its image in $V$. We set
\begin{equation}\label{Shufflegenerator}
\begin{split}
p: V \otimes \mathbb{C} \oplus \mathbb{C} \otimes V &\longrightarrow V \, , \\
v \otimes a + b \otimes w &\longmapsto av + b w \, .
\end{split}
\end{equation}
The space $ V \otimes \mathbb{C} \oplus \mathbb{C} \otimes V$ is identified as a subset of $T^c(V) \otimes T^c(V)$. We can then extend $p$ to $T^c(V) \otimes T^c(V)$ by zero on everything else. $p$ can then be lifted into a morphism into all of $T^c(V)$ as
\begin{equation}
\nabla = \sum_{n \ge 0} p^{\otimes n}\circ\Delta^n_{T^c(V) \otimes T^c(V)} \, .
\end{equation}
We denote by $\nabla_n$ the image in $V^{\otimes n}$ given by
\begin{equation}\label{nshuffle}
\nabla_n = p^{\otimes n} (1 \otimes \Delta^{n-1}_{T^c(V) \otimes T^c(V)})\circ \Delta_{T^c(V) \otimes T^c(V)} = (p \otimes \nabla_{n-1})\circ \Delta_{T^c(V) \otimes T^c(V)}\,. 
\end{equation}
This gives an inductive definition for $\nabla$. We compute $\nabla_n$  on $x_1\cdots x_k \otimes x_{k+1} \cdots x_n$ for $k$ arbitrary, where it is equal to $\nabla$. We first compute
\begin{equation}
\begin{split}
\Delta_{T^c(V) \otimes T^c(V)}(x_1\cdots x_k,x_{k+1}\cdots x_n) &= (-)^{x_{k+1}(x_1 + ...+ x_k)} 1 \otimes x_{k+1} \otimes x_1 \cdots x_k \otimes x_{k+2} \cdots x_n \\
&\quad + x_1 \otimes 1 \otimes x_2 \cdots x_{k} \otimes x_{k+1} \cdots x_n + ... \, ,
\end{split}
\end{equation}
where $...$ denotes all terms that do not contribute to \eqref{nshuffle}. These are all terms that are not of the form $(\mathbb{C} \otimes V \oplus V \otimes \mathbb{C}) \otimes ...$ and vanish when evaluated on $p \otimes ... \ $. We therefore find
\begin{equation}\label{InductiveShuffle}
\begin{split}
\nabla(x_1\cdots x_k,x_{k+1}\cdots x_n) &= x_1 \nabla(x_2\cdots x_k,x_{k+1}\cdots x_n) \\ & \quad + (-)^{x_{k+1}(x_1 +\ldots+ x_k)}x_{k+1}\nabla(x_1\cdots x_k,x_{k+2}\cdots x_n)\;. 
\end{split}
\end{equation}

The equation \eqref{InductiveShuffle} defines the shuffle product. The name stems from the fact that it literally shuffles its inputs: 
When $x_1 \cdots x_k$ and $x_{k+1} \cdots x_n$ are thought of as stacks of cards, $\nabla(x_1\cdots x_k,x_{k+1} \cdots x_n)$ is the sum of all ways to shuffle the two stacks into each other, without changing the order of the cards in each individual stack. We have the following formula
\begin{equation}
\nabla(x_1\cdots x_k,x_{k+1} \cdots x_n) = \sum_{\sigma \in {Sh}(k,n-k)} (-)^{\epsilon(\sigma,x_1,...,x_n)} x_{\sigma^{-1}(1)} \cdots x_{\sigma^{-1}(n)} \, ,
\end{equation}
where the sum runs over all permutations, such that $\sigma(1) < ... < \sigma(k)$ and $\sigma(k+1) < ... < \sigma(n)$. The sign is the natural one coming from the degrees of the $x_i$.

An important fact about the shuffle product is that it is commutative. This follows from the fact that its corestriction to $V \subseteq T^c(V)$ given by \eqref{Shufflegenerator} already is commutative in its inputs. Since lifts to coalgebra morphisms into $T^c(V)$ are unique, the lift $\nabla: T^c(V) \otimes T^c(V) \rightarrow T^c(V)$ of the linear map $p: T^c(V) \otimes T^c(V) \rightarrow V$ is also commutative.

\subsubsection*{Homotopy Commutative Algebras}
\vspace{-1.5ex}
A subset of associative algebras are the commutative (and associative) algebras. Because of that we could in principle just think of them as embedded in that framework when describing their homotopical version. However, in doing so we would lose some information about the homotopical properties of commutative algebras.

That said, a homotopy commutative algebra is a homotopy associative algebra $(A,\text{d},\mu_{k \ge 2})$ with some additional structure: We demand that the maps $\mu_n$ vanish on signed shuffles:
\begin{equation}
\sum_{\sigma \in {Sh}(k,n-k)} (-)^{\sigma + \epsilon(\sigma,x_1,...,x_n)} x_{\sigma^{-1}(1)} \cdots x_{\sigma^{-1}(n)} = 0 \, ,
\end{equation}
where $1 \le k \le n-1$. For example, $\mu_2$ satisfies
\begin{equation}
0 = \mu_2(a \otimes b - (-)^{ab} b \otimes a) = \mu_2(a,b) - (-)^{ab}\mu_2(b,a) \, .
\end{equation}
Therefore, $\mu_2$ is commutative. For $\mu_3$, there is also only a single relation due to commutativity of the shuffle product, which reads
\begin{equation}
\mu_3(a,b,c) - (-)^{ab} \mu_3(b,a,c) + (-)^{a(b+c)}\mu_3(b,c,a) = 0 \, .
\end{equation}

As for homotopy associative algebras, there is also a bar construction in the commutative case. We first consider the reduced tensor coalgebra
\begin{equation}
\bar{T}^c(sA) = \bigoplus_{k \ge 1} (sA)^{\otimes k} \, . 
\end{equation}
The reduced coproduct $\bar \Delta: \bar{T}^c(sA)\rightarrow \bar{T}^c(sA) \otimes \bar{T}^c(sA)$ induces a co-Lie bracket
\begin{equation}
\beta: \bar{T}^c(sA) \rightarrow \bar{T}^c(sA) \otimes \bar{T}^c(sA) 
\end{equation}
defined by $\beta = (1 - \tau)\bar{\Delta}$. The fact that the shuffle product is compatible with the  coproduct then can be used to show that $\beta$ reduces to a co-Lie bracket on
\begin{equation}
L^c(sA) := \bar T^c(sA)/\nabla(\bar T^c(sA),\bar T^c(sA)) \, ,
\end{equation} 
i.e. the quotient of the reduced tensor coalgebra with the elements given by shuffles. The $C_\infty$ products $\mu_n: A^{\otimes n} \rightarrow A$ then induce a coderivation of $\beta$ using the standard lift to coderivations on $T^c(sA)$ and then projecting to $L^c(sA)$. The fact that the result is well defined follows from \cite{CHENG20082535}. For this it is necessary that the $m_n: (sA)^{\otimes n} \rightarrow sA$ derived from the $\mu_n$ vanish on (unsigned) shuffles.

Given a graded vector space $V$, the co-Lie algebra $L^c(V)$ is the cofree co-Lie algebra over $V$, that is the morphisms into $L^c(V)$ have a similar lifting property with respect to all co-Lie algebras. The upshot is that the bar construction associates to a $C_\infty$ algebra $A$ the differential graded co-Lie algebra $L^c(sA)$. This is the opposite of what happens in the case of $L_\infty$ algebras, where the bar construction associates to an $L_\infty$ algebra $A$ a differential graded cocommutative coalgebra $S^c(sA)$.

$C_\infty$ algebras also have a homotopy transfer. The formula is the same as for $A_\infty$ algebras. It was shown in \cite{CHENG20082535} that the transferred morphism and products all vanish on shuffles, so that the homotopy transfer of a $C_\infty$ algebra yields again a $C_\infty$ algebra. In field theories, when the homotopy transfer is applied to Yang-Mills theory, the transferred maps $i_k$ perturbatively construct solutions. We saw in the discussion of the homotopy transfer theorem that these can be written as $i_k = h \circ f_k$, where the $f_k$ are the Berends-Giele currents. The fact that they vanish on shuffles is known as the Kleiss-Kuijf relations. These relations hold for any $C_\infty$ algebra, not only for Yang-Mills theory, where the $\mu_k$ are zero for $k \ge 4$.

\section{Conclusions and outlook}

In this paper we have spelled out and generalized the relation in terms of homotopy transfer between the cyclic $L_{\infty}$ algebra encoding a field theory 
and its tree-level scattering amplitudes. So far this relationship has been discussed in the literature exclusively 
as homotopy transfer to the minimal model given by the cohomology itself. Since in the minimal model  the differential trivializes this is insufficient in order 
to derive  such properties of scattering amplitudes as Ward identities, which express the invariance under linear gauge transformations 
of the polarization  vectors. Instead, here  we have defined the homotopy transfer to a larger space, with harmonic fields and gauge parameters, 
on which there is still a non-trivial 
differential and a non-trivial $L_{\infty}$ algebra governing the `on-shell' gauge structure encoded in Ward identities. 
To this end we worked on the space of finite sums of plane-wave solutions, for which in particular all homotopy maps are well-defined. 
We believe that this formulation properly captures, and thereby  axiomatizes,  the Feynman-diagrammatic techniques 
of passing from a Lagrangian to its (tree-level) scattering amplitudes. In particular, the Feynman rules are directly encoded in the 
multilinear $L_{\infty}$ brackets of the original theory, and the resulting scattering amplitudes are given by the homotopy transferred $L_{\infty}$ brackets. 
Furthermore, the framework of homotopy  algebras yields a  direct method for stripping off color factors in 
Yang-Mills theory, leading to a formulation in terms of  $C_{\infty}$ algebras in which, for instance, the so-called Kleiss-Kuijf
relations are built-in. 
Performing homotopy transfer  at the level of the $C_{\infty}$ algebra 
then yields directly the so-called color-ordered scattering amplitudes together with their Ward identities.

It will be important to generalize these results in various directions. The main motivation for this work was the investigation of 
color-kinematics duality in Yang-Mills theory on which the double copy construction of gravity is based 
and which suggests that there is a hidden `kinematic' Lie algebra. 
This duality was first observed `experimentally' by the study  of scattering amplitudes \cite{Bern:2008qj,Bern:2010yg} and later proved at tree-level 
by indirect methods from string theory \cite{Bjerrum-Bohr:2010pnr,Mafra:2011kj}. The question then 
arises how to manifest color-kinematics duality directly at the level of the original Yang-Mills  Lagrangian. 
For this it is important to understand the relationship between a Lagrangian and its (tree-level) scattering amplitudes algebraically, 
which is given by the  homotopy transfer of cyclic $L_{\infty}$ algebras or, in the color-ordered sector, by $C_{\infty}$ algebras.
Following the terminology of \cite{Bonezzi:2023ciu} we may refer to this as the `surface structure' because it is essentially an algebraic 
reformulation of familiar structures  in QFT, but in order to express and expose color-kinematics duality directly at the level of the original 
off-shell fields there needs to be hidden `deep structure'. Important progress was made in recent years, beginning with 
a proposal by Reiterer for a vast hidden algebra on the kinematic space of Yang-Mills theory in four Euclidean dimensions \cite{Reiterer:2019dys} and later 
work on general Yang-Mills theories \cite{Borsten:2022vtg,Bonezzi:2022bse,Bonezzi:2023pox,Borsten:2023reb,Borsten:2023ned}. 
In addition to the maps $m_2$ and $m_3$,  the $b$-operator discussed in sec.~3.2 plays a crucial role here, 
since $b$ can be viewed as a second differential that obeys $b^2=0$ but does not act via the Leibniz rule on $m_2$. 
Rather, the failure to do so defines a Lie-type bracket, up to homotopy and up to `$\B$ failures', leading to a generalization 
of a Batalin-Vilkovisky (BV) algebra. It remains to display this algebra to all orders in fields and to prove it for Yang-Mills theory. 
We hope that the homotopy transfer relation between a Lagrangian and its scattering amplitudes, 
for which a wealth of results exists thanks to the amplitude community, will facilitate this program.

From the point of view of finding a satisfying  axiomatization of even just perturbative QFT, undoubtedly the most important 
open problem is to extend these results to loop level. There is a notion of quantum or loop $L_{\infty}$ algebra \cite{Zwiebach:1992ie,Markl:1997bj}, 
and quite a few papers have used  them in order to treat loop amplitudes \cite{Doubek:2017naz,Jurco:2019yfd,Okawa:2022sjf,Konosu:2023pal}. However, the definition of loop $L_{\infty}$ algebras 
is subtle in the infinite-dimensional context of field theory and so such treatments often stay formal. 
Alternatively, one may pass over to the dual setting of functions (or rather functionals) on the space defining the homotopy algebra, 
on which one has a BV algebra. Based on earlier work in \cite{Gwilliam:PhD}, it was shown for quantum mechanics that the computation 
of quantum expectation values can be done using homological perturbation theory \cite{Chiaffrino:2021pob}. It is important  to generalize these results to the computation of higher-loop 
scattering amplitudes in genuine QFTs. We plan to return to these problems.  

\subsection*{Acknowledgements} 

We would like to thank Owen Gwilliam, Silvia Nagy, Jan Plefka, Ivo Sachs, Bruno Vallette and Barton Zwiebach for discussions and collaborations on 
related topics. CC thanks LMU Munich for hospitality during the final stage of this work. This work  is supported by the European Research Council (ERC) under the European Union's Horizon 2020 research and innovation program (grant agreement No 771862). The work of R.B.~is funded by the Deutsche Forschungsgemeinschaft (DFG, German Research Foundation)–Projektnummer 524744955. 
F.D.~is supported by the Deutsche Forschungsgemeinschaft (DFG, German Research Foundation) - Projektnummer 417533893/GRK2575 ``Rethinking Quantum Field Theory". 

\appendix

\section{Recursive formula for homotopy transfer of $L_\infty$ algebras}\label{Sec:Appendix Loo}

Here we give an explicit proof of the homotopy transfer theorem for $L_\infty$ algebras. We do so by providing a recursive formula for the transferred brackets, and showing that they obey the generalized Jacobi identities.

We start from a generic $L_\infty$ algebra $(\cX,\{B_n\})$, where all brackets have intrinsic degree $|B_n|=+1$ and are graded symmetric, namely obeying
\begin{equation}
B_n(\cdots,x_i,x_j,\cdots)=(-1)^{x_ix_j}B_n(\cdots,x_j,x_i,\cdots)\;,\quad x_i,x_j\in\cX\;,   
\end{equation}
under exchange of adjacent inputs. Every symbol in phase factors is meant as the degree of the corresponding object.
Throughout this section, all inputs of multilinear maps are implicitly meant to be graded symmetrized with strength one, which we remind with the symbol $\{12\cdots n\}$ on equal signs.
For instance, we denote
\begin{equation}
\begin{split}
B_2\big(B_2(x_1,x_2),x_3\big)&\stackrel{\{123\}}{=} \frac13\,\Big(B_2\big(B_2(x_1,x_2),x_3\big)\\
&\hspace{5mm}+(-1)^{x_1(x_2+x_3)}B_2\big(B_2(x_2,x_3),x_1\big)+(-1)^{x_3(x_1+x_2)}B_2\big(B_2(x_3,x_1),x_2\big)\Big)\;.   
\end{split}    
\end{equation}
This notation will allow us to deal explicitly with arbitrary numbers of inputs. The set of brackets $\{B_n\}$ obey the generalized Jacobi identities, which read
\begin{equation}\label{GenGenJac appendix}
\sum_{l=1}^n\binom{n}{l}\,B_{n+1-l}\big(B_l(x_1,\ldots,x_l),x_{l+1},\ldots,x_n\big)\stackrel{\{12\cdots n\}}{=}0 \;,  
\end{equation}
acting on $n$ inputs, where $\binom{n}{l}$ is the binomial coefficient.

We now consider a second chain complex $(\bar \cX,\bar B_1)$, together with projection and inclusion maps $p$ and $\iota$ of degree zero
obeying
\begin{equation}
\begin{split}
 p&:\cX\rightarrow\bar\cX\;,\quad \iota:\bar\cX\rightarrow\cX \;,\quad p\,\iota=1_{\bar\cX}\;,
\\
p\, B_1&=\bar B_1\, p\;,\quad \iota\,\bar B_1=B_1\, \iota\;. 
\end{split}    
\end{equation}
We further demand that there is a degree $-1$ homotopy $h:\cX\rightarrow\cX$ obeying
\begin{equation}\label{homotopyrel app}
B_1h+hB_1=\iota\, p-1_\cX\;.    
\end{equation}
Given this linear structure, we will now construct a set of multilinear brackets $\{\bar B_n\}$ on $\bar\cX$ recursively.

To this end, we define multilinear maps
\begin{equation}
\begin{split}
\Psi_n&:\bar\cX^{\otimes n}\rightarrow\cX \;,\quad |\Psi_n|=0\;,\quad n\geq1\;,\\ 
\cJ_n&:\bar\cX^{\otimes n}\rightarrow\cX \;,\quad |\cJ_n|=1\;,\quad n\geq2\;,\\ 
\end{split}    
\end{equation}
 by means of the following recursive relation:
\begin{equation}\label{Jpsi defined}
\begin{split}
\Psi_1(\bar x)&:=\iota(\bar x)\;,\quad \Psi_n(\bar x_1,\ldots,\bar x_n):=h\,\cJ_n(\bar x_1,\ldots,\bar x_n)\;,\quad n\geq2\;,\\
\cJ_n(\bar x_1,\ldots,\bar x_n)&\stackrel{\{12\cdots n\}}{:=}\sum_{k=2}^n\frac{1}{k!}\,\sum_{i_1+\cdots+i_k=n}B_k\big(\Psi_{i_1}(\bar x_1,\ldots,\bar x_{i_1}),\ldots,\Psi_{i_k}(\bar x_{n-i_k+1},\ldots,\bar x_{n})\big)\;.
\end{split}    
\end{equation}
Given these definitions, we claim that the transferred brackets $\bar B_1=p\, B_1\, \iota$ and
\begin{equation}
\bar B_n(\bar x_1,\ldots\bar x_n)=n!\,p\, \cJ_n(\bar x_1,\ldots\bar x_n)\;,\quad n\geq2  \;,
\end{equation}
where $p$ is the projector, obey the generalized Jacobi identities \eqref{GenGenJac appendix}. We will prove this by showing that the generalized currents $\cJ_n$ obey
\begin{equation}\label{RelationJs}
B_1\cJ_n(\bar x_1,\ldots,\bar x_n)+n\,\cJ_n(\bar B_1\bar x_1,\ldots,\bar x_n)+\!\!\sum_{k+l=n+1}k\,\cJ_k\big(p\, \cJ_l(\bar x_1,\ldots,\bar x_l),\bar x_{l+1},\ldots,\bar x_n\big)\stackrel{\{12\cdots n\}}{=}0\;.    
\end{equation}
If \eqref{RelationJs} holds the $\bar B_n$ indeed satisfy
\begin{equation}
\begin{split}
&\bar B_1\bar B_n(\bar x_1,\ldots,\bar x_n)+n\,\bar B_n(\bar B_1\bar x_1,\ldots,\bar x_n)\stackrel{\{12\cdots n\}}{=}n!\,p\, \Big(B_1\cJ_n(\bar x_1,\ldots,\bar x_n)+n\,\cJ_n(\bar B_1\bar x_1,\ldots,\bar x_n)\Big)\\
&=-n!\,p\, \Big(\sum_{k+l=n+1}k\,\cJ_k\big(p\, \cJ_l(\bar x_1,\ldots,\bar x_l),\bar x_{l+1},\ldots,\bar x_n\big)\Big)\\
&=-\sum_{l=2}^{n-1}\binom{n}{l}\,\bar B_{n+1-l}\big(\bar B_l(\bar x_1,\ldots,\bar x_l),\bar x_{l+1},\ldots,\bar x_n\big)\;,
\end{split}    
\end{equation}
for $n\geq2$, which is the identity \eqref{GenGenJac appendix}. 
We will prove \eqref{RelationJs} by induction.

In order to lighten the formulas, we will use a shorthand notation in which we omit the symmetrized inputs and denote by juxtaposition the nesting of maps from the left. For instance, we will write
\begin{equation}
\begin{split}
B_{n+1-l}B_l&\stackrel{\{12\cdots n\}}{:=}B_{n+1-l}\big(B_l(\bar x_1,\ldots,\bar x_l),\bar x_{l+1},\ldots,\bar x_n\big)\;,\\
B_{k}(\Psi_{i_1},\ldots,\Psi_{i_k})&\stackrel{\{12\cdots n\}}{:=}B_{k}\big(\Psi_{i_i}(\bar x_1,\ldots,\bar x_{i_i}),\ldots,\Psi_{i_k}(\bar x_{n+1-i_k},\ldots,\bar x_n)\big)\;,
\end{split}    
\end{equation}
and so on.
With this notation, the relation \eqref{RelationJs} we want to prove takes the form
\begin{equation}\label{RelationJs inputfree}
B_1\cJ_n+n\,\cJ_n\bar B_1+\sum_{l=2}^{n-1}(n+1-l)\,\cJ_{n+1-l}\,p\, \cJ_l=0\;,\quad n\geq2\;.  \end{equation}
Before starting with the inductive proof, let us derive a similar relation for the maps $\Psi_n$. Since $\Psi_1=\iota$ is a chain map, it obeys $B_1\Psi_1-\Psi_1\bar B_1=0 $.
Provided $\cJ_n$ obeys \eqref{RelationJs inputfree}, for $n\geq2$ one has
\begin{equation}\label{RelationPsis inputfree}
B_1\Psi_n-n\,\Psi_n\bar B_1=\sum_{l=2}^n(n+1-l)\,\Psi_{n+1-l}\,p\, \cJ_l-\cJ_n\;,  \end{equation}
which descends from $\Psi_n=h\,\cJ_n$ and the homotopy relation \eqref{homotopyrel app}. For future convenience, it is useful to give the sums appearing in \eqref{RelationJs inputfree} and \eqref{RelationPsis inputfree} as constrained sums:
\begin{equation}\label{ConstrSums}
\begin{split}
\sum_{l=2}^{n-1}(n+1-l)\,\cJ_{n+1-l}\,p\, \cJ_l&=
\sum_{k+l={n+1}}k\,\cJ_k\,p\, \cJ_l\;,\quad k,l\geq2\;,\\
\sum_{l=2}^{n}(n+1-l)\,\Psi_{n+1-l}\,p\, \cJ_l&=
\sum_{k+l={n+1}}k\,\Psi_k\,p\, \cJ_l\;,\quad k\geq1\;,l\geq2    
\end{split}    
\end{equation}

We start now proving \eqref{RelationJs inputfree} for the lowest number of inputs. Taking $n=2$, $\cJ_2$ is given by
\begin{equation}\label{J2}
\cJ_2=\frac12\,B_2(\Psi_1,\Psi_1)\;,\quad{\rm or}\quad \cJ_2(\bar x_1,\bar x_2)=\frac12\,B_2\big(\iota(\bar x_1), \iota(\bar x_2)\big)\;.    
\end{equation}
Using the Leibniz property of $B_1$ with respect to $B_2$ (the $n=2$ relation of \eqref{GenGenJac appendix}), we compute
\begin{equation}\label{Induction n=2}
\begin{split}
B_1\cJ_2+2\,\cJ_2\bar B_1&= \frac12\,B_1B_2(\Psi_1,\Psi_1)+B_2(\Psi_1\bar B_1,\Psi_1)\\
&=-B_2\big(B_1\Psi_1-\Psi_1\bar B_1,\Psi_1\big)=0\;,
\end{split}    
\end{equation}
thanks to the chain map property $B_1\Psi_1=\Psi_1\bar B_1$. In order to compute $\cJ_2\bar B_1$ above, we used that under graded symmetrization one explicitly has
\begin{equation}\label{n=2 trick}
\begin{split}
2\,\cJ_2\bar B_1 &\stackrel{\{12\}}{=}2\,\cJ_2\big(\bar B_1 \bar x_1,\bar x_2\big)=\frac12\,\Big(B_2\big(\Psi_1(\bar B_1\bar x_1),\Psi_1(\bar x_2)\big)+(-1)^{\bar x_1}B_2\big(\Psi_1(\bar x_1),\Psi_1(\bar B_1\bar x_2)\big)\Big)\\
&=\frac12\,\Big(B_2\big(\Psi_1(\bar B_1\bar x_1),\Psi_1(\bar x_2)\big)+(-1)^{\bar x_1\bar x_2}B_2\big(\Psi_1(\bar B_1\bar x_2),\Psi_1(\bar x_1)\big)\Big)\\
&=B_2\big(\Psi_1(\bar B_1\bar x_1),\Psi_1(\bar x_2)\big)=B_2(\Psi_1\bar B_1,\Psi_1)\;.
\end{split}    
\end{equation}
These kind of simplifications under total graded symmetrization will be used extensively in the following, in order to deal with expressions with arbitrary number of inputs.
The linear relation \eqref{Induction n=2} coincides with \eqref{RelationJs inputfree} for $n=2$, with the sum of nonlinear terms vanishing for $n-1<2\leq l$.

Since the $n=2$ case is somewhat degenerate, we prove \eqref{RelationJs inputfree} explicitly also for $n=3$. According to the definition \eqref{Jpsi defined}, $\cJ_3$ is given by
\begin{equation}
\begin{split}
\cJ_3&=B_2(\Psi_2,\Psi_1)+\frac{1}{3!}\,B_3(\Psi_1,\Psi_1,\Psi_1) \;,\quad{\rm or}\\
\cJ_3(\bar x_1,\bar x_2,\bar x_3)&\stackrel{\{123\}}{=}B_2\big(\Psi_2(\bar x_1,\bar x_2),\iota(\bar x_3)\big)+\frac{1}{3!}\,B_3\big(\iota(\bar x_1),\iota(\bar x_2),\iota(\bar x_3)\big)\;,
\end{split}  
\end{equation}
recalling that $\Psi_1=\iota$. In order to compute
 $B_1\cJ_3+3\,\cJ_3\bar B_1$ and to further display the use of graded symmetrization, we write $\cJ_3\bar B_1$ with explicit inputs:
\begin{equation}
\begin{split}
3\,\cJ_3\bar B_1&\stackrel{\{123\}}{=}3\,\cJ_3(\bar B_1\bar x_1,\bar x_2,\bar x_3)=B_2\big(\Psi_2(\bar B_1\bar x_1,\bar x_2),\Psi_1(\bar x_3)\big)\\&+(-1)^{\bar x_1}B_2\big(\Psi_2(\bar x_1,\bar B_1\bar x_2),\Psi_1(\bar x_3)\big)+(-1)^{\bar x_1+\bar x_2}B_2\big(\Psi_2(\bar x_1,\bar x_2),\Psi_1(\bar B_1\bar x_3)\big)\\&+\frac{1}{3!}\,\Big(B_3\big(\Psi_1(\bar B_1\bar x_1),\Psi_1(\bar x_2),\Psi_1(\bar x_3)\big)+(-1)^{\bar x_1}B_3\big(\Psi_1(\bar x_1),\Psi_1(\bar B_1\bar x_2),\Psi_1(\bar x_3)\big)\\&+(-1)^{\bar x_1+\bar x_2}B_3\big(\Psi_1(\bar x_1),\Psi_1(\bar x_2),\Psi_1(\bar B_1\bar x_3)\big) \Big)\;.  
\end{split}    
\end{equation}
Using graded symmetry of $B_2$, $\Psi_2$ and $B_3$ this can be rewritten as
\begin{equation}\label{n=3 trick1}
\begin{split}
3\,\cJ_3\bar B_1&\stackrel{\{123\}}{=}B_2\big(\Psi_2(\bar B_1\bar x_1,\bar x_2),\Psi_1(\bar x_3)\big)+(-1)^{\bar x_1\bar x_2}B_2\big(\Psi_2(\bar B_1\bar x_2,\bar x_1),\Psi_1(\bar x_3)\big)\\
&+(-1)^{\bar x_3(\bar x_1+\bar x_2)}B_2\big(\Psi_1(\bar B_1\bar x_3),\Psi_2(\bar x_1,\bar x_2)\big)\\&+\frac{1}{3!}\,\Big(B_3\big(\Psi_1(\bar B_1\bar x_1),\Psi_1(\bar x_2),\Psi_1(\bar x_3)\big)+(-1)^{\bar x_1\bar x_2}B_3\big(\Psi_1(\bar B_1\bar x_2),\Psi_1(\bar x_1),\Psi_1(\bar x_3)\big)\\&+(-1)^{\bar x_3(\bar x_1+\bar x_2)}B_3\big(\Psi_1(\bar B_1\bar x_3),\Psi_1(\bar x_1),\Psi_1(\bar x_2)\big) \Big)\;,  
\end{split}    
\end{equation}
which, under graded symmetrization in the inputs $(\bar x_1,\bar x_2,\bar x_3)$, simplifies to
\begin{equation}\label{n=3 trick2}
\begin{split}
3\,\cJ_3\bar B_1&\stackrel{\{123\}}{=}2\,B_2\big(\Psi_2(\bar B_1\bar x_1,\bar x_2),\Psi_1(\bar x_3)\big)+B_2\big(\Psi_1(\bar B_1\bar x_1),\Psi_2(\bar x_2,\bar x_3)\big)\\
&\hspace{7mm}+\frac{1}{2}\,B_3\big(\Psi_1(\bar B_1\bar x_1),\Psi_1(\bar x_2),\Psi_1(\bar x_3)\big)\\
&\stackrel{\{123\}}{=}2\,B_2\big(\Psi_2\bar B_1,\Psi_1\big)+B_2\big(\Psi_1\bar B_1,\Psi_2\big)+\frac{1}{2}\,B_3\big(\Psi_1\bar B_1,\Psi_1,\Psi_1\big)\;.
\end{split}    
\end{equation}
Using the same manipulations, $B_1\cJ_3$ yields
\begin{equation}
\begin{split}
B_1\cJ_3&=B_1B_2(\Psi_2,\Psi_1)+\frac{1}{3!}\,B_1B_3(\Psi_1,\Psi_1,\Psi_1)\\
&=-B_2\big(B_1\Psi_2,\Psi_1\big)-B_2\big(B_1\Psi_1,\Psi_2\big)+\frac{1}{3!}\,B_1B_3(\Psi_1,\Psi_1,\Psi_1)\;,
\end{split}    
\end{equation}
where, in particular, we used
\begin{equation}
\begin{split}
&B_2\big(B_1\Psi_2(\bar x_1,\bar x_2),\Psi_1(\bar x_3)\big)+(-1)^{\bar x_1+\bar x_2}B_2\big(\Psi_2(\bar x_1,\bar x_2),B_1\Psi_1(\bar x_3)\big)\\   &\stackrel{\{123\}}{=}B_2\big(B_1\Psi_2(\bar x_1,\bar x_2),\Psi_1(\bar x_3)\big)+(-1)^{ \bar x_3(\bar x_1+\bar x_2)}B_2\big(B_1\Psi_1(\bar x_3),\Psi_2(\bar x_1,\bar x_2)\big)\\
&\stackrel{\{123\}}{=}B_2\big(B_1\Psi_2(\bar x_1,\bar x_2),\Psi_1(\bar x_3)\big)+B_2\big(B_1\Psi_1(\bar x_1),\Psi_2(\bar x_2,\bar x_3)\big)\;.
\end{split}    
\end{equation}
Summing the two expressions we thus obtain
\begin{equation}
\begin{split}
B_1\cJ_3+3\,\cJ_3\bar B_1&=-B_2\big(B_1\Psi_2-2\Psi_2\bar B_1,\Psi_1\big)+\frac{1}{3!}\,\Big(B_1B_3(\Psi_1,\Psi_1,\Psi_1)+3\,B_3(B_1\Psi_1,\Psi_1,\Psi_1)\Big)\;,    
\end{split}    
\end{equation}
upon using $B_1\Psi_1=\Psi_1\bar B_1$. Since we proved the relation \eqref{RelationJs inputfree} for $n=2$, \eqref{RelationPsis inputfree} holds as $B_1\Psi_2-2\Psi_2\bar B_1=(\iota p-1)\,\cJ_2$. Combining this with the definition \eqref{J2} of $\cJ_2$, we can rewrite the above as
\begin{equation}
\begin{split}
&B_1\cJ_3+3\,\cJ_3\bar B_1=-B_2\big(\Psi_1\,p\, \cJ_2,\Psi_1\big)+B_2\big(\cJ_2,\Psi_1\big)+\frac{1}{3!}\,\Big(B_1B_3(\Psi_1,\Psi_1,\Psi_1)+3\,B_3(B_1\Psi_1,\Psi_1,\Psi_1)\Big)\\
&=-2\,\cJ_2\,p\, \cJ_2+\frac{1}{3!}\,\Big(3\,B_2\big(B_2(\Psi_1,\Psi_1),\Psi_1\big)+B_1B_3(\Psi_1,\Psi_1,\Psi_1)+3\,B_3(B_1\Psi_1,\Psi_1,\Psi_1)\Big)\\
&=-2\,\cJ_2\,p\, \cJ_2\;,
\end{split}    
\end{equation}
where we used the generalized Jacobi identity, thus proving \eqref{RelationJs inputfree} for $n=3$. In the above computation we recognized the nesting $\cJ_2\,p\, \cJ_2$ from
\begin{equation}
\begin{split}
\cJ_2\big(p\, \cJ_2(\bar x_1,\bar x_2),\bar x_3\big)\stackrel{\{123\}}{=}\frac12\,B_2\big(\Psi_1(p\, \cJ_2(\bar x_1,\bar x_2)),\Psi_1(\bar x_3)\big)\;.    
\end{split}    
\end{equation}

Having shown that \eqref{RelationJs inputfree}, and hence \eqref{RelationPsis inputfree}, hold for $n=2,3$, we now assume that they do up to $n-1$ inputs. With $n$ inputs, $\cJ_n$ is given by
\begin{equation}
\cJ_n=\sum_{k=2}^n\frac{1}{k!}\sum_{i_1+\cdots+i_k=n}B_k(\Psi_{i_1},\ldots,\Psi_{i_k}) \;.  
\end{equation}
Using graded symmetrization as in \eqref{n=3 trick1} and \eqref{n=3 trick2}, we compute
\begin{equation}\label{Generalproof1}
\begin{split}
B_1\cJ_n+n\,\cJ_n\bar B_1&=\sum_{k=2}^n\frac{1}{k!}\sum_{i_1+\cdots+i_k=n}\Big(B_1B_k(\Psi_{i_1},\ldots,\Psi_{i_k})+k\,i_1B_k(\Psi_{i_1}\bar B_1,\ldots,\Psi_{i_k})\Big)\;,    
\end{split}    
\end{equation}
where we relabeled the indices $i_j$, so that
\begin{equation}
\begin{split}
&\sum_{i_1+\cdots+i_k=n}\Big(i_1B_k(\Psi_{i_1}\bar B_1,\Psi_{i_2},\ldots,\Psi_{i_k})+i_2B_k(\Psi_{i_2}\bar B_1,\Psi_{i_1},\ldots,\Psi_{i_k})+\ldots\\
&\hspace{20mm}+i_kB_k(\Psi_{i_k}\bar B_1,\Psi_{i_1},\ldots,\Psi_{i_{k-1}})\Big)=\sum_{i_1+\cdots+i_k=n}k\,i_1B_k(\Psi_{i_1}\bar B_1,\Psi_{i_2},\ldots,\Psi_{i_k})\;.   
\end{split}    
\end{equation}
In order to use the generalized Jacobi identities of $B_k$ and \eqref{RelationPsis inputfree}, which hold by assumption for all the $\Psi_{i_j}$ appearing above, we add and subtract $k\,B_k(B_1\Psi_{i_1},\ldots,\Psi_{i_k})$ in the sums, obtaining
\begin{equation}\label{GeneralProof2}
\begin{split}
B_1\cJ_n+n\,\cJ_n\bar B_1&=\sum_{k=2}^n\frac{1}{k!}\sum_{i_1+\cdots+i_k=n}\Big(B_1B_k(\Psi_{i_1},\ldots,\Psi_{i_k})+k\,B_k(B_1\Psi_{i_1},\ldots,\Psi_{i_k})\Big)\\
&-\sum_{k=2}^n\frac{1}{(k-1)!}\sum_{i_1+\cdots+i_k=n}B_k\big(B_1\Psi_{i_1}-i_1\Psi_{i_1}\bar B_1,\ldots,\Psi_{i_k}\big)\;.
\end{split}    
\end{equation}
Let us focus on the second line. In the term $k=n$, the constrained sum over the labels $i_j$ implies that $i_1=i_2=\ldots=i_k=1$, which vanishes, since $B_1\Psi_1-\Psi_1\bar B_1=0$. The second sum over $k$ thus ends at $k=n-1$. We now use the relation \eqref{RelationPsis inputfree} for $\Psi_{i_1}$ in the form \eqref{ConstrSums}, and rewrite the second line above as 
\begin{equation}\label{GeneralProof3}
\begin{split}
&\sum_{k=2}^{n-1}\frac{1}{(k-1)!}\sum_{i_1+\cdots+i_k=n}B_k\big(B_1\Psi_{i_1}-i_1\Psi_{i_1}\bar B_1,\Psi_{i_2},\ldots,\Psi_{i_k}\big)= \sum_{k=2}^{n-1}\frac{1}{(k-1)!}\\
&\times\Big\{\sum_{i_1+\cdots+i_k+l=n+1}i_1B_k\big(\Psi_{i_1}\,p\cJ_l,\Psi_{i_2},\ldots,\Psi_{i_k}\big)-\sum_{i_2+\cdots+i_k+l=n}B_k\big(\cJ_l,\Psi_{i_2},\ldots,\Psi_{i_k}\big) \Big\}\;,
\end{split}    
\end{equation}
where $i_j\geq1$ and $l\geq2$, and we relabeled indices as to write a global constrained sum for the nested maps.

From the definition \eqref{Jpsi defined} for the $\cJ_n$ maps and graded symmetrization, singling out one input produces
\begin{equation}
\begin{split}
n\,\cJ_n( \bar y,\bar x_2,\ldots,\bar x_n)&\stackrel{\{2\cdots n\}}{=}\sum_{k=2}^n\frac{1}{k!}\,\sum_{i_1+\cdots+i_k=n}k\,i_1B_k\big(\Psi_{i_1}(\bar y,\bar x_2,\ldots,\bar x_{i_1}),\ldots,\Psi_{i_k}(\bar x_{n-i_k+1},\ldots,\bar x_{n})\big)\;.    
\end{split}    
\end{equation}
Taking $\bar y=p\, \cJ_l(\bar x_1,\ldots,\bar x_l)$ allows us to recognize the first sum in the second line of \eqref{GeneralProof3} as
\begin{equation}
\sum_{k=2}^{n-1}\frac{1}{k!}\sum_{i_1+\cdots+i_k+l=n+1}k\,i_1B_k\big(\Psi_{i_1}\,p\, \cJ_l,\Psi_{i_2},\ldots,\Psi_{i_k}\big)=\sum_{m+l=n+1}m\,\cJ_m\,p\, \cJ_l\;,\quad m,l\geq2\;, \end{equation}
which is the nonlinear part of the relation \eqref{RelationJs inputfree}
Using this and \eqref{GeneralProof3} in \eqref{GeneralProof2} we arrive at
\begin{equation}\label{GeneralProof4}
\begin{split}
&B_1\cJ_n+n\,\cJ_n\bar B_1+\sum_{k+l=n+1}k\,\cJ_k\,p\, \cJ_l\\
&=\sum_{k=2}^n\frac{1}{k!}\sum_{i_1+\cdots+i_k=n}\Big(B_1B_k(\Psi_{i_1},\ldots,\Psi_{i_k})+k\,B_k(B_1\Psi_{i_1},\ldots,\Psi_{i_k})\Big)
\\&+\sum_{k=2}^{n-1}\frac{1}{k!}\sum_{i_1+\cdots+i_k=n}k\,B_k\big(\cJ_{i_1},\ldots,\Psi_{i_k}\big)\;.    
\end{split}    
\end{equation}
We shall now show that the right-hand side above vanishes, thanks to the generalized Jacobi identities \eqref{GenGenJac appendix} of the original brackets $B_k$. To this end, we take the definition \eqref{Jpsi defined} for $\cJ_{i_1}$ and rewrite the last line in \eqref{GeneralProof4} as
\begin{equation}
\begin{split}
&\sum_{k=2}^{n-1}\frac{1}{k!}\sum_{i_1+\cdots+i_k=n}k\,B_k\big(\cJ_{i_1},\ldots,\Psi_{i_k}\big)\\
&=\sum_{k=2}^{n-1}\frac{1}{(k-1)!}\sum_{i_1+\cdots+i_k=n}\sum_{l=2}^{i_1}\frac{1}{l!}\sum_{j_1+\cdots+j_l=i_1}\,B_k\big(B_l(\Psi_{j_1},\ldots,\Psi_{j_l}),\Psi_{i_2},\ldots,\Psi_{i_k}\big)\\
&=\sum_{m=3}^{n}\frac{1}{m!}\sum_{i_1+\cdots+i_m=n}\Big\{\sum_{k+l=m+1}\frac{m!}{(k-1)!l!}\,B_k\big(B_l(\Psi_{i_1},\ldots,\Psi_{i_l}),\Psi_{i_{l+1}},\ldots,\Psi_{i_m}\big)\Big\}\;,
\end{split}    
\end{equation}
where $k,l\geq2$ and we denoted $m=k+l-1$ the total number of $\Psi$ inputs. Combining this with the terms involving $B_1$, the right-hand side of \eqref{GeneralProof4} can be written as
\begin{equation}\label{GeneralProof5}\sum_{m=2}^n\frac{1}{m!}\sum_{i_1+\cdots+i_m=n}\Big\{\sum_{k+l=m+1}^n\binom{m}{l}\,B_{k}\big(B_l(\Psi_{i_1},\ldots,\Psi_{i_l}),\Psi_{i_{l+1}},\ldots,\Psi_{i_m}\big)\Big\}\;,
\end{equation}
with $k,l\geq1$. 

This is nothing but the generalized Jacobi identity \eqref{GenGenJac appendix} applied to the composite inputs $\Psi_{i_1}\cdots\Psi_{i_m}$.
To see this, we have to show that the above expression is graded symmetric in the exchange of the arguments $\Psi_{i_j}(\bar x_{i_1+\cdots+i_{j-1}+1},\ldots,\bar x_{i_1+\cdots+i_j})$, thanks to graded symmetrization in the elementary inputs $\bar x_1,\ldots,\bar x_n$. Since the brackets $B_k$ and $B_l$ are graded symmetric, one has to show that this is the case only for the exchange of $\Psi_{i_l}$ with $\Psi_{i_{l+1}}$.
Let us thus consider the expression
\begin{equation}
\sum_{p+q}\cM\big(\ldots,\Psi_p(y_1,\ldots,y_p),\Psi_q(y_{p+1},\ldots,y_{p+q}),\ldots\big)\;,    
\end{equation}
where $\cM$ is an arbitrary multilinear map, with no symmetry properties, and all inputs $y_k$ are graded symmetrized. For the specific case above, $\cM=B_kB_l$, $p=i_l$, $q=i_{l+1}$ and $y_k=\bar x_{i_1+\cdots+i_{l-1}+k}$. Relabeling $p$ and $q$ and using graded symmetry in the $y$'s we obtain
\begin{equation}
\begin{split}
&\sum_{p+q}\cM\big(\ldots,\Psi_p(y_1,\ldots,y_p),\Psi_q(y_{p+1},\ldots,y_{p+q}),\ldots\big)\\
&\stackrel{\{12\cdots p+q\}}{=}\sum_{p+q}\cM\big(\ldots,\Psi_q(y_1,\ldots,y_q),\Psi_p(y_{q+1},\ldots,y_{p+q}),\ldots\big)\\
&\stackrel{\{12\cdots p+q\}}{=}\sum_{p+q}(-1)^{(y_1+\cdots+y_p)(y_{p+1}+\cdots+y_{p+q})}\cM\big(\ldots,\Psi_q(y_{p+1},\ldots,y_{p+q}),\Psi_p(y_{1},\ldots,y_{p}),\ldots\big) \;,   \end{split}   \end{equation}
which is the desired symmetry property, given that the maps $\Psi_i$ have intrinsic degree zero. This proves that \eqref{GeneralProof5} is graded symmetric in the composite arguments and thus vanishes by the generalized Jacobi identity. In turn, this shows that the right-hand side of \eqref{GeneralProof4} vanishes, which is the relation \eqref{RelationJs inputfree} for $n$ inputs, thereby concluding the proof.

\bibliography{HomAmp.bib}

\providecommand{\href}[2]{#2}\begingroup\raggedright\begin{thebibliography}{10}

\bibitem{costellorenormalization}
K.~Costello, {\em Renormalization and Effective Field Theory}.
\newblock Mathematical surveys and monographs. American Mathematical Soc.
\newblock \url{https://books.google.de/books?id=9iM3NSy\_xCUC}.

\bibitem{Costello:2021jvx}
K.~Costello and O.~Gwilliam,
  \href{http://dx.doi.org/10.1017/9781316678664}{{\em {Factorization Algebras
  in Quantum Field Theory}}}.
\newblock New Mathematical Monographs (41). Cambridge University Press, 9,
  2021.

\bibitem{Zwiebach:1992ie}
B.~Zwiebach, ``{Closed string field theory: Quantum action and the B-V master
  equation},'' \href{http://dx.doi.org/10.1016/0550-3213(93)90388-6}{{\em Nucl.
  Phys. B} {\bfseries 390} (1993) 33--152},
  \href{http://arxiv.org/abs/hep-th/9206084}{{\ttfamily arXiv:hep-th/9206084}}.

\bibitem{Zeitlin:2007ttl}
A.~M. Zeitlin, ``{Batalin-Vilkovisky Yang-Mills theory as a homotopy
  Chern-Simons theory via string field theory},''
  \href{http://dx.doi.org/10.1142/S0217751X09043031}{{\em Int. J. Mod. Phys. A}
  {\bfseries 24} (2009) 1309--1331},
  \href{http://arxiv.org/abs/0709.1411}{{\ttfamily arXiv:0709.1411 [hep-th]}}.

\bibitem{Zeitlin:2008cc}
A.~M. Zeitlin, ``{Conformal Field Theory and Algebraic Structure of Gauge
  Theory},'' \href{http://dx.doi.org/10.1007/JHEP03(2010)056}{{\em JHEP}
  {\bfseries 03} (2010) 056}, \href{http://arxiv.org/abs/0812.1840}{{\ttfamily
  arXiv:0812.1840 [hep-th]}}.

\bibitem{Zeitlin:2009zz}
A.~M. Zeitlin, ``{String field theory-inspired algebraic structures in gauge
  theories},'' \href{http://dx.doi.org/10.1063/1.3142964}{{\em J. Math. Phys.}
  {\bfseries 50} (2009) 063501},
  \href{http://arxiv.org/abs/0711.3843}{{\ttfamily arXiv:0711.3843 [hep-th]}}.

\bibitem{Hohm:2017pnh}
O.~Hohm and B.~Zwiebach, ``{$L_{\infty}$ Algebras and Field Theory},''
  \href{http://dx.doi.org/10.1002/prop.201700014}{{\em Fortsch. Phys.}
  {\bfseries 65} no.~3-4, (2017) 1700014},
  \href{http://arxiv.org/abs/1701.08824}{{\ttfamily arXiv:1701.08824
  [hep-th]}}.

\bibitem{Lada:1994mn}
T.~Lada and M.~Markl, ``{Strongly homotopy Lie algebras},''
  \href{http://dx.doi.org/10.1080/00927879508825335}{{\em Communications in
  Algebra} {\bfseries 23} no.~6, (1995) 2147--2161},
  \href{http://arxiv.org/abs/hep-th/9406095}{{\ttfamily arXiv:hep-th/9406095}}.

\bibitem{Lada:1992wc}
T.~Lada and J.~Stasheff, ``{Introduction to SH Lie algebras for physicists},''
  \href{http://dx.doi.org/10.1007/BF00671791}{{\em Int. J. Theor. Phys.}
  {\bfseries 32} (1993) 1087--1104},
  \href{http://arxiv.org/abs/hep-th/9209099}{{\ttfamily arXiv:hep-th/9209099}}.

\bibitem{Alexandrov:1995kv}
M.~Alexandrov, A.~Schwarz, O.~Zaboronsky, and M.~Kontsevich, ``{The Geometry of
  the master equation and topological quantum field theory},''
  \href{http://dx.doi.org/10.1142/S0217751X97001031}{{\em Int. J. Mod. Phys. A}
  {\bfseries 12} (1997) 1405--1429},
  \href{http://arxiv.org/abs/hep-th/9502010}{{\ttfamily arXiv:hep-th/9502010}}.

\bibitem{Crainic:2004bxw}
M.~Crainic, ``{On the perturbation lemma, and deformations},''
  \href{http://arxiv.org/abs/math/0403266}{{\ttfamily arXiv:math/0403266}}.

\bibitem{vallette2014algebra}
B.~Vallette, ``Homotopy theory of homotopy algebras,''
  \href{http://dx.doi.org/10.5802/aif.3322}{{\em Annales de l'Institut Fourier}
  {\bfseries 70} no.~2, (2020) 683--738}.
  \url{http://www.numdam.org/articles/10.5802/aif.3322/}.

\bibitem{Erbin:2020eyc}
H.~Erbin, C.~Maccaferri, M.~Schnabl, and J.~Vo\v{s}mera, ``{Classical algebraic
  structures in string theory effective actions},''
  \href{http://dx.doi.org/10.1007/JHEP11(2020)123}{{\em JHEP} {\bfseries 11}
  (2020) 123}, \href{http://arxiv.org/abs/2006.16270}{{\ttfamily
  arXiv:2006.16270 [hep-th]}}.

\bibitem{Koyama:2020qfb}
D.~Koyama, Y.~Okawa, and N.~Suzuki, ``{Gauge-invariant operators of open
  bosonic string field theory in the low-energy limit},''
  \href{http://arxiv.org/abs/2006.16710}{{\ttfamily arXiv:2006.16710
  [hep-th]}}.

\bibitem{Arvanitakis:2020rrk}
A.~S. Arvanitakis, O.~Hohm, C.~Hull, and V.~Lekeu, ``{Homotopy Transfer and
  Effective Field Theory I: Tree-level},''
  \href{http://arxiv.org/abs/2007.07942}{{\ttfamily arXiv:2007.07942
  [hep-th]}}.

\bibitem{Chiaffrino:2020akd}
C.~Chiaffrino, O.~Hohm, and A.~F. Pinto, ``{Gauge Invariant Perturbation Theory
  via Homotopy Transfer},''
  \href{http://dx.doi.org/10.1007/JHEP05(2021)236}{{\em JHEP} {\bfseries 05}
  (2021) 236}, \href{http://arxiv.org/abs/2012.12249}{{\ttfamily
  arXiv:2012.12249 [hep-th]}}.

\bibitem{Arvanitakis:2021ecw}
A.~S. Arvanitakis, O.~Hohm, C.~Hull, and V.~Lekeu, ``{Homotopy Transfer and
  Effective Field Theory II: Strings and Double Field Theory},''
  \href{http://arxiv.org/abs/2106.08343}{{\ttfamily arXiv:2106.08343
  [hep-th]}}.

\bibitem{Chiaffrino:2023wxk}
C.~Chiaffrino, T.~Ersoy, and O.~Hohm, ``{Holography as Homotopy},''
  \href{http://arxiv.org/abs/2307.08094}{{\ttfamily arXiv:2307.08094
  [hep-th]}}.

\bibitem{Kajiura:2003ax}
H.~Kajiura, ``{Noncommutative homotopy algebras associated with open
  strings},'' \href{http://dx.doi.org/10.1142/S0129055X07002912}{{\em Rev.
  Math. Phys.} {\bfseries 19} (2007) 1--99},
  \href{http://arxiv.org/abs/math/0306332}{{\ttfamily arXiv:math/0306332}}.

\bibitem{Munster:2011ij}
K.~Munster and I.~Sachs, ``{Quantum Open-Closed Homotopy Algebra and String
  Field Theory},'' \href{http://dx.doi.org/10.1007/s00220-012-1654-1}{{\em
  Commun. Math. Phys.} {\bfseries 321} (2013) 769--801},
  \href{http://arxiv.org/abs/1109.4101}{{\ttfamily arXiv:1109.4101 [hep-th]}}.

\bibitem{Doubek:2017naz}
M.~Doubek, B.~Jur\v{c}o, and J.~Pulmann, ``{Quantum $L_\infty$ Algebras and the
  Homological Perturbation Lemma},''
  \href{http://dx.doi.org/10.1007/s00220-019-03375-x}{{\em Commun. Math. Phys.}
  {\bfseries 367} no.~1, (2019) 215--240},
  \href{http://arxiv.org/abs/1712.02696}{{\ttfamily arXiv:1712.02696
  [math-ph]}}.

\bibitem{Nutzi:2018vkl}
A.~N\"utzi and M.~Reiterer, ``{Amplitudes in YM and GR as a Minimal Model and
  Recursive Characterization},''
  \href{http://dx.doi.org/10.1007/s00220-022-04339-4}{{\em Commun. Math. Phys.}
  {\bfseries 392} no.~2, (2022) 427--482},
  \href{http://arxiv.org/abs/1812.06454}{{\ttfamily arXiv:1812.06454
  [math-ph]}}.

\bibitem{Arvanitakis:2019ald}
A.~S. Arvanitakis, ``{The L$_\infty$-algebra of the S-matrix},''
  \href{http://dx.doi.org/10.1007/JHEP07(2019)115}{{\em JHEP} {\bfseries 07}
  (2019) 115}, \href{http://arxiv.org/abs/1903.05643}{{\ttfamily
  arXiv:1903.05643 [hep-th]}}.

\bibitem{Jurco:2019yfd}
B.~Jur\v{c}o, T.~Macrelli, C.~S\"amann, and M.~Wolf, ``{Loop Amplitudes and
  Quantum Homotopy Algebras},''
  \href{http://dx.doi.org/10.1007/JHEP07(2020)003}{{\em JHEP} {\bfseries 07}
  (2020) 003}, \href{http://arxiv.org/abs/1912.06695}{{\ttfamily
  arXiv:1912.06695 [hep-th]}}.

\bibitem{Lopez-Arcos:2019hvg}
C.~Lopez-Arcos and A.~Q. V\'elez, ``{L$_{\infty}$-algebras and the perturbiner
  expansion},'' \href{http://dx.doi.org/10.1007/JHEP11(2019)010}{{\em JHEP}
  {\bfseries 11} (2019) 010}, \href{http://arxiv.org/abs/1907.12154}{{\ttfamily
  arXiv:1907.12154 [hep-th]}}.

\bibitem{Saemann:2020oyz}
C.~Saemann and E.~Sfinarolakis, ``{Symmetry Factors of Feynman Diagrams and the
  Homological Perturbation Lemma},''
  \href{http://dx.doi.org/10.1007/JHEP12(2020)088}{{\em JHEP} {\bfseries 12}
  (2020) 088}, \href{http://arxiv.org/abs/2009.12616}{{\ttfamily
  arXiv:2009.12616 [hep-th]}}.

\bibitem{Okawa:2022sjf}
Y.~Okawa, ``{Correlation functions of scalar field theories from homotopy
  algebras},'' \href{http://arxiv.org/abs/2203.05366}{{\ttfamily
  arXiv:2203.05366 [hep-th]}}.

\bibitem{Konosu:2023pal}
K.~Konosu and Y.~Okawa, ``{Correlation functions involving Dirac fields from
  homotopy algebras I: the free theory},''
  \href{http://arxiv.org/abs/2305.11634}{{\ttfamily arXiv:2305.11634
  [hep-th]}}.

\bibitem{Hull:2009mi}
C.~Hull and B.~Zwiebach, ``{Double Field Theory},''
  \href{http://dx.doi.org/10.1088/1126-6708/2009/09/099}{{\em JHEP} {\bfseries
  09} (2009) 099}, \href{http://arxiv.org/abs/0904.4664}{{\ttfamily
  arXiv:0904.4664 [hep-th]}}.

\bibitem{Bonezzi:2023ced}
R.~Bonezzi, C.~Chiaffrino, F.~Diaz-Jaramillo, and O.~Hohm, ``{Weakly
  constrained double field theory: the quartic theory},''
  \href{http://arxiv.org/abs/2306.00609}{{\ttfamily arXiv:2306.00609
  [hep-th]}}.

\bibitem{Bonezzi:2023lkx}
R.~Bonezzi, C.~Chiaffrino, F.~Diaz-Jaramillo, and O.~Hohm, ``{Weakly
  Constrained Double Field Theory as the Double Copy of Yang-Mills Theory},''
  \href{http://arxiv.org/abs/2309.03289}{{\ttfamily arXiv:2309.03289
  [hep-th]}}.

\bibitem{Diaz-Jaramillo:2021wtl}
F.~Diaz-Jaramillo, O.~Hohm, and J.~Plefka, ``{Double field theory as the double
  copy of Yang-Mills theory},''
  \href{http://dx.doi.org/10.1103/PhysRevD.105.045012}{{\em Phys. Rev. D}
  {\bfseries 105} no.~4, (2022) 045012},
  \href{http://arxiv.org/abs/2109.01153}{{\ttfamily arXiv:2109.01153
  [hep-th]}}.

\bibitem{Bonezzi:2022yuh}
R.~Bonezzi, F.~Diaz-Jaramillo, and O.~Hohm, ``{The gauge structure of double
  field theory follows from Yang-Mills theory},''
  \href{http://dx.doi.org/10.1103/PhysRevD.106.026004}{{\em Phys. Rev. D}
  {\bfseries 106} no.~2, (2022) 026004},
  \href{http://arxiv.org/abs/2203.07397}{{\ttfamily arXiv:2203.07397
  [hep-th]}}.

\bibitem{Bonezzi:2022bse}
R.~Bonezzi, C.~Chiaffrino, F.~Diaz-Jaramillo, and O.~Hohm, ``{Gauge invariant
  double copy of Yang-Mills theory: The quartic theory},''
  \href{http://dx.doi.org/10.1103/PhysRevD.107.126015}{{\em Phys. Rev. D}
  {\bfseries 107} no.~12, (2023) 126015},
  \href{http://arxiv.org/abs/2212.04513}{{\ttfamily arXiv:2212.04513
  [hep-th]}}.

\bibitem{Bonezzi:2023ciu}
R.~Bonezzi, C.~Chiaffrino, F.~Diaz-Jaramillo, and O.~Hohm, ``{Gravity =
  Yang-Mills},''
\newblock 6, 2023.
\newblock \href{http://arxiv.org/abs/2306.14788}{{\ttfamily arXiv:2306.14788
  [hep-th]}}.

\bibitem{Borsten:2021hua}
L.~Borsten, H.~Kim, B.~Jur\v{c}o, T.~Macrelli, C.~Saemann, and M.~Wolf,
  ``{Double Copy from Homotopy Algebras},''
  \href{http://dx.doi.org/10.1002/prop.202100075}{{\em Fortsch. Phys.}
  {\bfseries 69} no.~8-9, (2021) 2100075},
  \href{http://arxiv.org/abs/2102.11390}{{\ttfamily arXiv:2102.11390
  [hep-th]}}.

\bibitem{Boulware:1968zz}
D.~G. Boulware and L.~S. Brown, ``{Tree Graphs and Classical Fields},''
  \href{http://dx.doi.org/10.1103/PhysRev.172.1628}{{\em Phys. Rev.} {\bfseries
  172} (1968) 1628--1631}.

\bibitem{Srednicki:2007qs}
M.~Srednicki, {\em {Quantum field theory}}.
\newblock Cambridge University Press, 1, 2007.

\bibitem{Monteiro:2011pc}
R.~Monteiro and D.~O'Connell, ``{The Kinematic Algebra From the Self-Dual
  Sector},'' \href{http://dx.doi.org/10.1007/JHEP07(2011)007}{{\em JHEP}
  {\bfseries 07} (2011) 007}, \href{http://arxiv.org/abs/1105.2565}{{\ttfamily
  arXiv:1105.2565 [hep-th]}}.

\bibitem{Arefeva:1974jv}
I.~Y. Arefeva, L.~D. Faddeev, and A.~A. Slavnov, ``{Generating Functional for
  the s Matrix in Gauge Theories},''
  \href{http://dx.doi.org/10.1007/BF01038094}{{\em Teor. Mat. Fiz.} {\bfseries
  21} (1974) 311--321}.

\bibitem{Berglund2009HomologicalPT}
A.~Berglund, ``Homological perturbation theory for algebras over operads,''
  {\em Algebraic \& Geometric Topology} {\bfseries 14} (2009) 2511--2548.
  \url{https://api.semanticscholar.org/CorpusID:16116039}.

\bibitem{Chuang2017OnTP}
J.~Chuang and A.~Lazarev, ``On the perturbation algebra,'' {\em Journal of
  Algebra} (2017) . \url{https://api.semanticscholar.org/CorpusID:119166530}.

\bibitem{Reiterer:2019dys}
M.~Reiterer, ``{A homotopy BV algebra for Yang-Mills and color-kinematics},''
  \href{http://arxiv.org/abs/1912.03110}{{\ttfamily arXiv:1912.03110
  [math-ph]}}.

\bibitem{Ben-Shahar:2021zww}
M.~Ben-Shahar and H.~Johansson, ``{Off-Shell Color-Kinematics Duality for
  Chern-Simons},'' \href{http://arxiv.org/abs/2112.11452}{{\ttfamily
  arXiv:2112.11452 [hep-th]}}.

\bibitem{Borsten:2022vtg}
L.~Borsten, B.~Jurco, H.~Kim, T.~Macrelli, C.~Saemann, and M.~Wolf,
  ``{Kinematic Lie Algebras From Twistor Spaces},''
  \href{http://arxiv.org/abs/2211.13261}{{\ttfamily arXiv:2211.13261
  [hep-th]}}.

\bibitem{Bonezzi:2023pox}
R.~Bonezzi, F.~Diaz-Jaramillo, and S.~Nagy, ``{Gauge independent kinematic
  algebra of self-dual Yang-Mills theory},''
  \href{http://dx.doi.org/10.1103/PhysRevD.108.065007}{{\em Phys. Rev. D}
  {\bfseries 108} no.~6, (2023) 065007},
  \href{http://arxiv.org/abs/2306.08558}{{\ttfamily arXiv:2306.08558
  [hep-th]}}.

\bibitem{Bern:2008qj}
Z.~Bern, J.~J.~M. Carrasco, and H.~Johansson, ``{New Relations for Gauge-Theory
  Amplitudes},'' \href{http://dx.doi.org/10.1103/PhysRevD.78.085011}{{\em Phys.
  Rev. D} {\bfseries 78} (2008) 085011},
  \href{http://arxiv.org/abs/0805.3993}{{\ttfamily arXiv:0805.3993 [hep-ph]}}.

\bibitem{DelDuca:1999rs}
V.~Del~Duca, L.~J. Dixon, and F.~Maltoni, ``{New color decompositions for gauge
  amplitudes at tree and loop level},''
  \href{http://dx.doi.org/10.1016/S0550-3213(99)00809-3}{{\em Nucl. Phys. B}
  {\bfseries 571} (2000) 51--70},
  \href{http://arxiv.org/abs/hep-ph/9910563}{{\ttfamily arXiv:hep-ph/9910563}}.

\bibitem{Bandiera:2020aqn}
R.~Bandiera and C.~R. Mafra, ``{A closed-formula solution to the color-trace
  decomposition problem},'' \href{http://arxiv.org/abs/2009.02534}{{\ttfamily
  arXiv:2009.02534 [math.CO]}}.

\bibitem{Berends:1987me}
F.~A. Berends and W.~T. Giele, ``{Recursive Calculations for Processes with n
  Gluons},'' \href{http://dx.doi.org/10.1016/0550-3213(88)90442-7}{{\em Nucl.
  Phys. B} {\bfseries 306} (1988) 759--808}.

\bibitem{Mafra:2015vca}
C.~R. Mafra and O.~Schlotterer, ``{Berends-Giele recursions and the BCJ duality
  in superspace and components},''
  \href{http://dx.doi.org/10.1007/JHEP03(2016)097}{{\em JHEP} {\bfseries 03}
  (2016) 097}, \href{http://arxiv.org/abs/1510.08846}{{\ttfamily
  arXiv:1510.08846 [hep-th]}}.

\bibitem{Macrelli:2019afx}
T.~Macrelli, C.~S\"amann, and M.~Wolf, ``{Scattering amplitude recursion
  relations in Batalin-Vilkovisky\textendash{}quantizable theories},''
  \href{http://dx.doi.org/10.1103/PhysRevD.100.045017}{{\em Phys. Rev. D}
  {\bfseries 100} no.~4, (2019) 045017},
  \href{http://arxiv.org/abs/1903.05713}{{\ttfamily arXiv:1903.05713
  [hep-th]}}.

\bibitem{KLEISS1989616}
R.~Kleiss and H.~Kuijf, ``Multigluon cross sections and 5-jet production at
  hadron colliders,''
  \href{http://dx.doi.org/https://doi.org/10.1016/0550-3213(89)90574-9}{{\em
  Nuclear Physics B} {\bfseries 312} no.~3, (1989) 616--644}.
  \url{https://www.sciencedirect.com/science/article/pii/0550321389905749}.

\bibitem{Garozzo:2018uzj}
L.~M. Garozzo, L.~Queimada, and O.~Schlotterer, ``{Berends-Giele currents in
  Bern-Carrasco-Johansson gauge for $F^3$- and $F^4$-deformed Yang-Mills
  amplitudes},'' \href{http://dx.doi.org/10.1007/JHEP02(2019)078}{{\em JHEP}
  {\bfseries 02} (2019) 078}, \href{http://arxiv.org/abs/1809.08103}{{\ttfamily
  arXiv:1809.08103 [hep-th]}}.

\bibitem{CHENG20082535}
X.~Z. Cheng and E.~Getzler, ``Transferring homotopy commutative algebraic
  structures,''
  \href{http://dx.doi.org/https://doi.org/10.1016/j.jpaa.2008.04.002}{{\em
  Journal of Pure and Applied Algebra} {\bfseries 212} no.~11, (2008)
  2535--2542}.
  \url{https://www.sciencedirect.com/science/article/pii/S0022404908000832}.

\bibitem{AlgebraicOperads}
J.-L. Loday and B.~Vallette,
  \href{http://dx.doi.org/10.1007/978-3-642-30362-3}{{\em Algebraic Operads}},
  vol.~346.
\newblock 01, 2012.

\bibitem{Stasheff1963HomotopyAO}
J.~Stasheff, ``Homotopy associativity of \$h\$-spaces. ii,'' {\em Transactions
  of the American Mathematical Society} {\bfseries 108} (1963) 293--312.

\bibitem{LodayPolytopes}
J.-L. Loday, ``Realization of the stasheff polytope,''
  \href{http://dx.doi.org/10.1007/s00013-004-1026-y}{{\em Archiv der
  Mathematik} {\bfseries 83} (01, 2003) }.

\bibitem{MappingCone}
D.~Fiorenza and M.~Manetti, ``L-infinity structures on mapping cones,'' {\em
  Algebra and Number Theory} {\bfseries 1} (01, 2006) .

\bibitem{vallette2012algebrahomotopyoperad}
B.~Vallette, ``Algebra+homotopy=operad,'' 2012.

\bibitem{Bern:2010yg}
Z.~Bern, T.~Dennen, Y.-t. Huang, and M.~Kiermaier, ``{Gravity as the Square of
  Gauge Theory},'' \href{http://dx.doi.org/10.1103/PhysRevD.82.065003}{{\em
  Phys. Rev. D} {\bfseries 82} (2010) 065003},
  \href{http://arxiv.org/abs/1004.0693}{{\ttfamily arXiv:1004.0693 [hep-th]}}.

\bibitem{Bjerrum-Bohr:2010pnr}
N.~E.~J. Bjerrum-Bohr, P.~H. Damgaard, T.~Sondergaard, and P.~Vanhove, ``{The
  Momentum Kernel of Gauge and Gravity Theories},''
  \href{http://dx.doi.org/10.1007/JHEP01(2011)001}{{\em JHEP} {\bfseries 01}
  (2011) 001}, \href{http://arxiv.org/abs/1010.3933}{{\ttfamily arXiv:1010.3933
  [hep-th]}}.

\bibitem{Mafra:2011kj}
C.~R. Mafra, O.~Schlotterer, and S.~Stieberger, ``{Explicit BCJ Numerators from
  Pure Spinors},'' \href{http://dx.doi.org/10.1007/JHEP07(2011)092}{{\em JHEP}
  {\bfseries 07} (2011) 092}, \href{http://arxiv.org/abs/1104.5224}{{\ttfamily
  arXiv:1104.5224 [hep-th]}}.

\bibitem{Borsten:2023reb}
L.~Borsten, B.~Jurco, H.~Kim, T.~Macrelli, C.~Saemann, and M.~Wolf,
  ``{Tree-Level Color-Kinematics Duality from Pure Spinor Actions},''
  \href{http://arxiv.org/abs/2303.13596}{{\ttfamily arXiv:2303.13596
  [hep-th]}}.

\bibitem{Borsten:2023ned}
L.~Borsten, B.~Jurco, H.~Kim, T.~Macrelli, C.~Saemann, and M.~Wolf, ``{Double
  Copy from Tensor Products of Metric BV${}^{\color{gray}
  \blacksquare}$-algebras},'' \href{http://arxiv.org/abs/2307.02563}{{\ttfamily
  arXiv:2307.02563 [hep-th]}}.

\bibitem{Markl:1997bj}
M.~Markl, ``{Loop homotopy algebras in closed string field theory},''
  \href{http://dx.doi.org/10.1007/PL00005575}{{\em Commun. Math. Phys.}
  {\bfseries 221} (2001) 367--384},
  \href{http://arxiv.org/abs/hep-th/9711045}{{\ttfamily arXiv:hep-th/9711045}}.

\bibitem{Gwilliam:PhD}
O.~Gwilliam, {\em Factorization algebras and free field theories}.
\newblock PhD Thesis. University of Massachussets.
\newblock \url{https://people.math.umass.edu/~gwilliam/thesis.pdf}.

\bibitem{Chiaffrino:2021pob}
C.~Chiaffrino, O.~Hohm, and A.~F. Pinto, ``{Homological Quantum Mechanics},''
  \href{http://arxiv.org/abs/2112.11495}{{\ttfamily arXiv:2112.11495
  [hep-th]}}.

\end{thebibliography}\endgroup
\bibliographystyle{utphys}

\end{document}